%% file: main.tex
\documentclass[prd,11pt,tightenlines,aps,letterpaper,amsmath,amssymb,preprintnumbers,showpacs,floatfix,longbibliography,nofootinbib,superscriptaddress]{revtex4-2}
\usepackage{graphicx}
\usepackage[export]{adjustbox}
\usepackage[caption=false]{subfig}
\usepackage{bm}  
\usepackage{makecell} 
\allowdisplaybreaks 
\usepackage[hypertexnames=true]{hyperref}
\usepackage[capitalize]{cleveref}
\hypersetup{
    colorlinks=true,       
    linkcolor=blue,          
    citecolor=blue,        
    filecolor=blue,      
    urlcolor=blue           
}

\usepackage{slashed}
\usepackage{mathtools}
\usepackage{listings}
\usepackage{pgf}
\usepackage{fancyvrb}
\usepackage{longtable}
\usepackage{siunitx}
\usepackage{braket}

\usepackage[normalem]{ulem} 
\usepackage{dsfont} 

\usepackage{tikz}
\usetikzlibrary{shapes,arrows,chains}
\usetikzlibrary{calc}
\usetikzlibrary{patterns}
\usetikzlibrary{patterns}
\usetikzlibrary{positioning,decorations.pathmorphing,decorations.markings}

\tikzset{>=latex}
\tikzset{every picture/.style={line width=0.75pt}}

\usepackage{array}
\newcolumntype{P}[1]{>{\centering\arraybackslash}p{#1}}

\usepackage{soul}

\DeclareMathAlphabet{\mathbbb}{U}{bbold}{m}{n}

\definecolor{listinggreen}{rgb}{0,0.6,0}
\definecolor{listinggray}{rgb}{0.5,0.5,0.5}
\definecolor{listingmauve}{rgb}{0.58,0,0.82}
\definecolor{listingkeywordcolor}{rgb}{1.0,0.4,0.0}
\definecolor{listinglightgray}{rgb}{0.8863,0.8863,0.8863}

\newcommand{\dNp}[0]{\delta E^{N^+}}
\newcommand{\dNm}[0]{\delta E^{N^-}}
\newcommand{\DLE}[0]{\Delta_\text{cut}}
\newcommand{\Id}[0]{\mathds{1}}

\newcommand{\lqcd}{\Lambda_{\rm QCD}}

\begin{document}

\preprint{FERMILAB-PUB-24-0126-T}
\preprint{MIT-CTP/5700}

\begin{figure}
  \vskip -1.5cm
  \leftline{\includegraphics[width=0.15\textwidth]{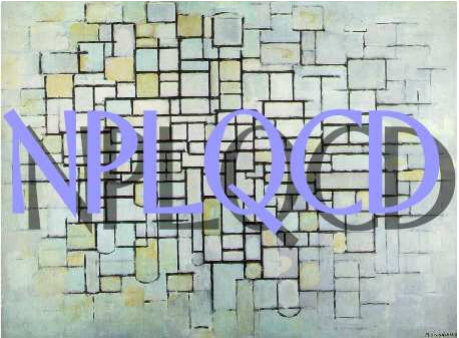}}
\end{figure}

\title{Constraints on the finite volume two-nucleon spectrum at $m_\pi \approx 806$ MeV
}

\author{William Detmold}
\affiliation{Center for Theoretical Physics, Massachusetts Institute of Technology, Cambridge, MA 02139, USA}
\affiliation{The NSF AI Institute for Artificial Intelligence and Fundamental Interactions}
\author{Marc Illa}
\affiliation{InQubator for Quantum Simulation (IQuS), Department of Physics, University of Washington, Seattle, WA 98195}
\author{William I. Jay}
\affiliation{Center for Theoretical Physics, Massachusetts Institute of Technology, Cambridge, MA 02139, USA}
\author{Assumpta Parre\~no}
\affiliation{Departament de F\'isica Qu\`antica i Astrof\'isica and Institut de Ci\`encies del Cosmos, Universitat de Barcelona,  E-08028 Barcelona, Spain}
\author{Robert J. Perry}
\affiliation{Departament de F\'isica Qu\`antica i Astrof\'isica and Institut de Ci\`encies del Cosmos, Universitat de Barcelona,  E-08028 Barcelona, Spain}
\author{Phiala E. Shanahan}
\affiliation{Center for Theoretical Physics, Massachusetts Institute of Technology, Cambridge, MA 02139, USA}
\affiliation{The NSF AI Institute for Artificial Intelligence and Fundamental Interactions}
\author{Michael L. Wagman}
\affiliation{Fermi National Accelerator Laboratory, Batavia, IL 60510, USA}
\collaboration{NPLQCD}

\begin{abstract}
The low-energy finite-volume
spectrum of the two-nucleon system at a quark mass corresponding to a pion mass of $m_\pi \approx 806~\si{MeV}$ is studied with lattice quantum chromodynamics (LQCD) using variational methods. The interpolating-operator sets used in \textbf{[Phys.Rev.D 107 (2023) 9, 094508]} are extended by
including a complete basis of local hexaquark operators, as well as plane-wave dibaryon operators built from products of both positive- and negative-parity nucleon operators.
Results are presented for the  isosinglet and isotriplet two-nucleon channels. In both channels, noticably weaker variational bounds on the lowest few energy eigenvalues are obtained from operator sets which contain only hexaquark operators or operators constructed from the product of two negative-parity nucleons, while other operator sets produce low-energy variational bounds which are consistent within statistical uncertainties.
The consequences of these studies for the LQCD understanding of the two-nucleon spectrum are investigated.  
\end{abstract}

\maketitle

\section{Introduction}
\label{sec:introduction}
Lattice quantum chromodynamics (LQCD) offers the tantalizing prospect of computing quantities in nuclear physics from first principles in a systematically-improvable manner. 
Besides the importance of these calculations to reveal the emergence of nuclear complexity from the Standard Model, such studies provide necessary theoretical inputs for experimental searches for physics beyond the Standard Model using nuclei, where precise theoretical understanding of the nuclear targets used in experiments is required to maximise sensitivity to new  physics~\cite{Detmold:2019ghl,Drischler:2019xuo,Davoudi:2020ngi,Tews:2022yfb,USQCD:2022mmc}. For example, theoretical constraints on nuclear matrix elements are required to interpret the results of dark-matter direct-detection experiments \cite{Prezeau:2003sv,Chang:2017eiq,Krebs:2020plh,deVries:2023hin}, to study lepton-number violation via neutrinoless double-beta decay measurements~\cite{Prezeau:2003xn,Cirigliano:2020dmx,Cirigliano:2022rmf,Davoudi:2024ukx}, and to determine neutrino oscillation parameters precisely from long-baseline neutrino scattering experiments~\cite{NuSTEC:2017hzk,Kronfeld:2019nfb,Ruso:2022qes}. In each case, LQCD can provide essential non-perturbative QCD information which, via matching to nuclear effective field theories (EFTs), can enable a low-energy description of nuclear processes. The constrained EFTs can be paired with many-body methods to make predictions for properties of heavy nuclei that are beyond the scope of present-era LQCD calculations. Grounding these nuclear-physics calculations in QCD is expected to lead to reduced systematic errors arising from nuclear-model uncertainty and thereby increased sensitivity in experimental searches.

For somewhat more than a decade, LQCD calculations of single-hadron masses have achieved few-percent statistical precision and careful control of systematic effects~\cite{Kronfeld:2012uk,Detmold:2019ghl}. 
The success of this program for stable single-hadron systems is aided by large energy gaps, $\delta E$, between the ground state and the lowest excitations because excited-state effects are suppressed by $e^{-\delta E\, t_{\text{max}}}$, where $t_\text{max}\gg \delta E^{-1}$ is the maximum Euclidean time extent for which statistically precise two-point correlation functions can be resolved with available resources.
The achievements in computing the single-hadron spectrum 
have motivated the spectroscopic study of multi-hadron systems, hadron resonances, and nuclei using LQCD~\cite{Beane:2010em,Briceno:2017max,Hansen:2019nir,Drischler:2019xuo,Davoudi:2020ngi,Horz:2022glt,Romero-Lopez:2022usb,Hanlon:2024fjd}. 

The focus of this work will be on two-nucleon systems.
The extraction of the physically-relevant scattering parameters for these systems requires knowledge of energy levels beyond the ground state, which are typically closely spaced.
Below inelastic thresholds, low-energy scattering in multi-hadron systems is described in terms of phase-shifts and mixing angles.
Demonstrating that the scattering amplitudes of multi-hadron systems can be reliably determined from LQCD is an important step towards grounding nuclear-physics calculations in QCD.
While a complete validation of the calculation of these quantities can only be achieved utilizing the physical values of the quark masses, calculations at
heavier-than-physical quark masses provide a useful testbed for developing methods for studying two-nucleon systems since the computational resources required to achieve a given statistical precision decrease exponentially with increasing quark mass~\cite{Parisi:1983ae,Lepage:1989hd,Beane:2009gs}. 
In addition, studies of the dependence of nuclear interactions on the quark masses are
interesting in their own right; this dependence has implications for Big-Bang nucleosynthesis and the stellar production mechanisms of carbon, oxygen, and other elements that are necessary for life~\cite{Beane:2002vs,Epelbaum:2002gb,Beane:2002xf,Braaten:2003eu,Flambaum:2007mj,Bedaque:2010hr,Chen:2010yt,Soto:2011tb,Epelbaum:2012iu}.
Understanding the dependence of resonances and bound-states in QCD-like theories on quark masses, the number of colors, and the number of flavors is also relevant for testing models of strongly-coupled dark matter~\cite{Detmold:2014qqa,Detmold:2014kba,DeGrand:2019vbx}.

Multiple studies of the finite-volume two-nucleon spectrum from LQCD have been undertaken at a range of quark masses~\cite{Fukugita:1994na,Fukugita:1994ve,Beane:2006mx,Beane:2006gf,Beane:2009py,Yamazaki:2009ua,NPLQCD:2011naw,NPLQCD:2012mex,NPLQCD:2013bqy,Wagman:2017tmp,Orginos:2015aya,NPLQCD:2020ozd,NPLQCD:2020lxg,Yamazaki:2012hi,Yamazaki:2015asa,Berkowitz:2015eaa,Francis:2018qch,Amarasinghe:2021lqa,Horz:2020zvv}.
These studies arrive at differing conclusions about the nature of this spectrum, resulting in significant uncertainty about the phase-shifts.
In particular, these calculations arrive at mixed conclusions regarding the existence of bound deuteron and dineutron states at larger-than-physical quark masses. 
Thus, while these proof-of-principle calculations have demonstrated that the application of LQCD to the few-baryon sector is possible, it remains an outstanding challenge to demonstrate that LQCD in the nuclear sector can achieve the statistical precision and systematic control already achieved in the single-hadron sector, particularly given the additional physical and technical subtleties arising in nuclear systems~\cite{Beane:2010em,Detmold:2019ghl}.

The majority of LQCD studies of two-nucleon systems~\cite{Fukugita:1994na,Fukugita:1994ve,Beane:2006mx,Beane:2006gf,Beane:2009py,Yamazaki:2009ua,NPLQCD:2011naw,NPLQCD:2012mex,NPLQCD:2013bqy,Wagman:2017tmp,Orginos:2015aya,NPLQCD:2020ozd,NPLQCD:2020lxg,Yamazaki:2012hi,Yamazaki:2015asa,Berkowitz:2015eaa} have been performed using asymmetric correlation functions in order to minimize the computational costs involved. The results of these calculations implied the presence of bound two-nucleon systems over a range of unphysically-large values of the light-quark masses.
During the same time period, the two-nucleon system was also studied using the potential method~\cite{Luscher:1986pf,Lin:2001ek,CP-PACS:2005gzm,Ishii:2006ec,Murano:2011nz,Aoki:2011gt,Ishii:2012ssm,HALQCD:2019wsz} and the resulting potentials extracted from this approach did not produce bound two-nucleon systems over a similar range of unphysically-heavy quark masses. 

As will be discussed in detail below, LQCD spectroscopy can also be cast as a variational problem from which information about the energy eigenvalues can be obtained~\cite{Fox:1981xz,Michael:1982gb,Luscher:1990ck}. 
While variational calculations to date have 
not revealed the existence of a bound dibaryon  in the two-nucleon system~\cite{Francis:2018qch,Horz:2020zvv,Amarasinghe:2021lqa}, it is important to emphasize that the variational method provides only upper bounds (up to statistical fluctuations) on the energy eigenvalues of the theory.
Thus, while no definitive evidence for two-nucleon bound states has been found in Refs.~\cite{Francis:2018qch,Horz:2020zvv,Amarasinghe:2021lqa}, these results do not, and by definition cannot, rule out the presence of bound dibaryons at larger-than-physical quark masses.

In Ref.~\cite{Amarasinghe:2021lqa}, the low-energy spectrum of two-nucleon systems was studied using a variational approach in a lattice geometry with a cubic spatial volume of side-length $L \approx4.5~\si{fm}$ ($L/a=32$, where $a$ is the lattice spacing) and degenerate light and strange quark masses corresponding to a pion mass of $m_\pi \approx 806~\si{MeV}$. 
The dependence of the low-energy variational bounds on the chosen operator set was studied. While operator sets which included momentum-projected dibaryon operators produced variational bounds which could be identified as shifted non-interacting energy levels, the removal of the $n$th operator from the variational set led to much weaker variational bounds, providing a clear demonstration of the risk in over-interpreting these variational bounds. If the ground-state variational bound is assumed to be saturated, then the results presented in Ref.~\cite{Amarasinghe:2021lqa} do no support the presence of a bound state in either the dineutron ($I=1$) or deuteron ($I=0$) systems.

In the study reported here,  two-nucleon systems are investigated using LQCD with the same lattice action and quark masses as in Ref.~\cite{Amarasinghe:2021lqa}, in a smaller spatial volume with side-length $L\approx 3.4~\si{fm}$ ($L/a=24$).
The variational approach is again used with a range of interpolating-operator sets containing as many as 46 and 31 operators in the isospin-zero and isospin-one channels, respectively.
Sets of interpolating operators are constructed that contain dibaryon operators built from products of plane-wave nucleon operators, local hexaquark operators, and quasi-local operators inspired by low-energy nuclear EFTs.
New operators not considered previously are constructed, including dibaryon operators involving negative-parity quark spinor components (``lower-spin components" in the Dirac basis) and those built from products of two negative-parity nucleon operators.
Complete bases of local hexaquark operators with isospin-zero and isospin-one are also constructed.
This extends the operator set considered in Ref.~\cite{Amarasinghe:2021lqa} by the inclusion of operators which cannot be written
as the local product of two color-singlet three-quark operators with the quantum numbers of baryons, and therefore have the possibility of probing so-called ``hidden color"~\cite{Brodsky:1983vf,Miller:1984twp,Bashkanov:2013cla,Miller:2013hla} components of dibaryon states in QCD. 
As will be seen, the additional operators considered in this work do not produce large differences in the observed low-energy variational bounds.

The remainder of the paper is organized as follows.
Section~\ref{sec:hadron_spectroscopy} introduces notation and discusses techniques for hadron spectroscopy in LQCD.
Section~\ref{sec:operators} summarizes the full set of interpolating operators appearing in this work.
Section~\ref{sec:numerical_study} presents numerical results for both the dineutron and deuteron channels from a selection of different operator sets.
Section~\ref{sec:discussion} discusses the results and Sec.~\ref{sec:conclusion} presents conclusions.

\section{Hadron Spectroscopy}
\label{sec:hadron_spectroscopy}
In this section, the general principles of hadron spectroscopy and the variational method in LQCD~\cite{Michael:1982gb,Luscher:1990ck} are summarized. 
The starting point for all studies of hadron spectroscopy using LQCD is a two-point correlation function, which may be generically defined as
\begin{equation}
\label{eq:corr_function}
C_{\chi\chi^\prime}(t)=\braket{\Omega|\mathcal{O}_\chi(t)\mathcal{O}_{\chi^\prime}^\dagger(0)|\Omega},
\end{equation}
where $\mathcal{O}_{\chi^\prime}^{\dagger}(0)$ and $\mathcal{O}_\chi(t)$ are referred to as the source and sink interpolating operators, respectively, 
and $\ket{\Omega}$ denotes the vacuum state.\footnote{For simplicity, an infinite temporal extent is assumed throughout this work. See Ref.~\cite{Schiel:2015kwa} for a discussion of finite temporal extent effects in variational calculations.} The indices $\chi$ and $\chi^\prime$ label the interpolating operators, which are chosen to possess the quantum numbers of the states of interest. 
Explicit labels corresponding to these quantum numbers are used below but suppressed for clarity in this section.
Operators can also be labeled by other non-conserved quantities such as the relative momentum or separation between two components of the operator.
By the orthogonality of the energy eigenstates, the operators only project onto states commensurate with the quantum numbers of both the source and sink interpolating operators. Consequently, correlation functions admit a spectral decomposition, which for theories with an Euclidean metric take the form of a sum of decaying exponentials,
\begin{equation}
  C_{\chi\chi^\prime}(t)=\sum_{\mathsf{n}=0}^\infty Z_{\mathsf{n}\chi}Z_{\mathsf{n}\chi^\prime}^{*}\, e^{-t E_{\mathsf{n}}}\ ,
\end{equation}
where the sum is over all energy eigenstates, $\ket{\mathsf{n}}$, with the requisite quantum numbers.
The index $\mathsf{n}$ orders the states such that the corresponding energy eigenvalues\footnote{In this work, the vacuum energy, $E_\Omega$ is set to zero.} satisfy $E_{\mathsf{n}}\leq E_{\mathsf{m}}$ 
for $\mathsf{n}<\mathsf{m}$. The overlap factors, $Z_{\mathsf{n}\chi}$, are given by
\begin{equation}
\label{eq:overlap}
Z_{\mathsf{n}\chi}=\braket{\Omega|\mathcal{O}_{\chi}(0)|\mathsf{n}}.
\end{equation}
Relative overlap factors,
\begin{equation}
\mathcal{Z}_{\mathsf{n}\chi}=\frac{|Z_{\mathsf{n}\chi}|}{\sum_{\chi^\prime}|Z_{\mathsf{n}\chi^\prime}|}\ ,
\end{equation}
can be used to identify the state or states with which a particular operator $\mathcal{O}_\chi$ has large overlap, although these quantities depend explicitly on the full set of operators that are considered in a given calculation.
In general, there is no constraint on the complex phase of the overlap factors, and thus the above correlation functions are not real in general. 
However, when the source and sink interpolating-operators are related by Hermitian conjugation, the resulting correlation function is real-valued and
positive,
\begin{equation}
C_{\chi\chi}(t)=\sum_{\mathsf{n}=0}^\infty |Z_{\mathsf{n}\chi}|^2e^{-t E_{\mathsf{n}} }>0\ .
\end{equation}

The Euclidean-time dependence of such correlation functions provides information about the low-energy spectrum in sectors of fixed quantum numbers.
In particular, at sufficiently large Euclidean time, the  correlation function in Eq.~\eqref{eq:corr_function} is dominated by the lowest-energy eigenstate with non-zero overlap with the chosen operators. 

\subsection{The Variational Method}
\label{sec:variational_method}

Due to the exponential degradation of the signal-to-noise ratio observed in most numerical LQCD calculations of two-point correlation functions at large Euclidean times~\cite{Lepage:1989hd,Parisi:1983ae}, and in particular for systems with non-zero baryon number~\cite{Beane:2009kya,Beane:2009gs,Beane:2009py,Endres:2011jm,NPLQCD:2012mex,Grabowska:2012ik,Beane:2014oea,Wagman:2016bam,Wagman:2017xfh}, the Euclidean time range
where the low-energy eigenstates provide the largest contributions is difficult to access. 
It is therefore desirable to construct interpolating operators which overlap strongly with a particular state in the spectrum (although large overlaps do not necessarily minimize statistical noise~\cite{Detmold:2014hla}).
To this end, the variational method~\cite{Michael:1982gb,Luscher:1990ck} begins by choosing $N$ operators with the quantum numbers of the system being studied, $\{\mathcal{O}_1,\dots,\mathcal{O}_N\}$. By computing the quantities given in Eq.~\eqref{eq:corr_function} for all ${\chi}, {\chi^\prime} \in \{1,\dots,N\}$, an $N\times N$ matrix of correlation functions with elements $C_{\chi \chi^\prime}(t)$ 
can be constructed. To investigate the spectrum, one solves the  generalized eigenvalue problem  (GEVP) given by
\begin{equation}
\sum_{\chi^\prime} C_{\chi\chi^\prime}(t)v_{\mathsf{n}\chi^\prime}(t,t_0)=\lambda_{\mathsf{n}}(t,t_0)\sum_{\chi^\prime} C_{\chi\chi^\prime}(t_0)v_{\mathsf{n}\chi^\prime}(t,t_0)\ , \label{eq:GEVP}
\end{equation}
where $t_0$ is a chosen reference time, $v_{\mathsf{n}\chi'}(t,t_0)$ are the components of the eigenvector corresponding to the $\mathsf{n}$th eigenvalue,
\begin{equation}\label{eq:eigenvalue}
\lambda_{\mathsf{n}}(t,t_0)= e^{-(t-t_0)E_{\mathsf{n}}(t,t_0)}\ ,
\end{equation}
and $E_{\mathsf{n}}(t,t_0)$ are time-dependent effective masses extracted from the logarithm of Eq.~\eqref{eq:eigenvalue}. 
The index $\mathsf{n}$ orders the eigenvalues such that $\lambda_{\mathsf{n}}\geq\lambda_{\mathsf{m}}$ (and hence that $E_{\mathsf{n}}(t,t_0)\leq E_{\mathsf{m}}(t,t_0)$) for $\mathsf{n}<\mathsf{m}$. Note that this labelling is chosen such that the largest eigenvalue corresponds to the smallest time-dependent effective mass.

The eigenvectors $\vec{v}_{\mathsf{n}}(t,t_0)$ can be used to define overlap-optimized sets of interpolating operators 
that provide $N$ variational bounds on the lowest energy eigenvalues.
In particular,
\begin{equation}
  \psi_{\mathsf{n}}(t,t_0,t_\text{ref})=\sum_{\chi} v_{\mathsf{n}\chi}(t_\text{ref},t_0)\mathcal{O}_\chi(t)\ , \quad \mathsf{n}\in \{0,1,\dots, N-1\} \ ,
\end{equation}
is an interpolating operator whose overlap onto the $\mathsf{n}$th energy eigenstate is maximized within the set of operators considered. Both $t_0$ and $t_\text{ref}$ are  Euclidean times which may be chosen freely. 
With this set of overlap-optimized operators, a set of $N$ correlation functions known as the ``principal correlation functions" can be computed~\cite{Fox:1981xz,Michael:1982gb,Luscher:1990ck},
\begin{equation}
\label{eq:principal_correlators}
\hat{C}_{\mathsf{n}}(t,t_0,t_\text{ref})=\braket{0|\psi_{\mathsf{n}}(t,t_0,t_\text{ref})\psi_{\mathsf{n}}^{\dagger}(0,t_0,t_\text{ref})|0} \ , \quad \mathsf{n}\in \{0,1,\dots, N-1\} \ .
\end{equation}
These can be expressed in terms of the original correlation matrix as
\begin{equation}
  \hat{C}_{\mathsf{n}}(t,t_0,t_\text{ref})=\sum_{\chi,\chi^\prime} v_{\mathsf{n}\chi}(t_\text{ref},t_0)C_{\chi\chi^\prime}(t)v_{\mathsf{n}\chi^\prime}^{\dagger}(t_\text{ref},t_0) \ . 
\end{equation}
The principal correlation functions also admit a spectral decomposition, which is guaranteed to be positive-definite and convex up to statistical fluctuations,
\begin{equation}
  \hat{C}_{\mathsf{n}}(t,t_0,t_\text{ref})=\sum_{\mathsf{m}=0}^\infty |Z_{\mathsf{m}\mathsf{n}}(t_0,t_\text{ref})|^2e^{-t E_{\mathsf{m}}} \ ,
\end{equation}
where $Z_{\mathsf{m}\mathsf{n}}(t_0,t_\text{ref})=\braket{\Omega|\psi_\mathsf{n}(0,t_0,t_\text{ref})|\mathsf{m}}$.
One important property of the above GEVP lies in the constraints resulting from the Cauchy interlacing theorem in the infinite statistics limit.\footnote{The relevance of the interlacing theorem 
\cite{Hwang:2024,Horn:1985,doi:10.1137/1.9781421407944,Hwang:2024}, which is equivalent to the Poincar\'e separation theorem, to LQCD-spectroscopy calculations was recently highlighted in Ref.~\protect\cite{Fleming:2023zml}, but is implicit in earlier discussions \cite{Fox:1981xz,Michael:1982gb,Luscher:1990ck}.}
In particular, it is possible to make a rigorous statement about the minimum number of energy eigenvalues below a particular effective mass, $E_{\sf n}(t,t_0)$, extracted from the GEVP.
As this is a key component of the results presented here, it is helpful to review the origin of these constraints.
In this discussion, it is useful to distinguish the GEVP eigenvalues from the energy eigenvalues of the QCD Hamiltonian.
In this section, the term ``eigenvalue" is used to refer to an eigenvalue of the GEVP, while ``energy eigenvalue" is used to refer to an eigenvalue of the LQCD Hamiltonian.\footnote{These energy eigenvalues may be defined as the negative logarithms of the eigenvalues of the transfer matrix.} 
In matrix-vector notation, the GEVP in \cref{eq:GEVP} above can be written as
\begin{equation}
\mathbf{C}(t)\vec{v}_{\mathsf{n}}(t,t_0)=\lambda_{\mathsf{n}}(t,t_0) \mathbf{C}(t_0)\vec{v}_{n}(t,t_0)\ ,
\end{equation}
where $\mathbf{C}(t)$ is an $N\times N$ matrix with components $C_{\chi\chi'}(t)$.
This may be transformed into an eigenvalue problem of the form
\begin{equation}
\mathbf{B}(t,t_0)\vec{v}_{\mathsf{n}}(t,t_0)=\lambda_{\mathsf{n}}(t,t_0) \vec{v}_{\mathsf{n}}(t,t_0)\ ,
\end{equation}
where $\mathbf{B}(t,t_0)=\mathbf{C}^{-1}(t_0)\mathbf{C}(t)$. 

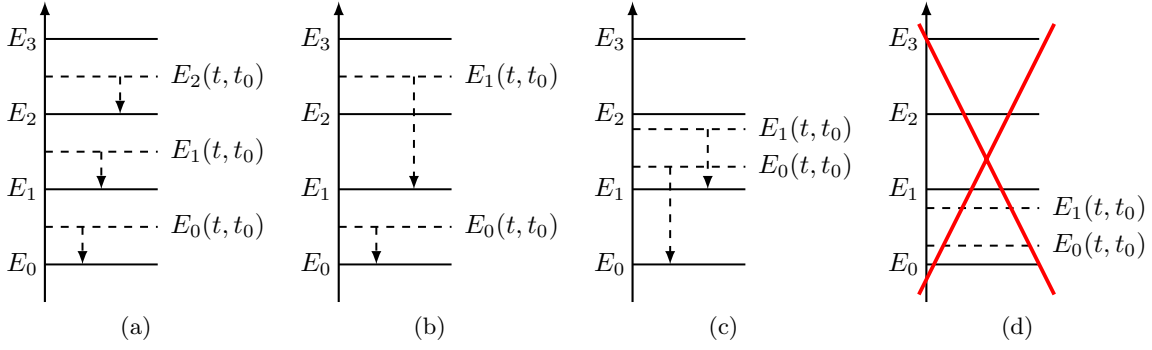
\begin{figure}
\centering
    \subfloat[]{
\begin{tikzpicture}
\draw[->,thick] (0,0) -- (0,4);
\draw (0,0.5) -- (1.5,0.5);
\draw (0,1.5) -- (1.5,1.5);
\draw (0,2.5) -- (1.5,2.5);
\draw (0,3.5) -- (1.5,3.5);
\node at (-0.3,3.5) {$E_3$};
\node at (-0.3,2.5) {$E_2$};
\node at (-0.3,1.5) {$E_1$};
\node at (-0.3,0.5) {$E_0$};
\draw[dashed] (0,1) -- (1.5,1);
\draw[dashed] (0,2) -- (1.5,2);
\draw[dashed] (0,3) -- (1.5,3);
\draw[dashed,->] (0.5,1) -- (0.5,0.5);
\draw[dashed,->] (0.75,2) -- (0.75,1.5);
\draw[dashed,->] (1,3) -- (1,2.5);
\node at (2.3,3) {$E_2(t,t_0)$};
\node at (2.3,2) {$E_1(t,t_0)$};
\node at (2.3,1) {$E_0(t,t_0)$};
\end{tikzpicture}
}
\subfloat[]{
\begin{tikzpicture}
\draw[->,thick] (0,0) -- (0,4);
\draw (0,0.5) -- (1.5,0.5);
\draw (0,1.5) -- (1.5,1.5);
\draw (0,2.5) -- (1.5,2.5);
\draw (0,3.5) -- (1.5,3.5);
\draw[dashed,->] (0.5,1) -- (0.5,0.5);
\draw[dashed,->] (1,3) -- (1,1.5);
\node at (-0.3,3.5) {$E_3$};
\node at (-0.3,2.5) {$E_2$};
\node at (-0.3,1.5) {$E_1$};
\node at (-0.3,0.5) {$E_0$};
\draw[dashed] (0,1) -- (1.5,1);
\draw[dashed] (0,3) -- (1.5,3);
\node at (2.3,3) {$E_1(t,t_0)$};
\node at (2.3,1) {$E_0(t,t_0)$};
\end{tikzpicture}
}
\subfloat[]{
\begin{tikzpicture}
\draw[->,thick] (0,0) -- (0,4);
\draw (0,0.5) -- (1.5,0.5);
\draw (0,1.5) -- (1.5,1.5);
\draw (0,2.5) -- (1.5,2.5);
\draw (0,3.5) -- (1.5,3.5);
\draw[dashed,->] (0.5,1.8) -- (0.5,0.5);
\draw[dashed,->] (1,2.3) -- (1,1.5);
\node at (-0.3,3.5) {$E_3$};
\node at (-0.3,2.5) {$E_2$};
\node at (-0.3,1.5) {$E_1$};
\node at (-0.3,0.5) {$E_0$};
\draw[dashed] (0,1.8) -- (1.5,1.8);
\draw[dashed] (0,2.3) -- (1.5,2.3);
\node at (2.3,2.3) {$E_1(t,t_0)$};
\node at (2.3,1.8) {$E_0(t,t_0)$};
\end{tikzpicture}
}
\subfloat[]{
\begin{tikzpicture}
\draw[->,thick] (0,0) -- (0,4);
\draw (0,0.5) -- (1.5,0.5);
\draw (0,1.5) -- (1.5,1.5);
\draw (0,2.5) -- (1.5,2.5);
\draw (0,3.5) -- (1.5,3.5);
\node at (-0.3,3.5) {$E_3$};
\node at (-0.3,2.5) {$E_2$};
\node at (-0.3,1.5) {$E_1$};
\node at (-0.3,0.5) {$E_0$};
\draw[dashed] (0,0.75) -- (1.5,0.75);
\draw[dashed] (0,1.25) -- (1.5,1.25);
\node at (2.3,1.25) {$E_1(t,t_0)$};
\node at (2.3,0.75) {$E_0(t,t_0)$};
\draw[red, line width=0.5mm] 
    (-0.1,0.1) -- (1.7,3.7)
    (-0.1,3.7) -- (1.7,0.1);
\end{tikzpicture}
}

\caption{Realizations of the interlacing theorem. Solid horizontal lines represent the true energy-eigenvalues of the LQCD system, while the dashed lines represent the locations of observed effective masses which are consistent (a, b, c) or inconsistent (d) with the interlacing theorem. Arrows indicate the largest energy-eigenvalue for which the effective mass serves as a variational bound for. Note that there exists at least one true energy-eigenvalue between each pair of variational bounds. The ``best case" scenario for the variational method is shown in (a), where there is exactly one energy eigenvalue between each variational bound, however, the situations shown in (b) or (c) are also possible. Panel (d) shows an impossible scenario in which there are two variational bounds below the first excited state energy.}
\label{fig:interlacing_theorem}
\end{figure}

If there are $K$ states in the spectrum of a given theory, solving this system for a correlation-function matrix constructed from $K$ independent operators would enable an extraction of the energies of the $K$ states. 
However, LQCD formally possesses an infinite dimensional Hilbert space~\cite{Kogut:1974ag}.
Even in practice, working with finite-precision floating-point numbers means that the (finite) number of states, $K$, numerically relevant to observables with a given set of quantum numbers is far larger than the size of any practicably-realizable matrix of correlation functions. Consequently, the eigenvalues obtained from an $N\times N$ correlation-function matrix do not correspond to the eigenvalues of the full $K\times K$ matrix for $N<K$.
To understand what can be learnt from the $N\times N$ correlation-function matrix, let $\mathbf{A}$ be a $K\times K$ correlation-function matrix constructed from $K$ independent interpolating operators with eigenvalues $\{\alpha_0, \dots, \alpha_{K-1}\}$.
In this case, the GEVP eigenvalues of $\mathbf{A}$ are related to the energy eigenvalues, $E_{\mathsf{n}}$, of the LQCD Hamiltonian as
\begin{equation}
\alpha_{\mathsf{n}}(t,t_0)= e^{-(t-t_0)E_{\mathsf{n}}}\ ,  
\end{equation}
where the eigenvalues are ordered  such that $\alpha_{\mathsf{n}}(t,t_0)\geq \alpha_{\mathsf{m}}(t,t_0)$ for $\mathsf{n}<\mathsf{m}$ and it is important to emphasize that the $E_{\mathsf{n}}$ are energy eigenvalues rather than the time-dependent effective masses in Eq.~\eqref{eq:eigenvalue}. 
Let $\mathbf{B}$ be a principal sub-matrix of $\mathbf{A}$ obtained by removing both the $i$th rows and $i$th columns for some values of $i\in\{i_1,\ldots,i_{K-N}\}$.
In this situation, the interlacing theorem states that the eigenvalues of $\mathbf{B}(t,t_0)$, $\lambda_{\mathsf{n}}(t,t_0)$ for $\mathsf{n}\in\{0,\ldots,N-1\}$ ordered such that $\lambda_{\mathsf{n}}(t,t_0)\geq \lambda_{\mathsf{m}}(t,t_0)$ for $\mathsf{n}<\mathsf{m}$,  obey the inequalities\footnote{Note that the lower bound in the right-hand inequality in Eq.~\eqref{eq:interlacing} is only usefully constraining if $K$ is not much larger than $N$.}
\begin{equation}
\alpha_{\mathsf{n}}(t,t_0)\geq \lambda_{\mathsf{n}}(t,t_0)\geq \alpha_{\mathsf{n}+K-N}(t,t_0)\ , \quad
\forall {\sf n}\in\{0,\ldots,N-1\} \ .
\label{eq:interlacing}
\end{equation}
From Eq.~\eqref{eq:eigenvalue}, it is clear that the largest eigenvalue, $\lambda_0(t,t_0)$, corresponds to the smallest effective-mass function. In this case, the bounds in Eq.~\eqref{eq:interlacing} read 
$\alpha_0(t,t_0)\geq \lambda_0(t,t_0)\geq \alpha_{K-N}(t,t_0)$.
From the left-hand inequality, there is \textit{at least} one eigenvalue greater than or equal to the variational eigenvalue $\lambda_0(t,t_0)$. 
By the monotonicity of the logarithm, this implies that there exists at least one energy-eigenvalue less than or equal to the lowest effective mass $E_0(t,t_0)$ for any $t$ or $t_0$.
The second largest eigenvalue of $\mathbf{B}(t,t_0)$, $\lambda_{1}(t,t_0)$, corresponds to a variational bound of the first excited energy-eigenvalue. In this case, the bounds read $\alpha_{1}(t,t_0)\geq \lambda_{1}(t,t_0)\geq \alpha_{K-N+1}(t,t_0)$. 
Since there is clearly one additional state with a larger eigenvalue, $\alpha_0(t,t_0)$, this implies that there are \textit{at least} two energy-eigenvalues less than or equal to the effective mass $E_1(t,t_0)$. 
This argument can be iterated to show that there exist \textit{at least} $(\mathsf{n}+1)$ energy-eigenvalues less than or equal to $E_{\mathsf{n}}(t,t_0)$ for any choice of $t$ and $t_0$.
Various realizations of this statement are shown in Fig.~\ref{fig:interlacing_theorem}. 
Due to the above eigenvalue orderings, the interlacing theorem provides a rigorous lower bound on the number of states at or below the $n$-th principle correlation-function effective mass. It is important to note that this is true regardless of the time-dependence of the effective masses, and that this counting theorem holds irrespective of whether the effective mass exhibits time-dependence or not.
Further, the interlacing theorem does not require any assumptions about how the spans of various interpolating-operator sets are embedded within Hilbert space.
The interlacing theorem (and  variational bounds in general) is valid only in the limit when the path integrals which define the correlation-function matrix are evaluated exactly. When Monte-Carlo importance-sampling is used to estimate these integrals, the bounds only hold in a statistical sense, with violations allowed at any finite statistical-precision.

\subsection{Principal correlation function definition}

Principal correlation functions can be defined either by using GEVP eigenvectors at a fixed reference time or by using time-dependent GEVP eigenvalues, with both definitions having distinct advantages.
The eigenvector-based definition in Eq.~\eqref{eq:principal_correlators} leads to principal correlation functions that are linear combinations of LQCD correlation functions and therefore have simple spectral representations, even when an interpolating-operator set spans a small subset of Hilbert space~\cite{Bulava:2010yg,Bulava:2016mks}.
In particular, they are symmetric, so they can be expressed rigorously as sums of exponentials with positive-definite coefficients and modelled by truncations of these sums.
However, it is the eigenvalues and their associated effective masses that constitute rigorous (stochastic) upper bounds on the energy eigenvalues.
To ensure that this property also holds for the principal correlation functions defined in Eq.~\eqref{eq:principal_correlators}, it suffices to choose $t_0$ and $t_{\text{ref}}$ so that the effective masses obtained using both definitions agree within uncertainties.
This condition can be achieved in (sufficiently precise) practical LQCD calculations because the eigenvectors become independent of $t_{\text{ref}}$ and $t_0$ when both parameters are taken sufficiently large.
It can therefore be used as a starting point for an algorithmic definition of $t_0$ and $t_{\text{ref}}$ for which $\hat{C}_{\mathsf{n}}(t,t_0,t_{\text{ref}})$ achieves the simultaneous benefits of having a positive-definite spectral representation and having eigenvalues that satisfy the interlacing theorem.

To make this definition precise, the effective mass function for the $\mathsf{n}$th eigenvector-based principal correlation function is defined as
\begin{equation}
  a E_{\mathsf{n}}(t,t_0,t_{\text{ref}}) = \ln \left[ \frac{\hat{C}_{\mathsf{n}}(t,t_0,t_{\text{ref}}) }{ \hat{C}_{\mathsf{n}}(t+1,t_0,t_{\text{ref}})} \right]\ ,
  \label{eq:emPC}
\end{equation}
and the effective mass obtained from the GEVP eigenvalues
is defined as~\cite{Blossier:2009kd}
\begin{equation}
a F_{\mathsf{n}}(t) =  \ln \left[ \frac{\lambda_{\mathsf{n}}(t,\lfloor t/2\rfloor ) }{ \lambda_{\mathsf{n}}(t+1,\lfloor t/2\rfloor)} \right]\ .
\label{eq:emEV}
\end{equation}
where $\lfloor\cdot\rfloor$ denotes the floor function evaluated for the argument expressed in lattice units. These definitions should coincide when an interpolating-operator set approximately spans the set of states which make statistically-resolvable contributions to correlation functions with separations $t_0$ and $t_{\text{ref}}$.
By varying $t_0$ and $t_{\text{ref}}$, a region of sufficiently large $t_0$ and $t_{\text{ref}}$ can be identified where $E_{\mathsf{n}}(t)$ and $F_{\mathsf{n}}(t)$ agree within statistical uncertainties.
Within this region, results are insensitive to $t_0$ and $t_{\text{ref}}$ by construction.
However, as $t_0$ and/or $t_{\text{ref}}$ increase, statistical noise will  eventually overwhelm the signal for the correlation function and results will become unreliable.
This noise can be diagnosed by failures of the central values of correlation functions to satisfy properties that must hold in the infinite statistics limit, in particular the positivity of GEVP eigenvalues and monotonicity of differences between $E_{\mathsf{n}}(t,t_0,t_{\text{ref}})$ and $F_{\mathsf{n}}(t)$ for large arguments.
An algorithm for choosing the largest $t_0$ and $t_{\text{ref}}$ where there is statistical agreement between $E_{\mathsf{n}}(t,t_0,t_{\text{ref}})$ and $F_{\mathsf{n}}(t)$ and the signal is larger than the noise
is presented in Appendix~\ref{app:t0tref}.

\section{Interpolating Operators}
\label{sec:operators}
\begin{figure}
    \centering
    \includegraphics[width=0.7\columnwidth]{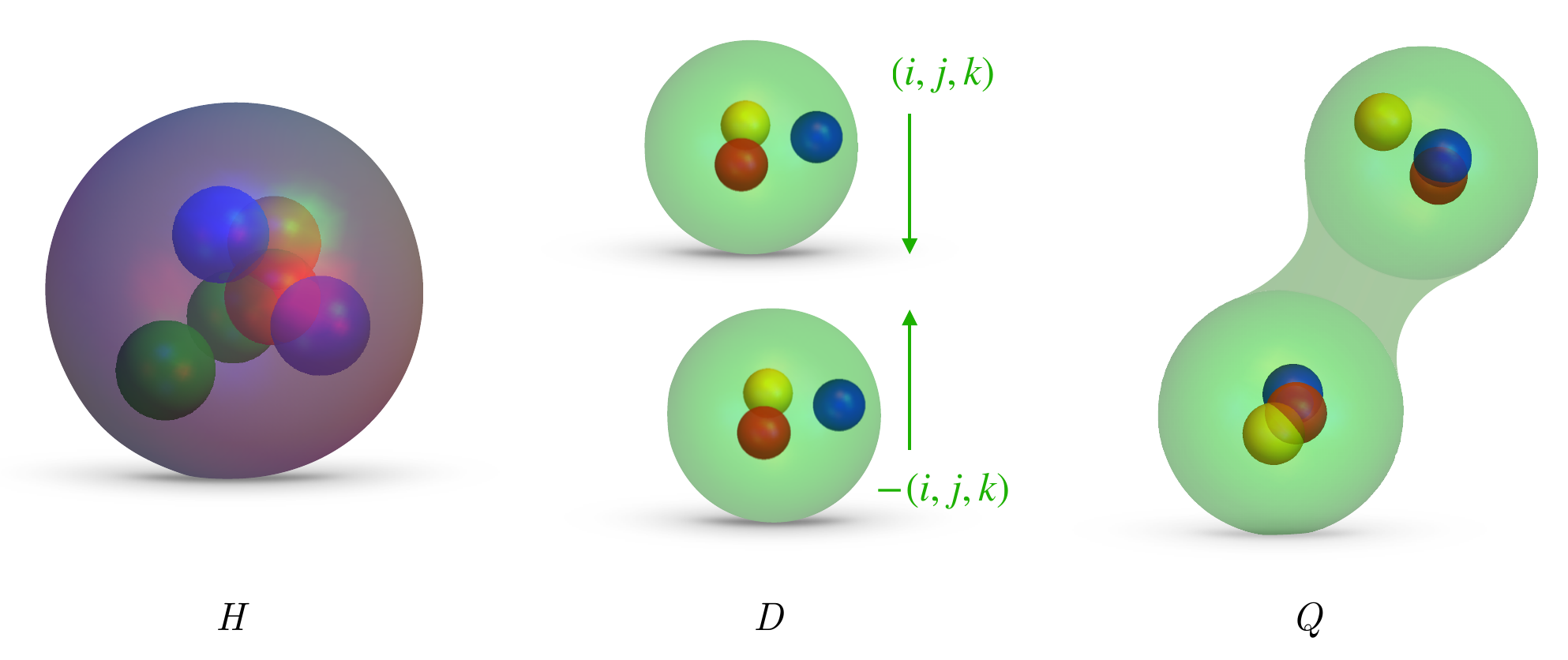}
    \caption{Graphical depiction of the three types of operators included in the numerical calculation. From left to right: hexaquark ($H$), dibaryon ($D$) and quasi-local ($Q$) operators. The relative momentum between the two nucleons in the dibaryon operators is labelled by $(i,j,k)$.}
    \label{fig:operators-cartoon}
\end{figure}

This section discusses extensions of the interpolating-operator sets used in Ref.~\cite{Amarasinghe:2021lqa} through the inclusion of one- and two-nucleon operators constructed using additional spin and color structures.
The spatial structures of the two-nucleon operators considered are the same as in Ref.~\cite{Amarasinghe:2021lqa}: hexaquark ($H$) operators are constructed from products of six quark fields centered at the same point, dibaryon ($D$) operators are constructed from products of plane-wave nucleon operators, and quasi-local ($Q$) operators are constructed from products of nucleon operators with relative wavefunctions that resemble bound-state wavefunctions in finite-volume EFT. These operator types are shown pictorially in Fig.~\ref{fig:operators-cartoon}.

\subsection{Single-Nucleon Operators}

The one- and two-nucleon operators used in this study 
can be described conveniently using diquark fields 
\begin{equation}
  \mathcal{D}^{ab}_{\Gamma,F}(x) =  \frac{1}{\sqrt{2}} q^{a T}(x) C \Gamma i\tau_2 F q^b(x)\ , \label{eq:diquark}
\end{equation}
where $q = (u, d)^T$ is an isodoublet quark field, $C = \gamma_2 \gamma_4$ is the Euclidean charge conjugation matrix,  $\tau_2 = \left(\begin{smallmatrix} 0& -i \\ i&0 \end{smallmatrix}\right)$ is a Pauli matrix in isospin space, $F$ and $\Gamma$ are flavor and spin matrices, and $a$ and $b$ are color indices. 
Proton and neutron operators can be built from isosinglet diquarks
\begin{equation}
  \mathcal{D}^{ab}_{\Gamma,\Id}(x) = \frac{1}{\sqrt{2}} \left[ u^{a T}(x) C \Gamma d^b(x) - d^{a T}(x) C \Gamma u^b(x) \right]\ ,
\end{equation}
where $\Id$ denotes the flavor identity matrix. In particular,
\begin{align}
  p_{\sigma}^{\Gamma}(x) = \epsilon_{abc}\mathcal{D}^{ab}_{\Gamma,\Id}(x) P_{\sigma} u^c(x) \ ,\\
  n_{\sigma}^{\Gamma}(x) = \epsilon_{abc}\mathcal{D}^{ab}_{\Gamma,\Id}(x) P_{\sigma} d^c(x) \ , 
\end{align}
where $P_{\sigma}$ projects the quark spin to a specific row $\sigma \in \{0,\ldots,3\}$ of the spinor representation of $SO(4)$.
In the Dirac basis, these projectors can be defined using parity projectors $P_{\pm} = ( \Id \pm \gamma_4)/2$ as
\begin{equation}
  \begin{split}
    P_\sigma = \begin{cases} P_{+} \left(  \Id - (-1)^\sigma i \gamma_1 \gamma_2 \right)/2, & \sigma \in \{0,1\}\ , 
    \\ 
    P_{-} \left( \Id - (-1)^\sigma i \gamma_1 \gamma_2 \right)/2, & \sigma \in \{2,3\}\ .
    \end{cases}
  \end{split}
\end{equation}
The isodoublet nucleon field $N = (p, n)^T$ is defined as 
\begin{equation}
  N_{\sigma}^{\Gamma}(x) = \epsilon_{abc}\mathcal{D}^{ab}_{\Gamma,\Id}(x) P_{\sigma} q^c(x) \ .
\end{equation}

In addition to using the isosinglet diquark shown above, $I=1/2$ nucleon operators can be constructed using an isovector diquark $\mathcal{D}^{ab}_{\Gamma,\tau_A}$, where $A \in \{1,2,3\}$ is an adjoint isospin index. 
However, as shown in Ref.~\cite{Basak:2005ir}, quark antisymmetry can be used to relate the isovector diquark operators (which correspond to mixed-symmetric operators in Ref.~\cite{Basak:2005ir}) to linear combinations of isoscalar diquark (mixed-antisymmetric) operators. 
A complete basis of local nucleon operators can therefore be obtained from $N_{\sigma}^\Gamma$ with all linearly independent choices of $\Gamma$ leading to nucleon quantum numbers, and is given by the three positive-parity and the three negative-parity operators presented in Ref.~\cite{Basak:2005ir}.
In the numerical calculations below, 
two positive-parity nucleon operators, given by
\begin{align}
    N_{\sigma}^{\gamma_5 P_+}(x) = \epsilon_{abc} \mathcal{D}^{ab}_{\gamma_5 P_+,\Id}(x) P_\sigma q^c(x)\ , \\
    N_{\sigma}^{\gamma_5 P_-}(x) = \epsilon_{abc} \mathcal{D}^{ab}_{\gamma_5 P_-,\Id}(x) P_\sigma q^c(x)\ ,
\end{align}
are included.
The operator $N_{\sigma}^{\gamma_5 P_+}$ involves only the Dirac basis upper components $\sigma \in \{0,1\}$  of the quark field and corresponds to the operator used in Ref.~\cite{Amarasinghe:2021lqa}, while the operator $N_{\sigma}^{\gamma_5 P_-}$ also involves the lower components $\sigma \in \{2,3\}$ in the diquark field (the nucleon spin is still restricted to $\sigma \in \{0,1\}$).
Products of these operators will be used to build dibaryon operators as described below.
Positive-parity dibaryon operators can also be constructed from products of two negative-parity nucleon operators such as 
\begin{align}
\label{eq:nucleon_negative_parity}
    N_{\sigma}^{\Id}(x) = \epsilon_{abc} \mathcal{D}^{ab}_{ \Id,\Id}(x) P_\sigma q^c(x)\ . 
\end{align}
The three linearly-independent nucleon operators that are not studied in this work are expected to overlap predominantly with excited states outside the low-energy region that is the focus of this work~\cite{Sasaki:2001nf,Melnitchouk:2002eg,Brommel:2003jm,Basak:2007kj}.

\subsection{Dibaryon Operators}
\label{sec:dibaryon_ops}
Dibaryon operators are constructed from products of two-nucleon operators that are individually projected to definite momentum.
Dibaryon operators with zero total three-momentum are defined by
\begin{equation}
\label{eq:dibaryon}
  D^{(2,I)\Gamma}_{\rho}(\vec{n},t) = \sum_{\vec{x}_1,\vec{x}_2 \in \Lambda_{\mathcal{S}}} 
  e^{i{2\pi\over L} \vec{n}\cdot(\vec{x}_1-\vec{x}_2) } \sum_{\sigma,\sigma',u,u'} v_{\rho}^{\sigma\sigma'}  P_{uu'}^{(I)} N^\Gamma_{\sigma u}(\vec{x}_1,t)  N^\Gamma_{\sigma' u'}(\vec{x}_2,t)\ ,
\end{equation} 
where $x=(\vec{x},t)$ are lattice coordinates whose components are integer multiple of the lattice spacing $a$, $L$ is the spatial extent of the lattice geometry, 
$\Lambda_{\mathcal{S}}$ is a sparse sublattice with $L/(a\mathcal{S})$ sites in each dimension that is introduced to make the volume sums computationally tractable as described in Refs.~\cite{Detmold:2019fbk,Li:2020hbj,Amarasinghe:2021lqa}.
The index
$\rho \in \{0,\ldots,3\}$ labels spin-singlet ($\rho=0$) and spin-triplet ($\rho \in \{1,2,3\}$) dibaryon operators, and $v_{\rho}^{\sigma\sigma'}$ denotes Clebsch-Gordan coefficients (explicitly presented in Ref.~\cite{Amarasinghe:2021lqa}) projecting the product of two spin-1/2 operators into the particular dibaryon spin state. 
Quantum numbers, here baryon number $B=2$ and total isospin $I$, are denoted by superscripts in parantheses.
Projection to operators with definite isospin is accomplished using $P^{(0)} = i\tau_2$ and $P^{(1)} = i\tau_2 \tau_3$, where $I_z=0$ is chosen for simplicity and $u,u'$ are flavor indices.

The set of dibaryon operators used here extends that of Ref.~\cite{Amarasinghe:2021lqa}
by including dibaryon operators with $\Gamma =  \gamma_5 P_-$ and $\Gamma = \Id$ in addition to $\Gamma =  \gamma_5 P_+$.
These dibaryon operators all have the same quantum numbers because products of two negative-parity nucleon operators ($\Gamma = \Id$) have positive parity.
Analogous dibaryon operators can also be constructed in cases where the two nucleons each have different spin structures, as well as from other products of two-baryon operators, such as $\Delta \Delta$ and $N\Delta$, but the construction of this larger operator set is beyond the scope of this work.

For each $\Gamma$, operators with relative momenta $\vec{k} = (2\pi/L)\vec{n}$ are included with
\begin{equation}
  \vec{n} \in \big\{ (0,0,0),\ (0,0,1),\ (0,1,1),\ (1,1,1),\ (0,0,2), \ldots \big\}\ ,
\end{equation}
where 
the ellipses denote momenta related to the ones that are shown by all  possible cubic group transformations.
Dibaryon operators that transform irreducibly under cubic transformations are obtained using appropriate averages of dibaryon operators with the same $|\vec{n}|$ but different $\vec{n}$.
Defining the momentum orbit $K^{(s)} = \{ \vec{n}\  \big|\  |\vec{n}|^2 = s \}$, projection to a cubic irrep $\Gamma_J$ with a row labeled by $J_z$ is achieved by forming linear combinations
\begin{equation}
\label{eq:dibaryon-cob}
  D^{(2,I,\Gamma_J,J_z)\Gamma}_{sm}(t) = \sum_{\vec{n} \in K^{(s)}} \sum_{\rho} G^{(\Gamma_J,J_z)}_{sm\vec{n}\rho} D^{(2,I)\Gamma}_{\rho}(\vec{n},t)\ ,
\end{equation}
where $m \in \{1,2,3,\ldots, N_s^{\Gamma_J} \}$ indexes the operators arising for a given $s$, $\Gamma_J$, $J_z$, and $\Gamma$. 
For $I=1$ operators, $m \in \{1\}$ is trivial, while for $I=0$, spin-orbit coupling leads to non-trivial multiplicities $N_s^{\Gamma_J}$ for some $s$ and $\Gamma_J$ as discussed in Refs.~\cite{Amarasinghe:2021lqa,Luu:2011ep} and below.
The change-of-basis coefficients $G^{(\Gamma_J,J_z)}_{sm\vec{n}\rho}$ 
are presented in Ref.~\cite{Amarasinghe:2021lqa} (see also Ref.~\cite{Detmold:2024ifm}).

\subsection{Quasi-local operators}

Quasi-local operators are constructed using spatial wavefunctions that are chosen to mimic the form of the asymptotic wavefunction for the deuteron determined from nuclear EFTs and phenomenological models~\cite{Luscher:1986pf,Luscher:1990ux,Konig:2011ti,Briceno:2013bda}, while having a factorizable form that enables efficient contraction calculations~\cite{Amarasinghe:2021lqa}.
They are defined as
\begin{equation}
  Q^{(2,I)\Gamma}_{\rho}(\kappa,t) = \sum_{\vec{x}_1,\vec{x}_2,R \in \Lambda_{\mathcal{S}}} e^{-\kappa |\vec{x}_1 - \vec{R}|} e^{-\kappa |\vec{x}_2 - \vec{R}|} \sum_{\sigma,\sigma'} v_{\rho}^{\sigma\sigma'}  P_{uu'}^{(I)} N^\Gamma_{\sigma u}(\vec{x}_2,t)  N^\Gamma_{\sigma' u'}(\vec{x}_1,t)\ ,
\end{equation}
where $\kappa$ is a parameter controlling the amount of correlation between the two nucleon positions, and $\vec{R}$ describes the center-of-mass of the two-nucleon system.
The analog of Eq.~\eqref{eq:dibaryon-cob} with $D^{(2,I)\Gamma}_{\rho}$ replaced by $Q^{(2,I)\Gamma}_{\rho}$ is used to project quasi-local operators onto rows of cubic irreps $\Gamma_J$ with definite $J_z$.
As for the dibaryon operators, the set of quasi-local operators considered in Ref.~\cite{Amarasinghe:2021lqa} is extended here to quasi-local operators with $\Gamma \in \{\gamma_5 P_+, \gamma_5 P_-, \Id \}$. 

\subsection{Hexaquark operators }
\label{sec:hex_construction}
In this section, a complete basis of local (single-site or smeared) hexaquark operators which project onto two-nucleon states is constructed. The general hexaquark operator used to construct this basis is
\begin{equation}
  \mathcal{H}_K^{(2,I,\Gamma_J,J_z)}(x)=\mathcal{H}^{C_1 C_2 C_3}_{G_1,F_1;G_2,F_2;G_3,F_3}(x) = T^{C_1 C_2 C_3}_{abcdef} \mathcal{D}^{ab}_{G_1,F_1}(x) \mathcal{D}^{cd}_{G_2,F_2}(x) \mathcal{D}^{ef}_{G_3,F_3}(x)\ ,   \label{eq:hexaquark}
\end{equation}
where $K=\{1,2,\dots\}$ corresponds to a particular spin-color-flavor structure. $T^{C_1 C_2 C_3}$ is a color tensor labeled by $C_1 C_2 C_3$ as described below, which projects the above operator to the color-singlet irrep, and diquarks $\mathcal{D}_{G_i,F_i}$ with Dirac and flavor structures $G_1,F_1,\ldots,G_3,F_3$ are defined in Eq.~\eqref{eq:diquark}.
In order to construct a complete basis of these operators, all possible color, spin and flavor labels must be enumerated. Only gauge invariant spin-singlet and spin-triplet operators with isospin zero and one are considered.

\subsubsection{Color}

Diquarks, being the product of two quarks, transform as $\mathbf{3}\otimes \mathbf{3}=\mathbf{6}\oplus \overline{\mathbf{3}}$ under $SU(3)_c$. 
Hexaquark operators formed from the product of three diquark operators therefore transform as
\begin{equation}
(\mathbf{3}\otimes \mathbf{3})\otimes (\mathbf{3}\otimes \mathbf{3})\otimes (\mathbf{3}\otimes \mathbf{3})=(\mathbf{6}\oplus \overline{\mathbf{3}})\otimes(\mathbf{6}\oplus \overline{\mathbf{3}})\otimes(\mathbf{6}\oplus \overline{\mathbf{3}})\ .
\end{equation}
Not all of these terms contain a color-singlet in this irrep decomposition, but it appears once in each of the five products
$\overline{\mathbf{3}}\otimes  \overline{\mathbf{3}}\otimes  \overline{\mathbf{3}}, \
\overline{\mathbf{3}}\otimes  \overline{\mathbf{3}}\otimes  {\mathbf{6}}, \
\overline{\mathbf{3}}\otimes  {\mathbf{6}}\otimes  \overline{\mathbf{3}}, \
{\mathbf{6}}\otimes  \overline{\mathbf{3}}\otimes  \overline{\mathbf{3}}$, and ${\mathbf{6}}\otimes  {\mathbf{6}}\otimes  {\mathbf{6}}$.
There are therefore five ways to combine the product of three diquarks into a color singlet.
The corresponding color tensors are given by
\begin{align}
T_{abcdef}^{AAA} &=\epsilon_{abe}\epsilon_{cdf}-\epsilon_{abf}\epsilon_{cde}\ ,
\nonumber \\
T_{abcdef}^{AAS} &=\epsilon_{abe}\epsilon_{cdf}+\epsilon_{abf}\epsilon_{cde}\ ,
\nonumber \\
T_{abcdef}^{ASA} &=\epsilon_{abc}\epsilon_{efd}+\epsilon_{abd}\epsilon_{efc}\ ,
\label{eq:color_Ts}
\\
T_{abcdef}^{SAA} &=\epsilon_{efa}\epsilon_{cdb}+\epsilon_{efb}\epsilon_{cda}\ ,
\nonumber \\
T_{abcdef}^{SSS} &=\epsilon_{ace}\epsilon_{bdf}+\epsilon_{acf}\epsilon_{bde}+\epsilon_{bce}\epsilon_{adf}+\epsilon_{bcf}\epsilon_{ade}\ , \nonumber
\end{align}
where the labels $C_1,C_2,C_3 \in \{A,S\}$ denote whether each tensor is antisymmetric or symmetric in each pair of indices $(a,b)$, $(c,d)$, and $(e,f)$, and hence is to be combined with a diquark in the $\overline{\mathbf{3}}$ or $\mathbf{6}$ representation for $A$ and $S$, respectively.

\subsubsection{Spin}\label{subsec:spin}

The operators introduced in Eq.~\eqref{eq:hexaquark} 
can be decomposed into direct sums of irreps of the relevant spatial symmetry group.
In a continuous infinite volume, this group is $Spin(3)=SU(2)$ (\textit{i.e.}, the double cover of the $SO(3)$ spatial rotation group), under which two-nucleon states transform in either the spin-triplet or spin-singlet representations.
On a periodic (hyper-)cubic lattice, the residual spatial symmetry is the double cover of the octahedral group, $O_h^D$.
The isovector (spin-singlet) two-nucleon states transform in the $A_1^+$  irrep, while the isosinglet (spin-triplet) states transform in the $T_1^+$ irrep.
In what follows, the continuum language of ``spin-singlet" and ``spin-triplet" will be used, as it unambiguoulsy specifies the internal spin degrees of freedom.

Recall from Eq.~\eqref{eq:diquark} above that each diquark contains a Dirac matrix of the form $CG_i$ where $C$ is the charge conjugation matrix. A possible basis for the matrices $G_i$ is given by 
\begin{equation}
G_i\in \{P_{R},P_{L},P_{R}\gamma^\mu,P_{L}\gamma^\mu,\sigma^{\mu\nu}\}\ ,    
\end{equation}
where $P_{R/L}=\frac{1}{2}(\Id\pm\gamma_5)$ and $\mu,\nu\in\{1,2,3,4\}$. In general, there are therefore $16^3=4096$ possible hexaquark spin structures. However, not all of these are independent. Fierz identities can be used to transform any product of two vector diquarks containing $\gamma^\mu$ or two tensor diquarks containing $\sigma^{\mu\nu}=\frac{i}{2}[\gamma^\mu,\gamma^\nu]$ into a product of scalar diquarks~\cite{Rao:1982gt}. Thus, when constructing the spin-singlet operator, it is sufficient to consider $G_i\in \{P_{R},P_{L}\}$. By a trivial change of basis, one can instead use the set $G_i\in \{\Id,\gamma_5\}$ to construct a complete basis of $Spin(4)$-invariant hexaquark operators. 
For interpolating-operator construction, where the relevant symmetry group is $O_h^D$, additional insertions of $\gamma_4$ do not change diquark transformation properties (note that the $P_\pm$ projectors are used to 
isolate the upper/lower quark components). 
The Dirac matrices required for a complete basis of diquarks that are singlets under $O_h^D$ are, after a change of basis,
\begin{equation}
G_i\in \{\Id,\gamma_5,\gamma_4,\gamma_4\gamma_5\}\ .
\end{equation}
This leads to $4^3=64$ independent spin-singlet  hexaquark operators. 

Using the parity projection operators $P_\pm$, the 64 combinations can be split into a set of 32 positive-parity combinations and a set of 32 negative-parity combinations, with the parity of the hexaquark operator equal to the product of the parities of the diquarks. 
Positive-parity diquarks correspond to 
\begin{equation}
G_i\in \{\gamma_5 P_+, \gamma_5 P_-\}\ ,
\end{equation}
while negative parity diquarks correspond to
\begin{equation}
G_i\in \{\Id, \gamma_4\}\ .
\end{equation}
This basis is convenient because it has definite symmetry properties for each diquark: diquarks with the Dirac matrix $C\gamma_4$ are symmetric under exchange of spin indices while diquarks with the $C$ and $C\gamma_5 P_{\pm}$ Dirac matrices 
are  antisymmetric. With this choice, operators that vanish due to quark antisymmetry can be easily identified.

For the spin-triplet hexaquark case, the only difference is that the construction must include one vector diquark whose Dirac structure includes one spin vector $S_i\equiv \tfrac{1}{2} \epsilon_{ijk} \gamma_j \gamma_k$ for $i\in\{1,2,3\}$.  
Since factors of $\gamma_4$ and $\gamma_5$ do not change $O_h^D$ transformation properties, these vector diquarks can involve four linearly independent Dirac matrices for a given spin index, $i$.
To ensure definite exchange symmetry, the four independent structures can be taken to be $S_i$, $S_i \gamma_4$, and $S_i \gamma_5 P_\pm$.
Further appearances of spin-vector diquarks can be removed using Fierz relations as above. There are therefore 64 linearly independent spin structures relevant for spin-triplet hexaquark operators for each spin index, $i$,
\begin{equation}
    G_1 \in \{S_i \gamma_5 P_+, S_i \gamma_5 P_-, S_i, S_i \gamma_4  \}\ , \hspace{10pt} G_2,G_3 \in \{\gamma_5 P_+, \gamma_5 P_-, \Id , \gamma_4 \}\ ,
    \label{eq:spin1}
\end{equation}
where the freedom to permute the diquarks to label the spin-vector diquark with $G_1$ has been used. As in the spin-singlet case, these operators can be split into sets of 32 operators with each parity. The positive-parity structures again correspond to operators with an odd number of structures involving $\gamma_5$.

\subsubsection{Flavor}
Assuming $SU(2)$ isospin symmetry is exact, products of two local nucleon operators form either an isovector spin-singlet state or an isoscalar spin-triplet state.
Although hexaquark operators with other transformation properties, for example isosinglet spin-singlet, can be constructed, they will not mix with operators in these two channels.

Each diquark can be projected into isosinglet and isovector flavor irreps as
\begin{equation}
\begin{split}
  \mathcal{D}_{G,\Id}^{ab}(x) &= q^{aT}(x) C G i \tau_2 q^b(x)\ , \\
  \mathcal{D}_{G,\tau_A}^{ab}(x) &= q^{aT}(x) C G i \tau_2 \tau_A q^b(x)\ ,
     \end{split}
\end{equation}
where $A\in\{1,2,3\}$.
Five linearly independent operators with $I=0$ can be constructed from these building blocks,
\begin{equation}
  \begin{split} \label{eq:I0}
  & \mathcal{D}_{G_1,\Id}^{ab}(x) \mathcal{D}_{G_2,\Id}^{cd}(x) \mathcal{D}_{G_3,\Id}^{ef}(x) \ , \\
  &  \mathcal{D}_{G_1,\tau_A}^{ab}(x) \mathcal{D}_{G_2,\tau_B}^{cd}(x) \mathcal{D}_{G_3,\Id}^{ef}(x) \delta_{AB}\ , \\
  & \mathcal{D}_{G_1,\tau_A}^{ab}(x) \mathcal{D}_{G_2,\Id}^{cd}(x) \mathcal{D}_{G_3,\tau_B}^{ef}(x) \delta_{AB}\ , \\
  & \mathcal{D}_{G_1,\Id}^{ab}(x) \mathcal{D}_{G_2,\tau_A}^{cd}(x) \mathcal{D}_{G_3,\tau_B}^{ef}(x) \delta_{AB}\ , \\
  &  \mathcal{D}_{G_1,\tau_A}^{ab}(x) \mathcal{D}_{G_2,\tau_B}^{cd}(x) \mathcal{D}_{G_3,\tau_C}^{ef}(x) \epsilon_{ABC}\ ,
\end{split}
\end{equation}
where $A,B,C \in \{1,2,3\}$. Note that the color and spin structures may differ on each diquark, making the second, third, and fourth combinations in Eq.~\eqref{eq:I0} distinct. 

A total of nine linearly independent isospin tensor operators with $I=1$ can be constructed analogously,
\begin{equation}
\begin{split}
   & \mathcal{D}_{G_1,\tau_A}^{ab}(x) \mathcal{D}_{G_2,\Id}^{cd}(x) \mathcal{D}_{G_3,\Id}^{ef}(x)\ , \\
   & \mathcal{D}_{G_1,\Id}^{ab}(x) \mathcal{D}_{G_2,\tau_A}^{cd}(x) \mathcal{D}_{G_3,\Id}^{ef}(x)\ , \\
  & \mathcal{D}_{G_1,\Id}^{ab}(x) \mathcal{D}_{G_2,\Id}^{cd}(x) \mathcal{D}_{G_3,\tau_A}^{ef}(x)\ , \\
   & \mathcal{D}_{G_1,\tau_B}^{ab}(x) \mathcal{D}_{G_2,\tau_C}^{cd}(x) \mathcal{D}_{G_3,\Id}^{ef}(x) \epsilon_{ABC}\ , \\
   & \mathcal{D}_{G_1,\tau_B}^{ab}(x) \mathcal{D}_{G_2,\Id}^{cd}(x) \mathcal{D}_{G_3,\tau_C}^{ef}(x) \epsilon_{ABC}\ , \\
   & \mathcal{D}_{G_1,\Id}^{ab}(x) \mathcal{D}_{G_2,\tau_B}^{cd}(x) \mathcal{D}_{G_3,\tau_C}^{ef}(x) \epsilon_{ABC}\ , \\
   & \mathcal{D}_{G_1,\tau_A}^{ab}(x) \mathcal{D}_{G_2,\tau_B}^{cd}(x) \mathcal{D}_{G_3,\tau_C}^{ef}(x)  \delta_{BC}\ , \\ 
   & \mathcal{D}_{G_1,\tau_B}^{ab}(x) \mathcal{D}_{G_2,\tau_A}^{cd}(x) \mathcal{D}_{G_3,\tau_C}^{ef}(x) \delta_{BC}\ , \\
   & \mathcal{D}_{G_1,\tau_B}^{ab}(x) \mathcal{D}_{G_2,\tau_C}^{cd}(x) \mathcal{D}_{G_3,\tau_A}^{ef}(x) \delta_{BC}\ . \\
\end{split}
\end{equation}

\subsubsection{Gram-Schmidt Reduction and Hexaquark Basis}

\begin{table}
\renewcommand{\arraystretch}{1.4}
\centering
\begin{tabular}{
P{0.02\textwidth}  
P{0.01\textwidth}  P{0.01\textwidth} 
P{0.01\textwidth} P{0.01\textwidth} P{0.01\textwidth}  
P{0.01\textwidth}  P{0.01\textwidth} 
P{0.06\textwidth} P{0.06\textwidth} P{0.06\textwidth}  
P{0.01\textwidth}  P{0.01\textwidth} 
P{0.01\textwidth} P{0.01\textwidth} P{0.01\textwidth}  
P{0.02\textwidth} | P{0.02\textwidth}  
P{0.02\textwidth}  
P{0.01\textwidth}  P{0.01\textwidth} 
P{0.01\textwidth} P{0.01\textwidth} P{0.01\textwidth}  
P{0.01\textwidth}  P{0.01\textwidth} 
P{0.06\textwidth} P{0.06\textwidth} P{0.06\textwidth}  
P{0.01\textwidth}  P{0.01\textwidth} 
P{0.01\textwidth} P{0.01\textwidth} P{0.01\textwidth}  
}
\hline\hline 
$K$ & & & \multicolumn{3}{c}{Color} & & & \multicolumn{3}{c}{Spin} & & & \multicolumn{3}{c}{Flavor} & & & 
$K$ & & & \multicolumn{3}{c}{Color} & & & \multicolumn{3}{c}{Spin} & & & \multicolumn{3}{c}{Flavor} \\ \hline
1 & & & $A$&$A$&$A$ & & & $\gamma_4$& $\gamma_5 P_+$&$\Id$ & & & $\tau$&$\Id$&$\Id$ & & & 9 & & &  $S$&$A$&$A$  & & &  $\gamma_5 P_-$&$\gamma_5 P_+$&$\gamma_5 P_+$  & & &  $\tau$&$\Id$&$\Id$ \\
2 & & & $A$&$A$&$A$ & & & $\gamma_4$& $\gamma_5 P_-$&$\Id$ & & & $\tau$&$\Id$&$\Id$ & & & 10 & & & $S$&$A$&$A$  & & & $\gamma_5 P_-$&$\gamma_5 P_-$&$\gamma_5 P_+$ & & & $\tau$&$\Id$&$\Id$ \\
3 & & & $S$&$A$&$A$ & & & $\gamma_5 P_+$&$\gamma_5 P_+$&$\gamma_5 P_+$ & & & $\tau$&$\Id$&$\Id$ & & & 11 & & & $S$&$A$&$A$ & & & $\gamma_5 P_-$&$\gamma_5 P_-$&$\gamma_5 P_-$ & & & $\tau$&$\Id$&$\Id$ \\
4 & & & $S$&$A$&$A$ & & & $\gamma_5 P_+$&$\gamma_5 P_-$&$\gamma_5 P_+$ & & & $\tau$&$\Id$&$\Id$ & & & 12 & & & $S$&$A$&$A$ & & & $\gamma_5 P_-$&$\Id$&$\Id$ & & & $\tau$&$\Id$&$\Id$ \\
5 & & & $S$&$A$&$A$ & & & $\gamma_5 P_+$&$\gamma_5 P_-$&$\gamma_5 P_-$ & & & $\tau$&$\Id$&$\Id$ & & & 13 & & & $S$&$A$&$A$ & & & $\gamma_5 P_-$&$\gamma_4$&$\gamma_4$ & & & $\tau'$&$\tau$&$\tau'$ \\
6 & & & $S$&$A$&$A$ & & & $\gamma_5 P_+$&$\Id$&$\Id$ & & & $\tau$&$\Id$&$\Id$ & & & 14 & & & $S$&$A$&$A$ & & &  $\gamma_5 P_-$&$\gamma_4$&$\gamma_4$ & & & $\tau$&$\tau'$&$\tau'$ \\
7 & & & $S$&$A$&$A$ & & & $\gamma_5 P_+$&$\gamma_4$&$\gamma_4$ & & & $\tau'$&$\tau$&$\tau'$ & & & 15 & & & $S$&$S$&$S$ & & & $\gamma_5 P_+$&$\gamma_5 P_-$&$\gamma_5 P_+$ & & & $\tau$&$\tau'$&$\tau'$ \\
8 & & & $S$&$A$&$A$ & & & $\gamma_5 P_+$&$\gamma_4$&$\gamma_4$ & & & $\tau$&$\tau'$&$\tau'$ & & & 16 & & & $S$&$S$&$S$ & & & $\gamma_5 P_+$&$\gamma_5 P_-$&$\gamma_5 P_-$ & & & $\tau'$&$\tau$&$\tau'$ \\ 
\hline \hline
\end{tabular}
  \caption{A complete basis of hexaquark operators with $I=1$ and spin zero,  $\mathcal{H}^{(2,1,A_1^+,J_z=0)}_K(x)$, enumerated by $K \in \{1,\dots,16\}$. Note that all but $K=3$ constitute operators with hidden color. Each operator takes the form $\mathcal{H}^{C_1 C_2 C_3}_{G_1,F_1;G_2,F_2;G_3,F_3}(x)$, where the color tensor labels $C_i$, Dirac matrices $G_i$, and flavor tensors $F_i$ appearing in each diquark as defined in the main text are indicated in the corresponding columns for each $K$. Here, $\tau$ always refers to an isovector diquark with a free isospin index, while $\tau'$ pairs indicate isovector diquarks whose indices are contracted as $\tau'_B \tau'_C \delta_{BC}$. }
\label{tab:spin0_hex}
\end{table}

\begin{table}
\renewcommand{\arraystretch}{1.4}
\centering
\begin{tabular}{
P{0.02\textwidth}  
P{0.01\textwidth}  P{0.01\textwidth} 
P{0.01\textwidth} P{0.01\textwidth} P{0.01\textwidth}  
P{0.01\textwidth}  P{0.01\textwidth} 
P{0.06\textwidth} P{0.06\textwidth} P{0.06\textwidth}  
P{0.01\textwidth}  P{0.01\textwidth} 
P{0.01\textwidth} P{0.01\textwidth} P{0.01\textwidth}  
P{0.02\textwidth} | P{0.02\textwidth}  
P{0.02\textwidth}  
P{0.01\textwidth}  P{0.01\textwidth} 
P{0.01\textwidth} P{0.01\textwidth} P{0.01\textwidth}  
P{0.01\textwidth}  P{0.01\textwidth} 
P{0.06\textwidth} P{0.06\textwidth} P{0.06\textwidth}  
P{0.01\textwidth}  P{0.01\textwidth} 
P{0.01\textwidth} P{0.01\textwidth} P{0.01\textwidth}  
}
\hline\hline 
$K$ & & & \multicolumn{3}{c}{Color} & & & \multicolumn{3}{c}{Spin} & & & \multicolumn{3}{c}{Flavor} & & & 
$K$ & & & \multicolumn{3}{c}{Color} & & & \multicolumn{3}{c}{Spin} & & & \multicolumn{3}{c}{Flavor} \\ \hline
1 & & & $A$&$A$&$A$ & & & $S_i \gamma_5 P_+$&$ \gamma_4$&$ \Id$ & & & $\tau$&$\tau$&$\Id$ & & &
9 & & & $S$&$A$&$A$ & & & $S_i\gamma_5 P_+$&$ \Id$&$ \Id$ & & & $\Id$&$ \Id$&$ \Id$ \\
2  & & & $A$&$A$&$A$ & & & $S_i \gamma_5 P_-$&$ \gamma_4$&$\Id$ & & & $\tau$&$ \tau$&$ \Id$  & & &
10  & & & $S$&$A$&$A$  & & & $S_i\gamma_5 P_-$&$ \gamma_5 P_-$&$ \gamma_5 P_+$  & & & $\Id$&$ \Id$&$ \Id$ \\
3  & & & $A$&$A$&$A$  & & & $S_i$&$ \gamma_4$&$ \gamma_5 P_+$  & & & $\tau$&$ \tau$&$ \Id$  & & &
11  & & & $S$&$A$&$A$   & & &$S_i\gamma_5 P_-$&$ \gamma_5 P_-$&$ \gamma_5 P_-$   & & & $\Id$&$ \Id$&$ \Id$ \\ 
4  & & & $A$&$A$&$A$  & & & $S_i\gamma_4$&$ \gamma_5 P_+$ &$ \Id$  & & & $\Id$&$ \Id$&$ \Id$  & & &
12 & & & $S$&$A$&$A$ & & & $S_i\gamma_5 P_-$&$ \Id$&$ \Id$ & & & $\Id$&$ \Id$&$ \Id$ \\
5   & & &$A$&$A$&$A$   & & &$S_i\gamma_4$&$\gamma_5 P_-$&$\Id$ & & &$\Id$&$ \Id$&$ \Id$   & & & 
13 & & &$S$&$S$&$S$ & & & $S_i\gamma_5 P_+$&$ \gamma_4$&$ \gamma_4$ & & & $\Id$&$ \Id$&$ \Id$ \\
6 & & & $S$&$A$&$A$ & & &$S_i\gamma_5 P_+$&$\gamma_5 P_+$&$\gamma_5 P_+$ & & & $\Id$&$ \Id$&$ \Id$ & & &
14 & & & $S$&$S$&$S$ & & & $S_i\gamma_5 P_-$&$ \gamma_4$&$ \gamma_4$ & & & $\Id$&$ \Id$&$ \Id$ \\
7 & & & $S$&$A$&$A$ & & & $S_i\gamma_5 P_+$&$ \gamma_5 P_-$&$\gamma_5 P_+$ & & & $\Id$&$ \Id$&$ \Id$  & & & 
15  & & & $S$&$S$&$S$  & & & $S_i\gamma_4$&$ \gamma_4$&$ \gamma_5 P_+$  & & & $\tau$&$ \Id$&$ \tau$ \\
8 & & & $S$&$A$&$A$ & & & $S_i\gamma_5 P_+$&$ \gamma_5 P_-$&$ \gamma_5 P_-$ & & &  $\Id$&$\Id$&$\Id$ & & & 
16  & & & $A$&$A$&$S$  & & & $S_i\gamma_5 P_+$&$ \gamma_5 P_-$&$ \gamma_5 P_-$  & & & $\tau$&$ \Id$&$ \tau$ \\ \hline \hline
\end{tabular}
  \caption{A complete basis of hexaquark operators with $I=0$ and spin one, $\mathcal{H}_K^{(2,0,T_1^+,J_z=i)}(x)$, enumerated by $K \in \{1,\dots,16\}$. Note that all but $K=6$ constitute operators with hidden color. Color, spin, and flavor labels are as in Table~\ref{tab:spin0_hex}. Here, the $\tau$ flavor structures correspond to contracted indices $\tau_A\tau_B\delta_{AB}$.
  }
\label{tab:spin1_hex}
\end{table}

Spin-color-flavor tensor hexaquark operators are obtained by contracting the flavor tensor operators above with one of the five color tensors shown in Eq.~\eqref{eq:color_Ts} and choosing $G_1$, $G_2$, and $G_3$ to correspond to the choices of spin operators described in Sec.~\ref{subsec:spin}.
However, the resulting operators will not all be linearly independent because of quark antisymmetry. 
The reduction of this overcomplete set to complete bases of positive-parity spin-singlet hexaquark operators with $I=1$ and positive-parity spin-triplet hexaquark operators with $I=0$ is discussed in this section.

The five color tensors and 32 positive-parity spin-singlet tensors discussed above can be combined with the nine $I=1$ flavor tensors in $5 \times 32 \times 9 = 1440$ ways.
Each of these spin-color-flavor tensor operators can be described as a contraction of six quark fields with a weight tensor that has six spin, color, and flavor indices.
The resulting rank-18 weight tensors are sparse and can be represented efficiently as lists of the non-zero weights and their corresponding index values.
To make the constraints from quark antisymmetry manifest, the quark fields appearing in every term with non-zero weights can be permuted into a fiducial flavor ordering, such as $uuuddd$ for the $I_z=0$ case, as described in Refs.~\cite{Detmold:2012eu,Amarasinghe:2021lqa}.
Many terms in the original weight tensor correspond to the same tensor structure after antisymmetrization and can be combined together to build a reduced rank-12 spin-color  weight tensor corresponding to the fiducial ordering.

The reduced weight tensors associated with linearly independent spin-color-flavor tensor operators are not linearly independent if some operators are related by quark antisymmetry.
An orthonormal basis of reduced-weight tensors is constructed using a Gram--Schmidt process;
this basis is a complete basis of hexaquark operators without redundancies from quark antisymmetry or Fierz relations.
The orthogonalization isolates 16 linearly independent operators from the full set of 1440 isovector spin-singlet  hexaquark operators.
Note that accounting for the quark antisymmetry of each individual diquark reduces the 1440 operators to 101; however, many of the remaining redundancies can be understood as arising from combined color-spin-flavor Fierz identities that complicate a group theoretic determination of the number of linearly independent operators~\cite{Rao:1982gt,Buchoff:2015qwa}.
These 16 orthonormal operators $H_i^{(2,1,A_1^+,J_z)}(x)$ are linear combinations of the 16 spin-color-flavor tensor operators $\mathcal{H}^{(2,1,A_1^+,J_z)}_K(x)$ shown in Table~\ref{tab:spin0_hex},
\begin{equation}
  H_i^{(2,1,A_1^+,J_z)}(x)=\sum_{K=1}^{i} w_{iK}^{(2,1,A_1^+)} \sum_{\vec{x}}  \mathcal{H}^{(2,1,A_1^+,J_z)}_K(x)\ .
  \label{eq:dineutron_hexaquark_gram_schmidt}
\end{equation}
Hexaquark operators with zero momentum are defined as $H_i^{(2,1,A_1^+,J_z)}(t) = \sum_{\vec{x}} H_i^{(2,1,A_1^+,J_z)}(\vec{x},t)$.
It is noteworthy that normalized isovector hexaquark operators constructed from products of color-singlet upper-spin-component baryon operators of the forms $NN$, $N\Delta$, and $\Delta\Delta$ are all identical to the basis operator $H_3^{(2,1,A_1^+,J_z)}$.
This can be explained by the fact that baryon-product operators of the forms $NN$, $N\Delta$, and $\Delta\Delta$ all include two diquarks that are antisymmetric in color and are comprised of only upper-spin-component quark fields --- 
$H_3^{(2,1,A_1^+,J_z)}$ is the only basis operator, or linear combination of basis operators, meeting this description.

The same considerations apply to positive-parity spin-triplet hexaquarks with $I=0$, where five flavor tensors are available.
In this case, a total of $5\times 32 \times 5 = 800$ spin-color-flavor tensor operators can be constructed.
As in the preceding spin-singlet case, reduced weights are constructed for each of these 800 operators.
Finally, the Gram--Schmidt algorithm isolates 16 orthonormal spin-triplet hexaquark operators using linear combinations of the operators shown in Table~\ref{tab:spin1_hex}.
Normalized isosinglet hexaquark operators constructed from products of color-singlet upper-spin-component  baryon operators of the forms $NN$ and $\Delta\Delta$ are both identical to the basis operator $H_6^{(2,0,T_1^+,J_z)}$, which can be explained analogously to the $I=1$ case.

\section{Numerical Study with $m_\pi\approx 806~\si{MeV}$}
\label{sec:numerical_study}

This section presents a variational study 
of the spectrum of two-nucleon systems for $N_f=3$ degenerate quarks with a common mass corresponding to a pion mass of $m_\pi\approx 806~\si{MeV}$. This calculation uses gauge fields also employed in previous studies of two-nucleon spectroscopy in Refs.~\cite{NPLQCD:2012mex,NPLQCD:2013bqy,Berkowitz:2015eaa,Wagman:2017tmp,Amarasinghe:2021lqa}. They were generated using the tadpole-improved L\"uscher-Weisz gauge field action~\cite{Luscher:1984xn} with a single level of stout smearing~\cite{Morningstar:2003gk} and the Wilson-clover fermion action~\cite{Wilson:1974sk} with a tadpole-improved tree-level clover coefficient $c_{SW}= 1.2493$~\cite{Sheikholeslami:1985ij}.  Relevant details are presented in Table.~\ref{tab:gauge_field}.

Quark propagators computed for all source points in a $6^3$ sparse sub-lattice of the $(L/a)^3 = 24^3$ spatial volume at a fixed time are used to construct sparsened quark propagators~\cite{Detmold:2019fbk} with sparseining factor 
$\mathcal{S} = 4$.
The quark sources employ gauge-invariant Gaussian smearing~\cite{Gusken:1989ad,Gusken:1989qx} with a \verb!Chroma! smearing parameter 2.1, which corresponds to a Gaussian smearing width of $\approx 0.18$ fm.
Interpolating operators constructed from these quark propagators are therefore identical to the ``thin"-smearing operators described in Ref.~\cite{Amarasinghe:2021lqa} except that an ensemble with larger $L$ is used in that work. Variational bounds on the one- and two-nucleon systems are obtained by performing multi-exponential fits to the principal correlation functions determined from the GEVP, following the algorithm described previously in Refs.~\cite{NPLQCD:2020ozd,Amarasinghe:2021lqa} augmented by the algorithm for choosing reference times $t_0$ and $t_{\rm ref}$ discussed in Appendix \ref{app:t0tref}.

\begin{table}[t]
\centering
\renewcommand{\arraystretch}{1.2}
\setlength{\tabcolsep}{0.75em}
\begin{tabular}{c c c c c c c c c c}
\hline \hline
$(L/a)^3\times (T/a)$ & $\beta$ & $am_q$ & $a$ [\si{fm}] & $L$ [\si{fm}] & $T$ [\si{fm}] & $m_\pi L$ & $m_\pi T$ & $N_\text{cfg}$ & $N_\text{src}$ \\ \hline 
$24^3\times 48$ &  6.1 & -0.2450 & 0.1453(16)  &  3.4 & 6.7 &   14.3  & 28.5 & 469 & 216 \\
\hline \hline
\end{tabular}
\caption{Details of the ensemble of gauge field configurations used for the numerical calculations. $L$ and $T$ are the spatial and temporal extents of the lattice, $\beta = 6/g^2$ is the inverse bare coupling, $m_q$ is the bare quark mass, $a$ is the lattice spacing, $N_\text{cfg}$ is the total number of gauge field configurations used, and $N_\text{src}$ is the number of source locations employed on each configuration.}
\label{tab:gauge_field}
\end{table}

\subsection{Interpolating-operator sets}

A range of different variational operator sets are considered.
Sets are chosen to study the operator dependence of variational bounds on the energy eigenvalues.
Understanding this variation is critical because a computation employing a complete basis of operators for the full Hilbert space of QCD is intractable. 

Shorthand notation is useful for discussing interpolating-operator sets and variational bounds.
The positive- and negative-parity single-nucleon channels have quantum numbers $Q=(B, I, \Gamma_J)$ that will be denoted
\begin{align}
N^+ &\equiv (1, \tfrac{1}{2}, G_1^+)\ , \label{eq:nucleon_Q}\\
N^- &\equiv (1, \tfrac{1}{2}, G_1^-)\ .\label{eq:nucleon_Q_negative_parity}
\end{align}
The energy spectrum is independent of $J_z$; numerical calculations average $N^+$ and $N^-$ correlation functions over $J_z \in \{1,2\}$.
The quantum numbers $Q = (B, I, \Gamma_J)$ for the two-baryon systems are
\begin{align}
    nn  &\equiv (2, 1, A_1^+)\ , \label{eq:dineutron_Q} \\
    d   &\equiv (2, 0, T_1^+)\ . \label{eq:deuteron_Q} 
\end{align}
Note that $nn$ is used to label the ``dineutron" $I=1$ spectrum, which is independent of $I_z$; numerical calculations are explicitly performed using $I_z=0$.
Similarly, the label $d$ is used to denote the isoscalar spin-triplet spectrum, which is independent of $J_z$; numerical calculations average correlation functions over $J_z \in \{1,2,3\}$.
Notation for energy gaps in these channels is summarized in Table~\ref{tab:notation}.

The sets of operators that are studied in this work can be divided broadly into three main categories:
\begin{enumerate}
    \item \emph{Dibaryon operator sets} include a range of different dibaryon operators, including those with lower-spin components and negative parity nucleon operators.
    Dibaryon operator sets are denoted $\mathbb{S}_{N_DD}^Q$, where $Q\in\{nn,d\}$ denotes the quantum numbers of the system and $N_D$ is the number of dibaryon operators in the set. Several operator sets are considered which contain the same total number of dibaryon operators, but contain a different selection of operators. In this case, the $\prime$ and $\prime\prime$ symbols are used to differentiate these operator sets.
    
    \item \emph{Hexaquark operator sets} include a variety of hexaquark operators introduced in Sec.~\ref{sec:hex_construction}.
    Hexaquark operator sets are denoted $\mathbb{S}_{N_HH}^Q$, with $Q$ as above and $N_H$ the number of hexaquark operators that are included.
    
    \item \emph{Dibaryon and hexaquark operator sets} combine operators from the previous two sets to examine the combined effect of both types of operators.
    Combined operator sets are denoted $\mathbb{S}_{N_D D}^Q \cup \,\mathbb{S}_{N_H H}^Q$.
    The integers $N_D$ and $N_H$ correspond to the number of dibaryon and hexaquark operators, respectively. 
    Thus, a total of $N_D$ + $N_H$ operators appear in the combined set.

\end{enumerate}

\begin{table}[t]
    \centering
    \begin{tabular}{l l}
        \hline\hline
        Quantity    &   Description \\
        \hline
        $M_N \equiv E_0^{N^+}$ & Nucleon mass \\\\
        $\dNp \equiv  E_1^{N^+} - M_N$  & Energy gap in channel $N_+$\\\\
        $\dNm \equiv  E_0^{N^-} - M_N$  & \makecell[l]{Energy gap in channel $N_-$ between lowest-energy negative-parity\\state and positive-parity ground state}\\\\
        $\Delta E_{\mathsf{n}}^Q \equiv E_{\mathsf{n}}^Q - 2 M_N$ & \makecell[l]{$\mathsf{n}$th energy gap from two-nucleon threshold in channel $Q$,\\ computed using fit results for each energy level}\\\\
        $\Delta E_{\mathsf{n}}^Q(t) \equiv E_{\mathsf{n}}^Q(t) - 2 E_0^{N^+}(t)$ & \makecell[l]{Time-dependent $\mathsf{n}$th energy gap from two-nucleon threshold in\\channel $Q$, computed using the effective energies defined in \cref{eq:emPC}}\\\\
        $\DLE \equiv 2\sqrt{M_N^2 + 5\left(\frac{2\pi}{L}\right)^2} - 2M_N$ &
        \makecell[l]{Energy gap from two-nucleon threshold to the first non-interacting \\  energy level with larger momentum than the operators used here
        }
        \\
    \hline\hline
    \end{tabular}
    \caption{Summary of spectral quantities used to characterize the low-energy one- and two-nucleon spectra.
    Single-nucleon splittings are denoted by $\delta$, while two-nucleon splittings use $\Delta$.}
    \label{tab:notation}
\end{table}

Sets of dibaryon operators constructed using upper-spin-component diquarks are defined by
\begin{equation}
\label{eq:dibaryon_operators}
\begin{split} 
\mathbb{S}_{5D}^{nn} &\equiv \bigg\{ D^{nn\Gamma}_{sm}(t)  \,\bigg|\, s\in\{0,1,2,3,4\},  m\in\{1,\ldots,N_s^{A_1^+}\}, \ \Gamma\in\{\gamma_5 P_+\} \bigg
  \}\ , \\
\mathbb{S}_{10D}^{d} &\equiv \bigg\{ D^{d\Gamma}_{sm}(t)  \,\bigg|\, s\in\{0,1,2,3,4\},  m\in\{1,\ldots,N_s^{T_1^+}\}, \ \Gamma\in\{\gamma_5 P_+\} \bigg
  \}\ .
  \end{split}
\end{equation}
On the right-hand side, $s = |\vec{n}|^2$ is the magnitude of the plane-wave momentum appearing in \cref{eq:dibaryon} and $m$ indexes the multiplicity of each shell. These sets are analogous to the interpolating-operator set denoted $\mathbb{S}_0$ in Ref.~\cite{Amarasinghe:2021lqa}, except that operators with a second Gaussian quark field smearing and operators with larger values of $s$ were also included in Ref.~\cite{Amarasinghe:2021lqa}.
Operator sets which explore the variational bounds obtained from only the novel dibaryon (negative-parity and lower spin component) operators introduced in Sec.~\ref{sec:dibaryon_ops} can also be constructed:
\begin{equation}
\begin{split} 
\mathbb{S}_{5D^\prime}^{nn} &\equiv \bigg\{ D^{nn\Gamma}_{sm}(t)  \,\bigg|\, s\in\{0,1,2,3,4\},  m\in\{1,\ldots,N_s^{A_1^+}\}, \ \Gamma\in\{\Id\} \bigg \}\ , 
\\
\mathbb{S}_{5D^{\prime\prime}}^{nn} &\equiv \bigg\{ D^{nn\Gamma}_{sm}(t)  \,\bigg|\, s\in\{0,1,2,3,4\},  m\in\{1,\ldots,N_s^{A_1^+}\}, \ \Gamma\in\{\gamma_5 P_- \} \bigg
  \}\ , \\
\mathbb{S}_{10D^{\prime}}^{d} &\equiv \bigg\{ D^{d\Gamma}_{sm}(t)  \,\bigg|\, s\in\{0,1,2,3,4\},  m\in\{1,\ldots,N_s^{T_1^+}\}, \ \Gamma\in\{\Id \} \bigg
  \}\ , \\
\mathbb{S}_{10D^{\prime\prime}}^{d} &\equiv \bigg\{ D^{d\Gamma}_{sm}(t)  \,\bigg|\, s\in\{0,1,2,3,4\},  m\in\{1,\ldots,N_s^{T_1^+}\}, \ \Gamma\in\{\gamma_5 P_-\} \bigg
  \}\ .
  \end{split}
\end{equation}
The meaning of $s$ and $m$ are the same as in the previous equations.

Sets of dibaryon operators involving products of two negative-parity nucleons and involving nucleon operators built from lower-spin-component diquarks are defined by
\begin{equation}
  \begin{split}
    \mathbb{S}_{10D}^{nn} \equiv& \bigg\{ D^{nn\Gamma}_{sm}(t)  \,\bigg|\, s\in\{0,1,2,3,4\}, m\in\{1,\ldots,N_s^{A_1^+}\}, \ \Gamma\in\{\gamma_5 P_+, \Id \} \bigg\}\ ,
    \\
  \mathbb{S}_{15D}^{nn} \equiv& \bigg\{ D^{nn\Gamma}_{sm}(t)  \,\bigg|\, s\in\{0,1,2,3,4\}, m\in\{1,\ldots,N_s^{A_1^+}\}, \ \Gamma\in\{\gamma_5 P_+, \Id, \gamma_5 P_-\} \bigg\}\ ,
\\
  \mathbb{S}_{20D}^{d} \equiv& \bigg\{ D^{d\Gamma}_{sm}(t)  \,\bigg|\, s\in\{0,1,2,3,4\}, m\in\{1,\ldots,N_s^{T_1^+}\}, \ \Gamma\in\{\gamma_5 P_+, \Id \} \bigg\}\ ,
    \\
  \mathbb{S}_{30D}^{d} \equiv& \bigg\{ D^{d\Gamma}_{sm}(t)  \,\bigg|\, s\in\{0,1,2,3,4\}, m\in\{1,\ldots,N_s^{T_1^+}\}, \ \Gamma\in\{\gamma_5 P_+, \Id, \gamma_5 P_-\} \bigg\}\ .
\end{split}
\end{equation}
The meanings of $s$, and $m$ are the same as in Eq.~\cref{eq:dibaryon_operators}.

Compared to dibaryon operators, spatially localized hexaquark operators might be expected to have larger overlap with compact bound states and smaller overlap with scattering states.
Sets containing one, two, and sixteen hexaquark operators are considered in order to study the variational bounds from hexaquark operators alone. 
For the dineutron, these sets are defined as
\begin{equation}
  \begin{split}
\mathbb{S}_{1H}^{nn}\equiv&\bigg\{H_i^{nn}(t) \,\bigg|\, i\in \{3\}\bigg\}\ ,
\\
\mathbb{S}_{2H}^{nn}\equiv&\bigg\{H_i^{nn}(t) \,\bigg|\, i\in \{1, 3\}\bigg\}\ , 
\\
\mathbb{S}_{16H}^{nn}\equiv&\bigg\{H_i^{nn}(t) \,\bigg|\, i\in \{1,2,\dots,16\}\bigg\}\ .
  \end{split}
\end{equation}
The dineutron operators $H^{nn}_i(t)$ appearing on the right-hand side are defined in \cref{eq:dineutron_hexaquark_gram_schmidt} as orthogonal linear combinations of the basic hexaquark operators in \cref{tab:spin0_hex}.
The operator $H_3^{nn}(t)$ appearing in all three sets is identical, apart from an irrelevant constant, to the product of color-singlet nucleon operators centered at the same point that was studied in Ref.~\cite{Amarasinghe:2021lqa}.
The two operators appearing in $\mathbb{S}_{2H}^{nn}$ are both unaffected by the Gram-Schmidt orthogonalization of the hexaquark operators described above, explicitly $H_1^{nn}(x) = \mathcal{H}_1^{nn}(x)$ and $H_3^{nn}(x) = \mathcal{H}_3^{nn}(x)$.
For the deuteron, the corresponding operator sets are defined as
\begin{equation}
    \begin{split}
    \mathbb{S}_{1H}^{d}\equiv&\bigg\{H_i^{d}(t) \,\bigg|\, i\in \{6\}\bigg\}\ ,
    \\
    \mathbb{S}_{2H}^{d}\equiv&\bigg\{H_i^{d}(t) \,\bigg|\, i\in \{3,6\}\bigg\}\ ,
    \\
    \mathbb{S}_{16H}^{d}\equiv&\bigg\{H_i^{d}(t) \,\bigg|\, i\in \{1,2,\dots,16\}\bigg\}\ ,
    \end{split}
\end{equation}
where $H_6^{d}(x) = \mathcal{H}_6^{d}(x)$ and $H_3^{d}(x) = \mathcal{H}_3^{d}(x) - \frac{1}{11} \mathcal{H}_1^{d}(x)$.
As in the dineutron channel, the operator $H_6^{d}(t)$ appearing in all three sets is identical to a product of color-singlet nucleon operators.
The second operators included in $\mathbb{S}_{2H}^{Q}$ for each channel are chosen because of their relatively large overlaps with low-energy states in the numerical results below.

Four additional interpolating-operator sets are chosen for each channel to study the combined effects of including both dibaryon and hexaquark operators,
\begin{equation}
\begin{split}
&\mathbb{S}_{5D}^{nn}\cup \mathbb{S}_{1H}^{nn}\ ,\quad \mathbb{S}_{5D}^{nn}\cup \mathbb{S}_{2H}^{nn}\ ,\quad \mathbb{S}_{5D}^{nn}\cup \mathbb{S}_{16H}^{nn}\ ,\quad \mathbb{S}_{15D}^{nn}\cup \mathbb{S}_{16H}^{nn} \ ,
\\
&\mathbb{S}_{10D}^{d}\cup \mathbb{S}_{1H}^{d}\ ,\quad \mathbb{S}_{10D}^{d}\cup \mathbb{S}_{2H}^{d}\ ,\quad \mathbb{S}_{10D}^{d}\cup \mathbb{S}_{16H}^{d}\ ,\quad \mathbb{S}_{30D}^{d}\cup \mathbb{S}_{16H}^{d} \ .
\end{split}
\end{equation}
It is possible to define additional sets including quasi-local operators.
In practice, such sets give results for the low-energy spectra which are consistent with one or more of the operator sets reported here, albeit with larger uncertainties.
Results for operator sets including quasi-local operators are therefore not presented below.

Dibaryon operators with all three choices of $\Gamma \in\{\Id, \gamma_5 P_+,\gamma_5 P_-\}$ are also computed for cubic irreps that are associated with $D$-wave phase shifts in infinite volume. 
For the $I=1$ case, such dibaryon operators are constructed for $\Gamma_J = E^+$ with $s\in \{1,2,4\}$ and for $\Gamma_J = T_2^+$ with $s \in \{2,3\}$. 
For the $I=0$ case, dibaryon operators are constructed for $\Gamma_J =  A_2^+$ with $s \in \{2,3\}$, for $\Gamma_J = E^+$ with $s \in \{2,3\}$, and for $\Gamma_J =  T_2^+$ with $s \in \{1,2,3,4\}$.
Although there are no hexaquarks built from products of color-singlet nucleons that transform in these representations, hexaquarks built from $N\Delta$ and $\Delta\Delta$ operators, as well as hidden-color operators, can be constructed that transform in the same representations.
The inclusion of such operators is left to future work.

\subsection{The Single-Nucleon Channel}
\label{sec:nuc}
\begin{figure}[!t]
	\includegraphics[width=0.47\columnwidth]{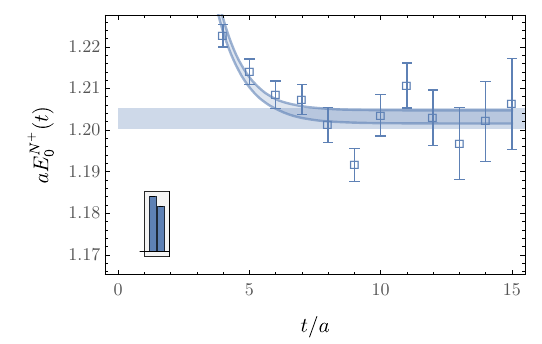}
	\includegraphics[width=0.47\columnwidth]{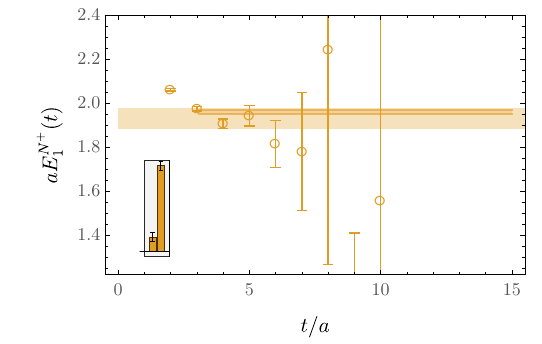}
   \caption{The positive parity single-nucleon channel is studied using the variational method utilizing a set of two operators constructed from the upper- and lower-spin components. The relative weights of the upper and lower spin components are shown as the left and right columns in the histogram (inset). The lower variational bound constrains the nucleon mass in lattice units. Lightly shaded colored bands show the total statistical plus fitting systematic uncertainties added in quadrature while the outlined regions show the statistical uncertainty of the highest-weight fit.
   Note that for the $\mathsf{n}=1$ case, fit results provide a variational bound even though there may still be significant curvature in the effective energy for all $t$ where signals can be resolved from noise. 
}
   \label{fig:single_nucleon_positive}
\end{figure}

The positive-parity single-nucleon channel can be studied using the interpolating-operator set 
$\mathbb{S}^{N^+} = \{ N^\Gamma \, | \, \Gamma \in \gamma_5 P_\pm \}$, with the quantum numbers $N^+$ defined in \cref{eq:nucleon_Q}.
Applying the algorithm for selecting $t_0$ and $t_{\text{ref}}$ discussed in Appendix~\ref{app:t0tref} leads to $t_0 = 7$ and $t_{\text{ref}} = 11$. The effective masses and fit results using the multi-state fitting algorithm detailed in Refs.~\cite{NPLQCD:2020ozd,Amarasinghe:2021lqa} are shown in Fig.~\ref{fig:single_nucleon_positive}.
The variational bound on the nucleon mass obtained from fits to the $\mathsf{n}=0$ principal correlation function is 
\begin{equation}
aM_N \equiv a E_0^{N^+} = 1.2036(15) \ , \label{eq:MN}
\end{equation}
which is consistent with previous analyses of this ensemble~\cite{NPLQCD:2012mex,Wagman:2017tmp}.
The variational bound on the first excited state in this channel obtained from analogous fits to the $\mathsf{n}=1$ principal correlation function is 
\begin{equation}
a E_1^{N^+} = 1.928(47) \ . 
\end{equation}
This bound is weaker than the corresponding bound $aE_1^{N^+}=1.770(14)$ obtained using a set of two interpolating operators with different Gaussian smearing widths on a larger spatial volume in Ref.~\cite{Amarasinghe:2021lqa}.
The presence of significant excited-state contamination is unsurprising, since it is the second eigenvalue from a $2\times 2$ matrix of correlation functions, and further since Fig.~\ref{fig:single_nucleon_positive} shows there may still be significant time-dependence in the effective energy at imaginary times where the signal is lost to noise.
The insets in \cref{fig:single_nucleon_positive} show the overlaps with the ground and first excited states with upper- and lower-component operators.
The ground state has similar overlaps, 0.548(1) and 0.452(1), with the respective upper- and lower-component operators.
The first excited state overlaps dominantly with the lower-spin-component operator.

\begin{figure}[!t]
 \includegraphics[width=0.47\columnwidth]{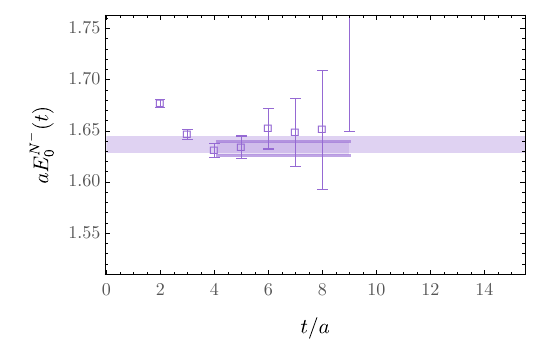}
   \caption{The negative parity nucleon channel is studied using the single negative parity nucleon operator defined in Eq.~\eqref{eq:nucleon_negative_parity}. }
   \label{fig:single_nucleon_negative}
\end{figure}

The negative-parity single-nucleon sector is orthogonal and is studied using the interpolating-operator set with a single operator
$\mathbb{S}^{N^-} = \{ N^{\Id} \}$.
The resulting diagonal correlation function is positive definite and provides a variational bound on the lowest-energy negative-parity state,
\begin{equation}
a E_0^{N^-} =  1.6357(76) \ ,
\end{equation} 
as shown in Fig.~\ref{fig:single_nucleon_negative}.

Constraints on the single-nucleon energy gaps defined in \cref{tab:notation}, $a\dNp=0.73(5)$ and $a\dNm=0.43(1)$, can inform interpretations of the two-nucleon energy spectrum.\footnote{These values only furnish variational bounds on the energy gaps under the assumption that the ground-state bound $a M_N$ has been saturated.}
In particular, two-nucleon states with energies near $a\dNp$ might be associated with $N^+ (N^+)^*$ scattering states.
Since positive-parity two-nucleon states include either zero or two negative-parity nucleon operators, states with energies near $a\dNm$ are not expected to appear in the positive-parity two-nucleon sector, while states with energies near $2a\dNm$ might be associated with $N^- N^-$ scattering states.

\subsection{Two-Nucleon Spectroscopy}

Since one- and two-nucleon correlation functions are computed on the same gauge-field configurations, statistical fluctuations are strongly correlated between them.
Consequently, correlated differences often lead to reduced statistical uncertainties.
In particular, correlated energy differences (e.g., from the differences between separate fits to one- and two-nucleon correlation functions) are more precisely determined than total two-nucleon energies.
Results are presented below for the effective energy gap $a \Delta E_{\mathsf{n}}^Q(t)$ defined in \cref{tab:notation},
where $t_0$ and $t_\text{ref}$ have been chosen as detailed in Appendix~\ref{app:t0tref}. 
Critically, variational bounds are obtained from correlated fits to the one- and two-nucleon principal correlation functions individually and \textit{not} to $a\Delta E_{\mathsf{n}}^{(2,I,\Gamma_J)}(t)$, which is not a convex sum of exponentials. 
Individual variational bounds from fits to one- and two-nucleon correlation functions also furnish results for the energy gap $a \Delta E_{\mathsf{n}}^Q$ defined in \cref{tab:notation}.
Energy gaps $a\Delta E_{\mathsf{n}}^Q$ are computed using the bootstrap methods described in Ref.~\cite{Amarasinghe:2021lqa}.
\cref{app:tabs} collects the strongest variational bounds on $a\Delta E_{\mathsf{n}}^{Q}$ for all $Q=(2,I,\Gamma_J)$ studied in the present work.

Before presenting results, it is useful to consider where the near-threshold variational bounds would be expected to appear in the absence of hadronic interactions.
In the present work, dibaryon operators are projected to all plane-wave momenta with $|\vec{n}|^2 < 5$. 
Therefore it may be expected that these operators 
will overlap predominately with finite-volume scattering states with energy gaps
\begin{equation}
a\Delta E \lesssim a\DLE \equiv 2\sqrt{(aM_N)^2 + 5(2\pi a/L)^2} - 2 a M_N=0.27\ ,
\end{equation}
where $aM_N$ is given in \cref{eq:MN} and $\DLE$ is the non-interacting energy difference corresponding to $|\vec{n}|^2 = 5$; see \cref{tab:notation}.
This energy region aligns with the energy region $a\Delta E \lesssim 0.24$ studied using plane-wave momentum-projected operators with $|\vec{n}|^2 <8$
on an ensemble with larger volume in Ref.~\cite{Amarasinghe:2021lqa}. 
Additional multi-hadron states that have small overlap with these dibaryon operators may be present at similar or somewhat lower energies than $\DLE$, in particular $N\Delta$ states with $I=1$ and $\Delta\Delta$ states with $I\in\{0,1\}$.
Using the mass of the $\Delta$ baryon (stable at these values of the quark masses) computed using this gauge-field ensemble in Ref.~\cite{Chang:2015qxa}, $aM_{\Delta} = 1.3321(21)$, the threshold for $S$-wave $\Delta\Delta$ states 
with $nn$ or $d$ quantum numbers 
is $a \Delta E^{\Delta\Delta} \equiv 2aM_{\Delta} -2 aM_N = 0.26$.
The corresponding threshold for $D$-wave $N\Delta$ states with $nn$ quantum numbers 
is\footnote{An $N\Delta$ system at rest transforms as $G_1^+ \otimes H^+ = E^+ \oplus T_1^+ \oplus T_2^+$ and can lead to $\Gamma_J = A_1^+$ when combined with $|\vec{n}|=1$ spatial wavefunctions transforming in the $E^+$ irrep~\cite{Basak:2005ir}. In the deuteron channel, isospin symmetry forbids  $N\Delta$ contributions.}
$a\Delta E^{N\Delta} \equiv \sqrt{(aM_\Delta)^2+(2\pi a/L)^2}+ \sqrt{(aM_N)^2+(2\pi a/L)^2}- 2 aM_N = 0.18$.
Variational bounds above these non-interacting thresholds still provide valid bounds satisfying the interlacing theorem.
However, such bounds seem very unlikely to have saturated due to small, though exponentially growing in $t$, contributions to the associated principal correlation functions from $N\Delta$ and $\Delta\Delta$ states. These thresholds will be indicated in the numerical results below. 

\subsubsection{The Dineutron Channel}

\begin{figure}
	\includegraphics[width=0.47\columnwidth]{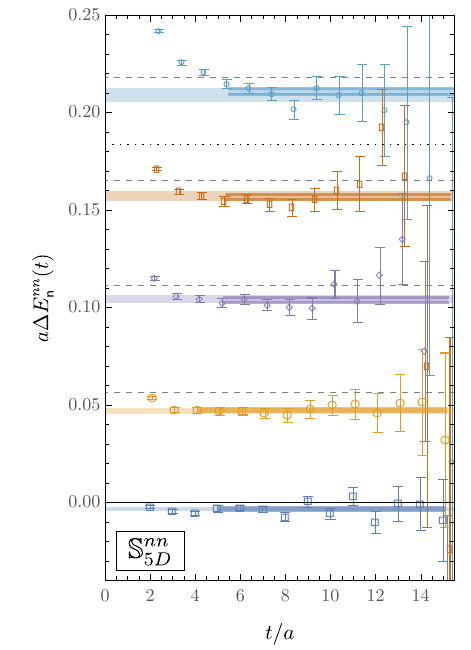}
 	\includegraphics[width=0.47\columnwidth]{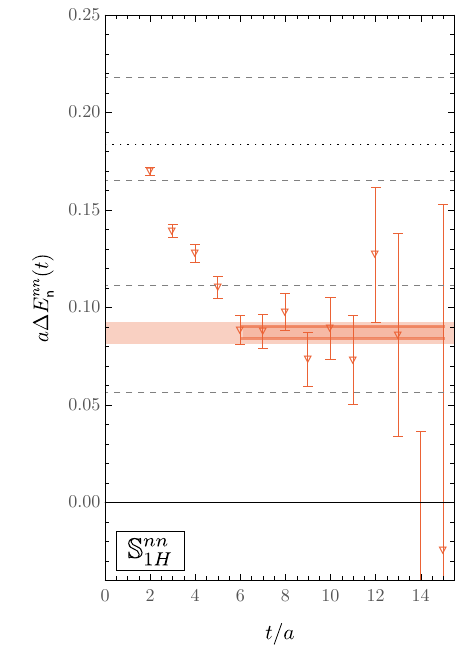} \\
 	\includegraphics[width=0.47\columnwidth]{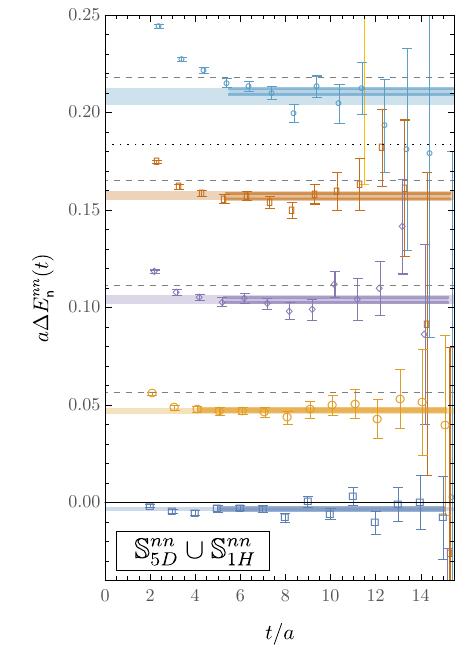}
	\includegraphics[width=0.47\columnwidth]{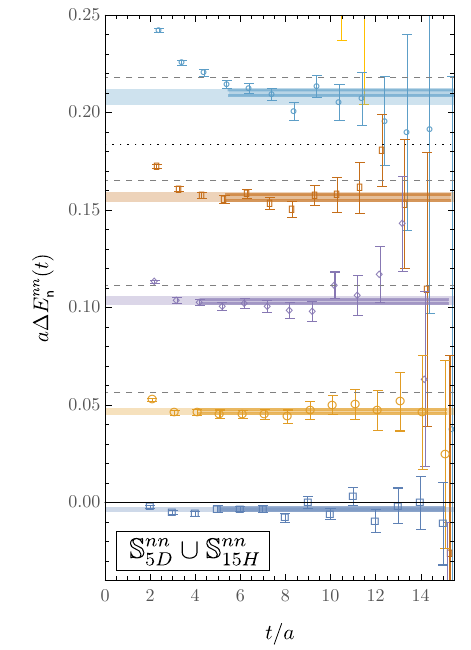}
   \caption{Effective mass functions and spectral bounds  for operator sets $\mathbb{S}_{5D}^{{nn}}$ (upper left), $\mathbb{S}_{1H}^{{nn}}$ (upper right), $\mathbb{S}_{5D}^{{nn}}\cup\mathbb{S}_{1H}^{{nn}}$ (lower left), and $\mathbb{S}_{5D}^{{nn}}\cup\mathbb{S}_{16H}^{{nn}}$ (lower right) for the dineutron, $(B,I,\Gamma_J)=(2,1,A_1^+)$. Non-interacting $NN$ and $N\Delta$ energy levels are represented as dashed and dotted horizontal lines, respectively. 
   \label{fig:dineutron_effmass}}  
\end{figure}

\begin{figure}
	\includegraphics[width=0.47\columnwidth]{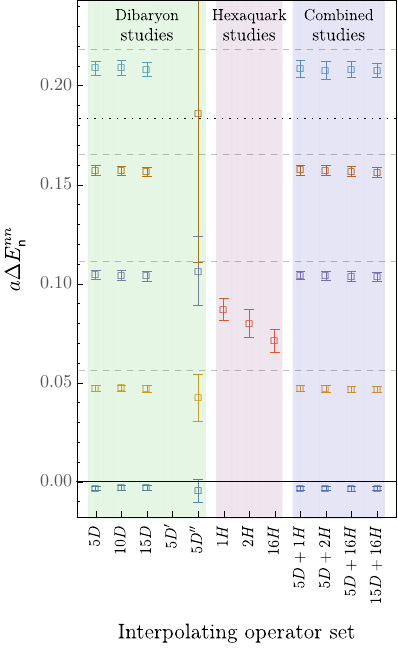}
\caption{Summary of variational bounds on the low-lying dineutron $(B,I,\Gamma_J)=(2,1,A_1^+)$ spectrum obtained from all operator sets considered in this work. Non-interacting $NN$ and $N\Delta$ energy levels are represented as dashed and dotted horizontal lines, respectively. For operator set $\mathbb{S}_{5D^{\prime}}^{nn}$, the lowest energy variational bound, $a\Delta E_0^{nn}$ is above the plotted region, and therefore no variational bounds appear. In the case of the operator set $\mathbb{S}_{5D^{\prime\prime}}^{nn}$, the uncertainty in the variational bound for $n=4$ state was larger than the plotted region, and was therefore removed for clarity.}
\label{fig:dineutron_comparison}
\end{figure}

The energy spectrum for two-nucleon systems with $I=1$ is independent of $I_z$ in the isospin-symmetric limit considered here.
Therefore the results for $nn$, $pp$, and spin-singlet $pn$ systems are equivalent.
This channel will be referred to as the dineutron channel in order to distinguish it from the $I=0$, spin-triplet deuteron channel. 
The most important contributions to the near-threshold scattering amplitude come from the $\ell=0$ partial-wave, as higher partial-waves are kinematically suppressed.
In this section, results are given for the $A_1^+$ irrep, which corresponds to states with orbital angular momentum, 
$\ell \in \{0,4,6,\ldots\}$ in the continuum and infinite-volume limit~\cite{Mandula:1982us,Mandula:1983ut,Luscher:1990ux,Luu:2011ep}.
Effective masses from four representative operator sets are shown in Fig.~\ref{fig:dineutron_effmass}.
Variational bounds from each of the operator sets defined in the previous section are presented in Fig.~\ref{fig:dineutron_comparison}.

In each of the dibaryon operator sets $\mathbb{S}_{5D}^{nn}$, $\mathbb{S}_{10D}^{nn}$, $\mathbb{S}_{15D}^{nn}$ and $\mathbb{S}_{5D^{\prime\prime}}^{nn}$ shown in Fig.~\ref{fig:dineutron_comparison}, there exists a variational bound which lies near the non-interacting two-nucleon energy levels for $\mathsf{n}\in\{0,1,2,3,4\}$. This placement suggests that these bounds may be associated with attractive finite-volume $NN$ scattering states. Results for the sets $\mathbb{S}_{10D}^{nn}$, $\mathbb{S}_{15D}^{nn}$ and $\mathbb{S}_{5D^{\prime\prime}}^{nn}$ are consistent with those from the $\mathbb{S}_{5D}^{nn}$ set for all variational bounds below $\DLE$. The operator set $\mathbb{S}_{5D^\prime}^{nn}$ contains operators constructed from the product of two negative parity nucleon operators. It is interesting to note that the lowest variational bound is greater than $\DLE$, and thus no variational bounds are shown on Fig.~\ref{fig:dineutron_comparison} for this operator set.
The lowest energy bound for each dibaryon-only operator set except $\mathbb{S}_{5D^{\prime}}^{nn}$ is just below the two-nucleon threshold.
For all operator sets consided, the lowest variational bound is higher than the previous estimates of the ground state energy on this ensemble using asymmetric source and sink interpolating operators
and is therefore consistent in the sense of a variational bound.\footnote{Under the assumption that this bound has been saturated, the analysis of these operator sets does not suggest a bound state in contrast to studies using asymmetric source and sink interpolating operators. See Ref.~\cite{Amarasinghe:2021lqa} for an extensive discussion of the meaning of this contrast.}

\begin{figure}[!t]
	\includegraphics[width=0.47\columnwidth]{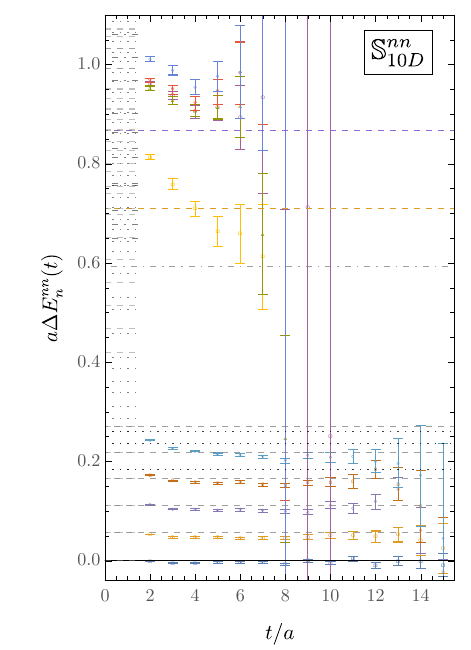}
	\includegraphics[width=0.47\columnwidth]{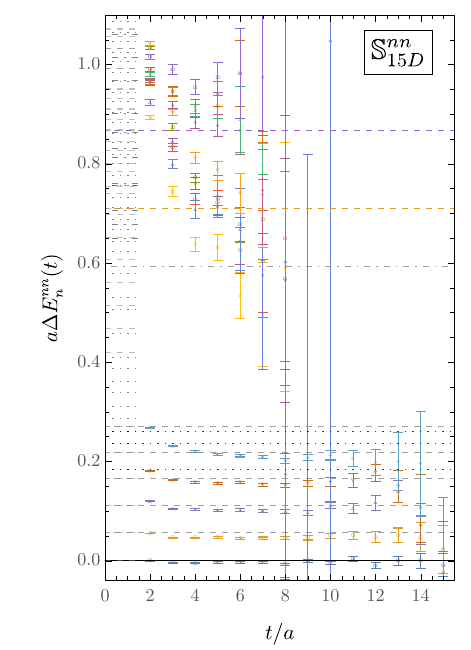}
	\includegraphics[width=0.47\columnwidth]{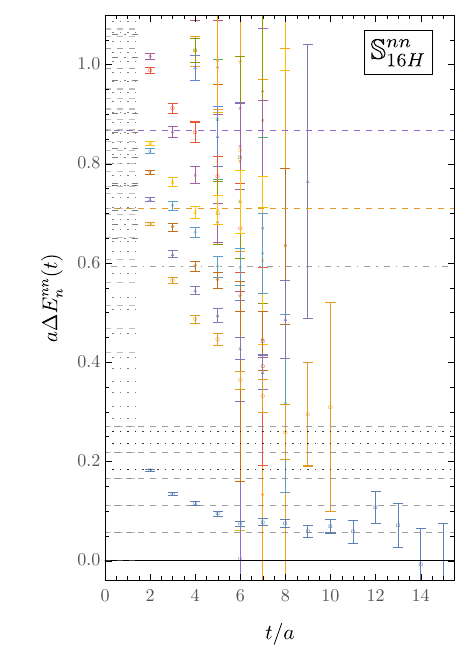}
	\includegraphics[width=0.47\columnwidth]{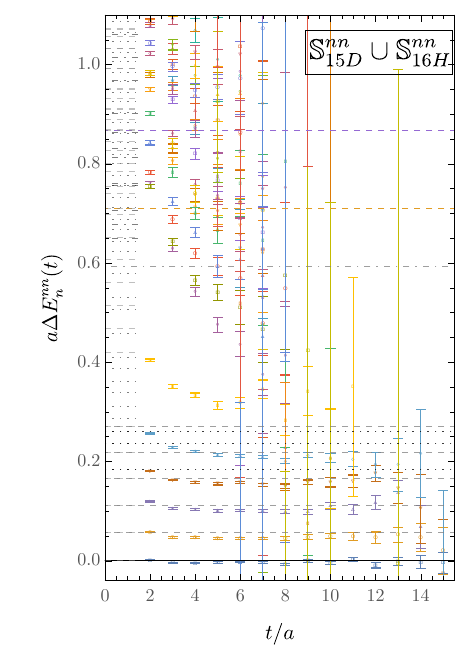}
   \caption{The effective mass spectra showing all of the variational bounds obtained with the interpolating-operator sets indicated. Non-interacting energy levels are represented as follows: 
   dashed gray lines for $NN$; dotted black lines for $N\Delta$ and $\Delta\Delta$; and dot-dashed gray lines for $NN\pi$.
   Lines for non-interacting energy levels above $a\DLE$ are cut off for visual clarity. 
   Thresholds related to single-nucleon excitations are also indicated: 
   dashed orange line for $a\dNp$ and dashed purple line for $a\dNm$.
   \label{fig:dineutron_higher} }
\end{figure}

Additional bounds not present for $\mathbb{S}_{5D}^{nn}$ appear in results for $\mathbb{S}_{10D}^{nn}$, $\mathbb{S}_{15D}^{nn}$, $\mathbb{S}_{5D^\prime}^{nn}$ and $\mathbb{S}_{5D^{\prime \prime}}^{nn}$ at higher energies, as shown in Fig.~\ref{fig:dineutron_higher}.
For $\mathbb{S}_{10D}^{nn}$, which contains dibaryon operators constructed from products of negative-parity nucleon operators, all but one of the additional bounds are above the threshold $2a\dNm \approx  0.87$.  
For $\mathbb{S}_{15D}^{nn}$, several additional bounds appear around $a\dNp \approx 0.74 $.
This suggests that these operators predominantly overlap with $N^- N^-$ and  $N^+ (N^+)^*$ scattering states, although the presence of one bound below $a\dNp$ shows that the additional operators do not overlap exclusively with such states.

The sets $\mathbb{S}_{1H}^{nn}$, $\mathbb{S}_{2H}^{nn}$ and $\mathbb{S}_{16H}^{nn}$ furnish variational bounds for energy eigenstates obtained from operator sets which contain only hexaquark operators.
While these sets may seem unnatural since they lack the dibaryon operators, which have been observed to have significant overlap with
low-energy two-nucleon-like states, the interlacing theorem implies that  $\mathbb{S}_{16H}$ nevertheless provides valid variational bounds on the lowest sixteen energy eigenvalues in the theory. 
The variational bounds determined from these interpolating-operator sets are shown in Figs.~\ref{fig:dineutron_effmass} - \ref{fig:dineutron_higher}. 
For each hexaquark operator set studied here, it is possible to observe a plateau at late Euclidean times in the $\mathsf{n}=0$ principal correlator. 
In each operator set, the variational bound on the lowest energy is $a \Delta E_0^d \approx 0.07$ above the two-nucleon threshold.
For each of these hexaquark-only operator sets, the other variational bounds are above $\DLE$ and do not exhibit clear plateaus, as shown in Fig.~\ref{fig:dineutron_higher}. 

For the operator sets which include the upper-spin-component positive-parity dibaryon operators and either one, two, or sixteen hexaquark operators ($\mathbb{S}_{5D}^{nn} \cup \mathbb{S}_{1H}^{nn}$, $\mathbb{S}_{5D}^{nn}\cup\mathbb{S}_{2H}^{nn}$ and $\mathbb{S}_{5D}^{nn}\cup\mathbb{S}_{16H}^{nn}$) and for the operator set which includes all dibaryon and hexaquark operators ($\mathbb{S}_{15D}^{nn}\cup \mathbb{S}_{16H}^{nn}$), the resulting lowest five variational bounds are consistent at $1\sigma$ with the results from $\mathbb{S}_{5D}^{nn}$. Examining the overlap factors shown in Fig.~\ref{fig:B2I1A1_FV_fits}, $H_3$ (which can be shown to be equivalent to the product of two color-singlet nucleon operators) overlaps with states which also have strong overlap with dibaryon operators. It is interesting to note that while this operator has strongest overlap with $\mathsf{n}=5$, it also has a sizable overlap with $\mathsf{n}=0,1,\dots,4$. Note that the operator $H_3$ is equivalent to a sum over all dibaryon operators of all possible lattice momenta, but the $H_3$-$H_3$ correlation function is not the sum of all dibaryon correlation functions. 

\begin{figure}[!t]
	\includegraphics[width=0.8\columnwidth]{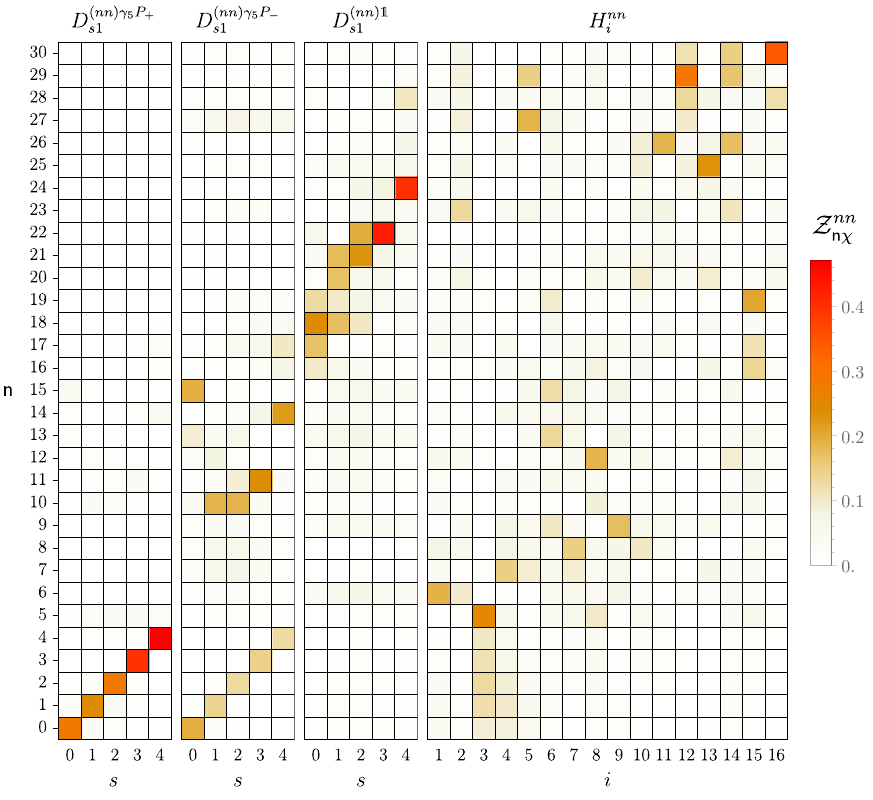}
   \caption{\label{fig:B2I1A1_Zplot} Results for relative overlap factors 
   $\mathcal{Z}_{\mathsf{n} \chi}^{nn}$
   for operator set $\mathbb{S}_{15D}^{nn}\cup\mathbb{S}_{16H}^{nn}$. Uncertainties are not shown here but are shown by error bars in histograms for the 6 lowest-energy states in Fig.~\ref{fig:B2I1A1_FV_fits}.  
   }
\end{figure}

The set $\mathbb{S}_{16H}^{nn}$ contains many operators which cannot be written as the local product of two color-singlet three-quark operators. Such operators potentially probe the hidden-color components of the two-nucleon wavefunction. 
While these operators do not provide strong variational bounds on the low-energy dineutron spectrum and exhibit large statistical fluctuations, evidence exists for states higher in the spectrum having statistically-significant overlaps with these novel operators. This can be seen in Fig.~\ref{fig:B2I1A1_Zplot}, where the relative overlaps of each GEVP eigenvector onto each operator is shown.
In particular, the $\mathsf{n}=5$ dineutron level has largest overlaps with hexaquark operators with $T^{AAS}$ color structures arising from products of two color-singlet baryon operators and includes an admixture of upper-spin-component and lower-spin-component operators of this form.
Operators involving the color tensors $T^{AAA}$ and $T^{SSS}$ that cannot be constructed from products of color-singlet baryons are associated with the appearance of variational bounds above $\DLE$.
It is noteworthy that several of these variational bounds appear between $\DLE$ and $a\dNp$, which suggests that hidden-color operators in this channel predominantly overlap with lower-energy states than operators involving single-nucleon excitations.
The structure of these states may exhibit novel features associated with the presence of hidden color that should be explored in future work.

\subsubsection{The Deuteron Channel}
\begin{figure}
	\includegraphics[width=0.47\columnwidth]{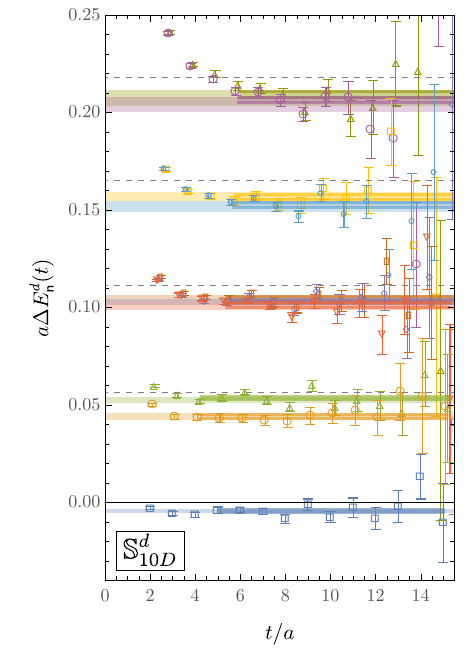}
 	\includegraphics[width=0.47\columnwidth]{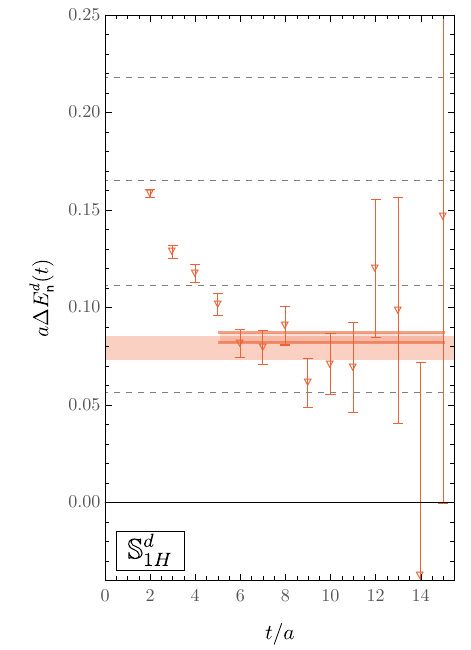} \\
 	\includegraphics[width=0.47\columnwidth]{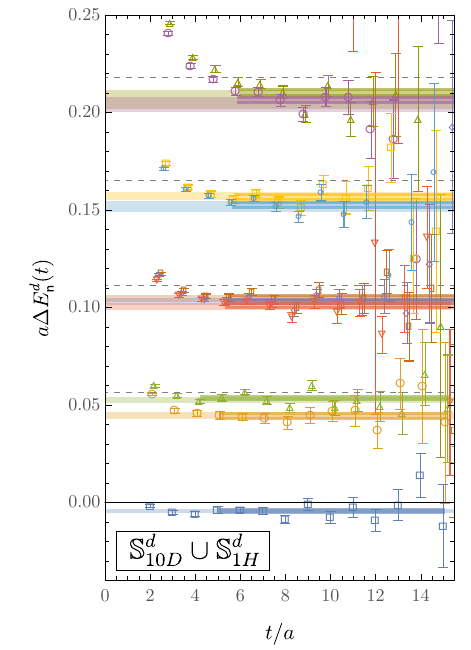}
	\includegraphics[width=0.47\columnwidth]{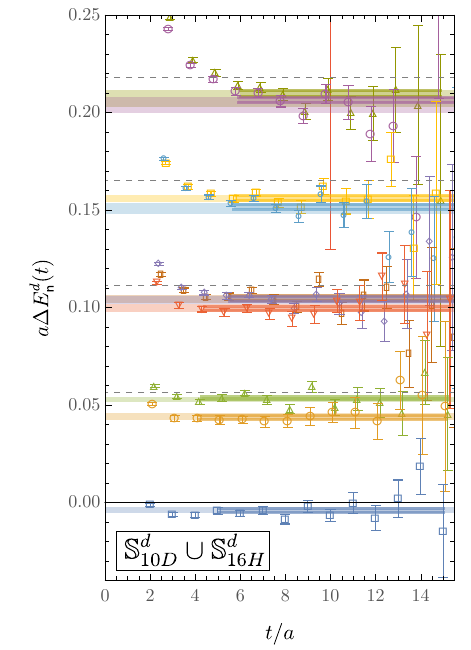}
   \caption{Effective mass spectra for operator sets $\mathbb{S}_{{10D}}^{{d}}$ (top left), $\mathbb{S}_{1H}^{{d}}$ (top right), $\mathbb{S}_{{10D}}^{{d}}\cup\mathbb{S}_{1H}^{d}$ (bottom left), and $\mathbb{S}_{{10D}}^{{d}}\cup\mathbb{S}_{16H}^{{d}}$ (bottom right) for the deuteron, 
   {$(B,I,\Gamma_J)=(2,0,T_1^+)$}.
   Non-interacting $NN$ energy levels are represented as dashed horizontal lines.
   \label{fig:deuteron_effmass}}  
\end{figure}

\begin{figure}
	\includegraphics[width=0.47\columnwidth]{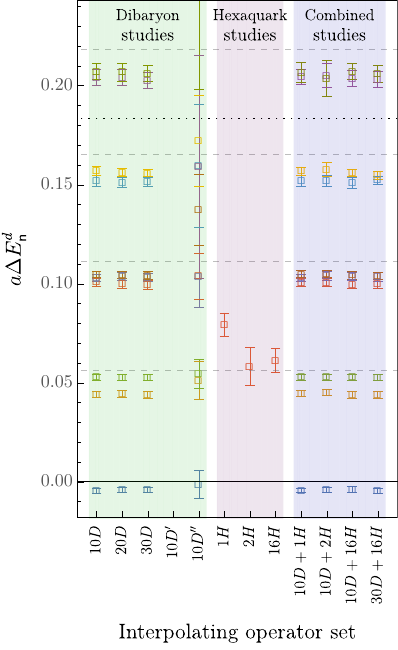}
\caption{Summary of low-lying deuteron 
{$(B,I,\Gamma_J) = (2,0,T_1^+)$}
spectra obtained from all operator sets considered in this work. Non-interacting $NN$ energy levels are represented as dashed horizontal lines. For operator set $\mathbb{S}_{5D^{\prime}}^{nn}$, the lowest energy variational bound, $a\Delta E_0^{nn}$ is larger than the plot window, and therefore no variational bounds appear.
}
\label{fig:deuteron_comparison}
\end{figure}

Results are presented for the $J_z=0$ row of the $T_1^+$ irrep, which contains the $\ell=0$ partial-wave contribution in the infinite volume limit.
The resulting variational bounds on the spectra are shown in Figs.~\ref{fig:deuteron_effmass} and \ref{fig:deuteron_comparison}. It is important to note that unlike the dineutron channel, multiple linearly independent dibaryon interpolating operators in a given row of the cubic group representation can be constructed. 
These arise because the total angular momentum irrep is a tensor product of the ``orbital" angular momentum and the spin, $\Gamma_J=\Gamma_{\ell} \otimes \Gamma_{\rm spin}$.
The total-angular-momentum irrep containing the deuteron ($\Gamma_J=T_1^+$) includes contributions from spatial wavefunctions with $\Gamma_{\ell} \in \{A_1^+,E^+, T_2^+\}$ (other irreps are only relevant for higher-momentum dibaryon operators. For example, $\Gamma_{\ell} = T_1^+$ contributes for $s\geq 5$). 
Thus, these operators are sensitive to the energy splittings arising from the orbital angular momenta of the states.
The multiplicities of operators with total angular momentum irrep $\Gamma_J$ are given in Table~III of Ref.~\cite{Amarasinghe:2021lqa}. In particular, for $\Gamma_J=T_1^+$, the multiplicities are $N_s^{(0,T_1^+)}=\{1,2,3,2,2\}$ for $s=\{0,1,2,3,4\}$, respectively.
Consequently, if the low-energy deuteron spectrum could be described purely in terms of  non-interacting energy levels, then one should expect  $N_s^{(0,T_1^+)}$ nearby variational bounds for each non-interacting energy level.

\begin{figure}[!t]
	\includegraphics[width=0.47\columnwidth]{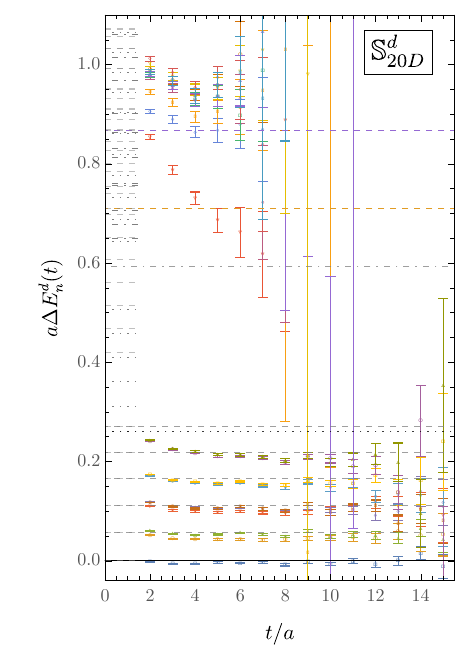}
	\includegraphics[width=0.47\columnwidth]{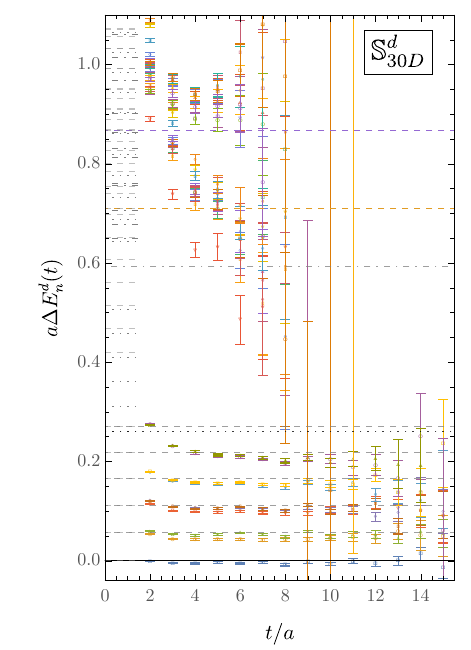}
	\includegraphics[width=0.47\columnwidth]{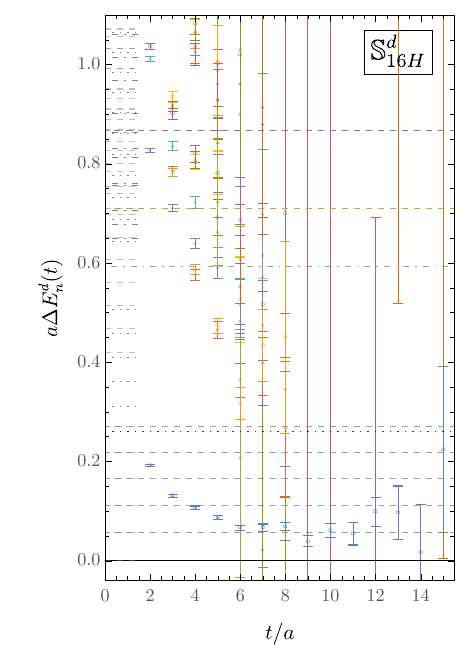}
	\includegraphics[width=0.47\columnwidth]{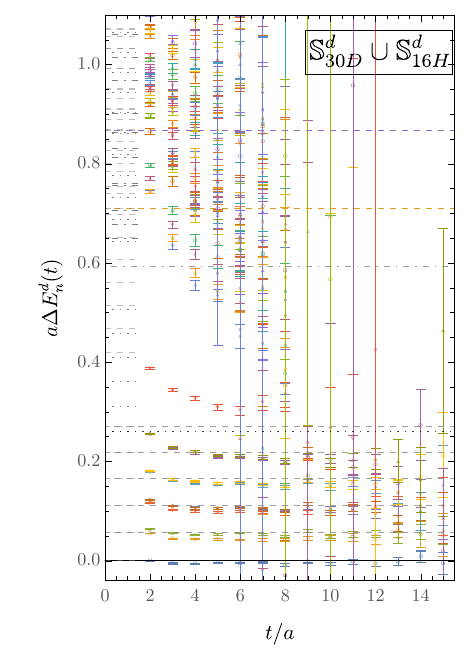}
   \caption{The effective mass spectra showing all of the variational bounds obtained with the interpolating-operator sets indicated.  
   Non-interacting energy levels are represented as in Fig.~\ref{fig:dineutron_higher} except that $N\Delta$ states are excluded from the deuteron channel by isospin.
    \label{fig:deuteron_higher}}
\end{figure}

For the operator sets $\mathbb{S}_{{10D}}^{d}$, $\mathbb{S}_{{20D}}^{d}$, $\mathbb{S}_{{30D}}^{d}$ and $\mathbb{S}_{{10D^{\prime \prime}}}^{d}$, exactly $N_s^{(0,T_1^+)}$ nearby variational bounds for each $s$ are observed, in agreement with the expected counting of non-interacting degenerate copies. The variational bounds for the two $\mathsf{n}=1$ states obtained from the sets $\mathbb{S}_{{10D}}^{d}$, $\mathbb{S}_{{20D}}^{d}$ and $\mathbb{S}_{{30D}}^{d}$ differ by several standard deviations. 
If these variational bounds were saturated, these differences could be used to study $S$-$D$ partial-wave mixing~\cite{Briceno:2013bda,Orginos:2015aya}.
Additional variational bounds appearing only for $\mathbb{S}_{{20D}}^{d}$ and $\mathbb{S}_{{30D}}^{d}$ are present at higher energies  that, as in the dineutron channel, are consistent with $2 a\dNm$ and $a\dNp$, respectively, as shown in Fig.~\ref{fig:deuteron_higher}.

The sets $\mathbb{S}_{1H}^{d}$, $\mathbb{S}_{2H}^{d}$ and $\mathbb{S}_{16H}^{d}$ again provide much less-constraining variational bounds on the low-energy eigenvalues of the deuteron system than the dibaryon operator-sets, but the ${\sf n}=0$ effective mass function exhibits a plateau at 
$a\Delta E^{d} \approx 0.06$. 
Similarly to the dineutron channel, operator sets which contain both dibaryon and hexaquark operator sets produce low-energy variational bounds which are consistent with the variational bounds obtained from $\mathbb{S}_{{10D}}^{d}$ within one standard deviation. Therefore, in both the dineutron and deuteron channels, the addition of novel hexaquark or dibaryon operators does not appear to produce a qualitatively different set of low-energy variational bounds at the current level of statistical precision.

\begin{figure}[!t]
	\includegraphics[width=\columnwidth]{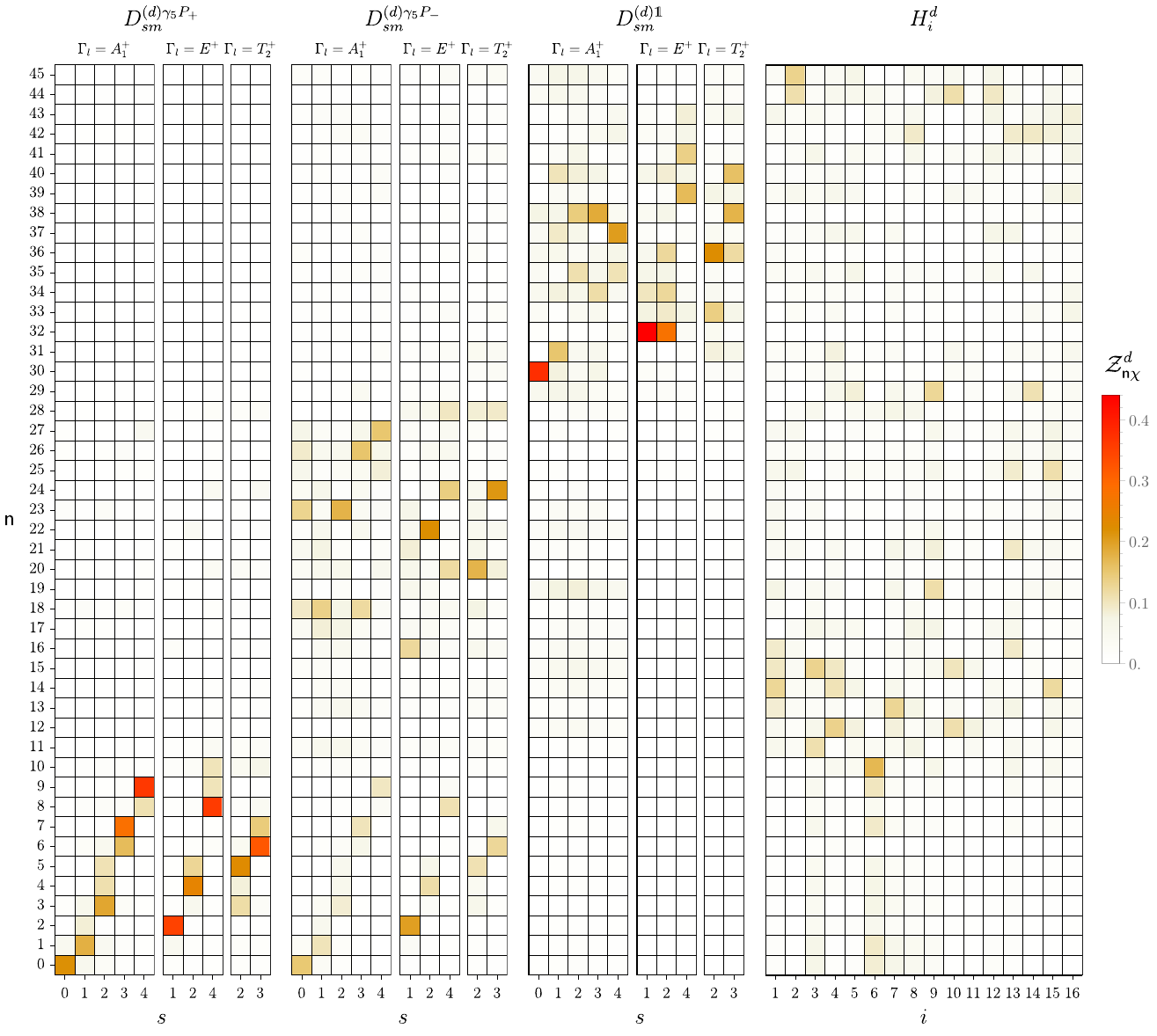}
   \caption{ Results for relative overlap factors 
   $\mathcal{Z}_{\mathsf{n} \chi}^{d}$
   for the operator set $\mathbb{S}_{{30D}}^{d}\cup \mathbb{S}_{16H}^{d}$. Uncertainties are not shown here but are shown for the 11 lowest-energy states by the error bars in the histograms in Figs.~\ref{fig:B2I0T1_FV_fits}-\ref{fig:B2I0T1_FV_fits_2}.
\label{fig:B2I0T1_Zplot}
}\end{figure}

Further understanding of these results can be obtained by examining the relative overlaps, shown in Fig.~\ref{fig:B2I0T1_Zplot} and Appendix~\ref{app:energies}. As in the dineutron channel, it can be seen that the dibaryon operators have large relative overlaps with GEVP eigenvectors whose eigenvalues are close to the  non-interacting two-nucleon energy levels ($\mathsf{n}\in \{0,1,3,4,5\}$), while hexaquark operators have larger overlap with states higher in the spectrum. 

\section{Discussion}
\label{sec:discussion}

In order to have confidence in the results of variational calculations, it is important to explore the physically relevant regions of Hilbert space. Since these are not known \textit{a priori}, studies of the operator dependence of the obtained variational bounds are an important step towards gaining confidence in the employed operator sets. From the studies presented in this work, no evidence is found for the importance of the novel operators that have been introduced here in constraining low-energy variational bounds in the dineutron and deuteron channels. Instead, these operators are seen to have larger overlap with states higher in the spectrum. It is important to note that these conclusions are valid only at the current statistical prevision, and could change with more precise numerical data. In addition, since the Hilbert space is formally infinite, the absence of large changes in variational bounds should not be seen as confirmation that operator sets which contain only positive-parity dibaryon operators are sufficient to describe the low-energy two-baryon spectrum. This point is particularly relevant for these systems because previous calculations which studied the spectrum using asymmetric correlators observed evidence for the presence of bound states in both the dineutron and deuteron channels.

In order to make physical statements, quantities determined from LQCD calculation must be extrapolated to the continuum limit. 
Due to the use of a perturbatively-improved action in this study, the leading lattice artifacts in energies and energy-differences are expected to arise at $\mathcal{O}(g^2 a \lqcd)$ and $\mathcal{O}(a^2\lqcd^2)$ (here, $g$ is the strong coupling at scale $\mu\sim1/a$ and $\lqcd\approx0.3$ GeV is the typical QCD scale), both of which are $\alt0.1$.
However, since the current study is performed at a single lattice spacing,  the magnitude of the lattice artifacts that are present has not been quantified. Given this, the possibility that the variational bounds observed in both two-nucleon channels  have large lattice artifacts which would shift their energies to significantly different values than in the continuum cannot be ruled out. 
A quantitative study of lattice artifacts in the obtained variational bounds is left for future work. It should also be noted that significant changes between variational bounds at different lattice spacings do not necessarily imply significant differences between energy eigenvalues.

\begin{figure}
    \centering
 	\includegraphics[width=0.45\columnwidth]{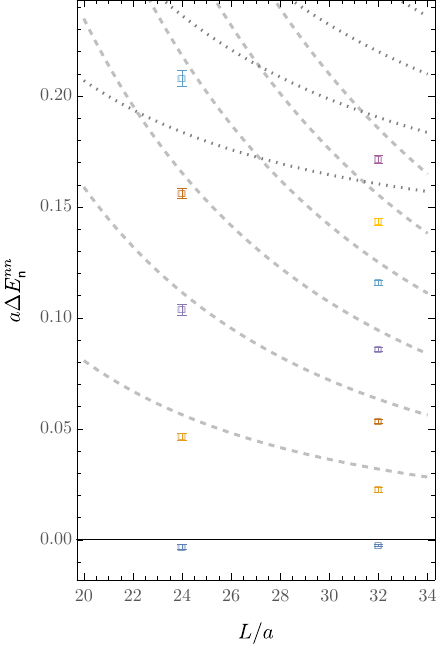}
	\includegraphics[width=0.45\columnwidth]{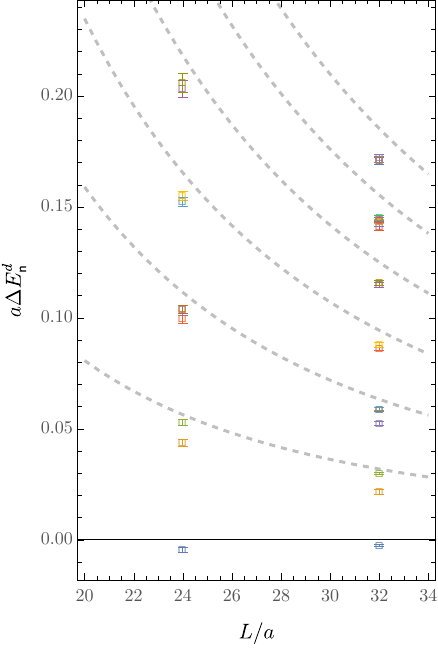}
    \caption{The strongest variational  bounds achieved here in the dineutron and deuteron channels, compared to best variational bounds obtained at a larger volume in Ref.~\cite{Amarasinghe:2021lqa}. Non-interacting $NN$ and $N\Delta$ energy levels are represented as dashed and dotted lines, respectively. 
    \label{fig:volume_dependence} }
\end{figure}

By combining the results obtained from operator set $\mathbb{S}_{0}$ in Ref.~\cite{Amarasinghe:2021lqa} (corresponding to 14 positive-parity, upper-spin-component dibaryon operators, plus a single $T^{AAS}$ hexaquark operator) with the results from this analysis, it is possible to study the volume dependence of variational bounds for the finite-volume spectra as depicted in Fig.~\ref{fig:volume_dependence}. For both ensembles considered, there exists $N_s^{(I,\Gamma)}$ variational bounds near each non-interacting two-nucleon energy level.  Although these variational bounds are negatively shifted with respect to the non-interacing levels, the magnitude of this shift is not large enough to provide evidence for bound states in either channel.

\begin{figure}
\centering
\includegraphics[width=\textwidth]{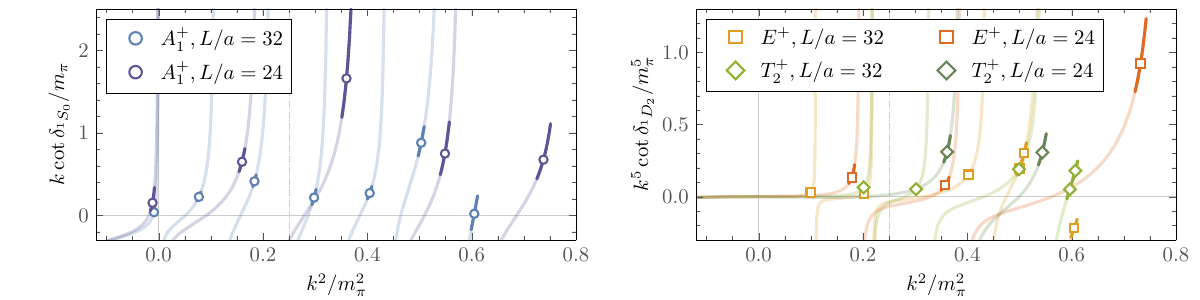}
\caption{Values of the phase-shift from different irreps for the dineutron channel combining the results from this work ($L/a=24$) and Ref.~\cite{Amarasinghe:2021lqa} ($L/a=32$). The opaque lines represent the statistical and systematic uncertainties from the fit to the correlation matrix, while the translucent lines represent the bounds from the interlacing theorem. The vertical dot-dashed line indicates the radius of convergence of the effective range expansion due to the $t$-channel cut.}
\label{fig:I1_phase_shift}
\end{figure}

Having extracted variational bounds for the low-energy dineutron and deuteron spectra in two physical volumes, it is natural to attempt to compute the associated phase-shifts using the finite-volume quantization-conditions described in Refs.~\cite{Luscher:1986pf,Luu:2011ep,Briceno:2013lba}.
However, the Euclidean time extents over which statistically-significant signals for energy shifts have been extracted in the current study are smaller than the inverse of the energy gap between the lowest two finite-volume energy levels of non-interacting nucleons and are not sufficient to have confidence that any of the variational bounds have been saturated.
One should therefore not assume that variational bounds are in one-to-one correspondence with energy eigenvalues.
The interlacing theorem rigorously guarantees that at least $n$ energy eigenvalues sit below the $n$th variational bound, and these one-sided intervals can be mapped to one-sided intervals in the $(k^2,k\cot\delta)$-plane using the quantization conditions.
However, due to the singularity structure of the quantization conditions, these intervals do not provide meaningful constraints on the phase shifts.
Precise constraints can only be obtained by making the assumption that variational bounds are saturated and correspond to stochastic estimates of energy eigenvalues. Assuming that only the lowest partial-wave contributes to a given cubic irrep (\textit{e.g.}, ${}^1S_0$ in $I=1$, $\Gamma=A_1^+$), one can obtain the corresponding phase-shifts, as shown in Figs.~\ref{fig:I1_phase_shift} and \ref{fig:I0_phase_shift}. 
Further details on the quantization conditions used here are discussed in Appendix G of Ref.~\cite{Amarasinghe:2021lqa}. 
\begin{figure}
\centering
\includegraphics[width=\textwidth]{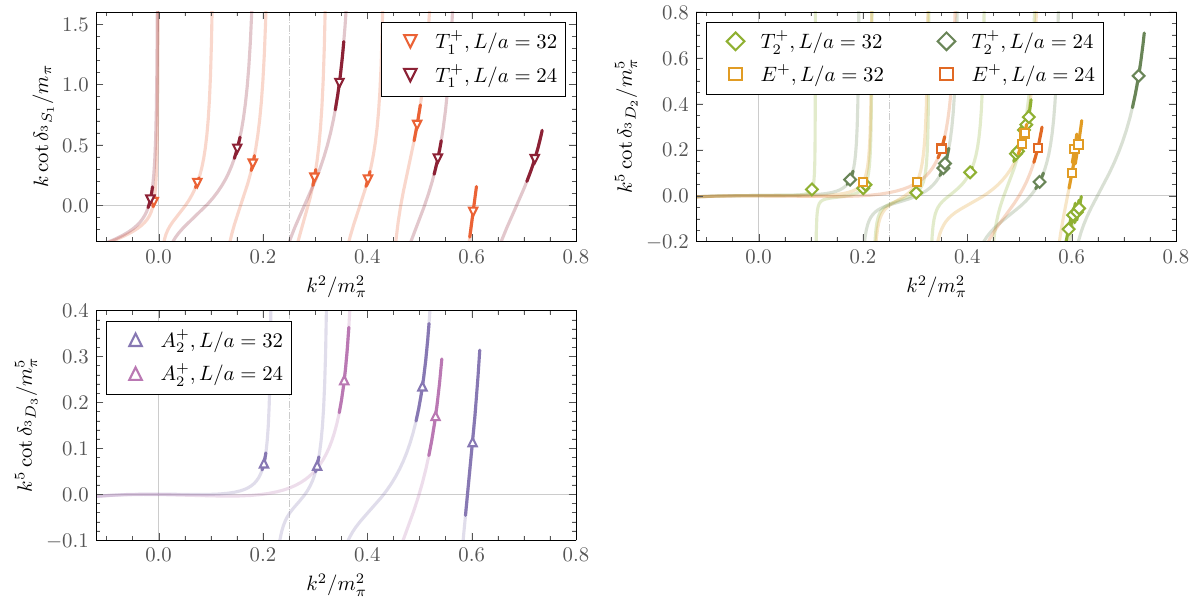}
\caption{Values of the phase-shift from different irreps for the deuteron channel combining the results from this work ($L/a=24$) and Ref.~\cite{Amarasinghe:2021lqa} ($L/a=32$). Details are as in Fig.~\ref{fig:I1_phase_shift}.}
\label{fig:I0_phase_shift}
\end{figure}

\section{Summary and Outlook}
\label{sec:conclusion}

In this work, the two-nucleon spectrum is studied using LQCD at quark masses corresponding to a pion mass of $m_\pi\approx 806~\si{MeV}$ using the variational method, leading to bounds on the finite-volume energy eigenstates. The effects on the variational bounds for the two-nucleon spectrum from various choices of dibaryon and local hexaquark operators with both $I=0$ and $I=1$ are quantified.
In particular, this study analyzes 
for the first time the 
impact of including products of negative-parity nucleon operators as well as dibaryon operators containing the lower-spin-components of quark fields.
These operators lead to minimal changes to the low-energy variational bounds obtained using only upper-spin-component dibaryon operators.
Operator sets which contain only hexaquark operators produce variational bounds of the low-energy spectrum that are much less constraining than those from operator sets which contain dibaryon operators. 
Operator sets which contain both dibaryon and hexaquark operators produce low-energy variational bounds consistent within uncertainties with operator sets which contain only dibaryon operators constructed from the positive-parity, upper-spin components. 
Hidden-color hexaquark operators make significant contributions to the overlap factors associated with some of these bounds, which suggests that there may be two-nucleon excited states with novel structure related to hidden-color components.

Since the calculations presented here are performed at a single lattice spacing, it is possible that the variational bounds contain significant lattice artifacts, and bounds could move in either direction as they approach the continuum limit. 
Further studies will explore the magnitude of the lattice artifacts in these systems and study the continuum limit of the extracted variational bounds.

\section*{Acknowledgements}
We are grateful to Zohreh Davoudi, George Fleming, Anthony Grebe, and Martin Savage for helpful discussions.
Computations in this work were carried out using the Chroma~\cite{Edwards:2004sx}, Qlua~\cite{qlua},
QUDA~\cite{Clark:2009wm,Babich:2011np,Clark:2016rdz}, and Tiramisu~\cite{tiramisu} software libraries. Data analysis used NumPy~\cite{harris2020array} and Julia~\cite{Julia-2017,mogensen2018optim}, and figures were produced using Mathematica \cite{Mathematica} and Matplotlib \cite{Hunter:2007}. 

This research used resources of the Oak Ridge Leadership Computing Facility at the Oak Ridge National Laboratory, which is supported by the Office of Science of the U.S. Department of Energy under Contract number DE-AC05-00OR22725
and the resources of the National Energy Research Scientific Computing Center (NERSC), a Department of Energy Office of Science User Facility using NERSC award NP-ERCAPm747. 
The authors thankfully acknowledge the computer resources at MareNostrum and the technical support provided by Barcelona Supercomputing Center (RES-FI-2022-1-0040). 
The research reported in this work made use of computing facilities of the USQCD Collaboration, which are funded by the Office of Science of the U.S. Department of Energy.

WD, WJ and PES are supported in part by the U.S. Department of Energy, Office of Science under grant Contract Number DE-SC0011090 and by the SciDAC5 award DE-SC0023116.
PES is additionally supported by the U.S. DOE Early Career Award DE-SC0021006 and by Simons Foundation grant 994314 (Simons Collaboration on Confinement and QCD Strings).
WD and PES are additionally supported by the National Science Foundation under Cooperative Agreement PHY-2019786 (The NSF AI Institute for Artificial Intelligence and Fundamental Interactions, http://iaifi.org/). 
MI is partially supported by the Quantum Science Center (QSC), a National Quantum Information Science Research Center of the U.S.\ Department of Energy. 
AP and RJP have been supported by the projects CEX2019-000918-M (Unidad de Excelencia “María de Maeztu”), RED2022-134428-T, PID2020-118758GB-I00, financed by MICIU/AEI/10.13039/501100011033/ and FEDER, UE, as well as by the EU STRONG-2020 project, under the program H2020-INFRAIA-2018-1 Grant Agreement No. 824093.
This manuscript has been authored by Fermi Research Alliance, LLC under Contract No. DE-AC02-07CH11359 with the U.S. Department of Energy, Office of Science, Office of High Energy Physics.

\appendix

\section{Choosing $t_0$ and $t_{\rm ref}$}
\label{app:t0tref}

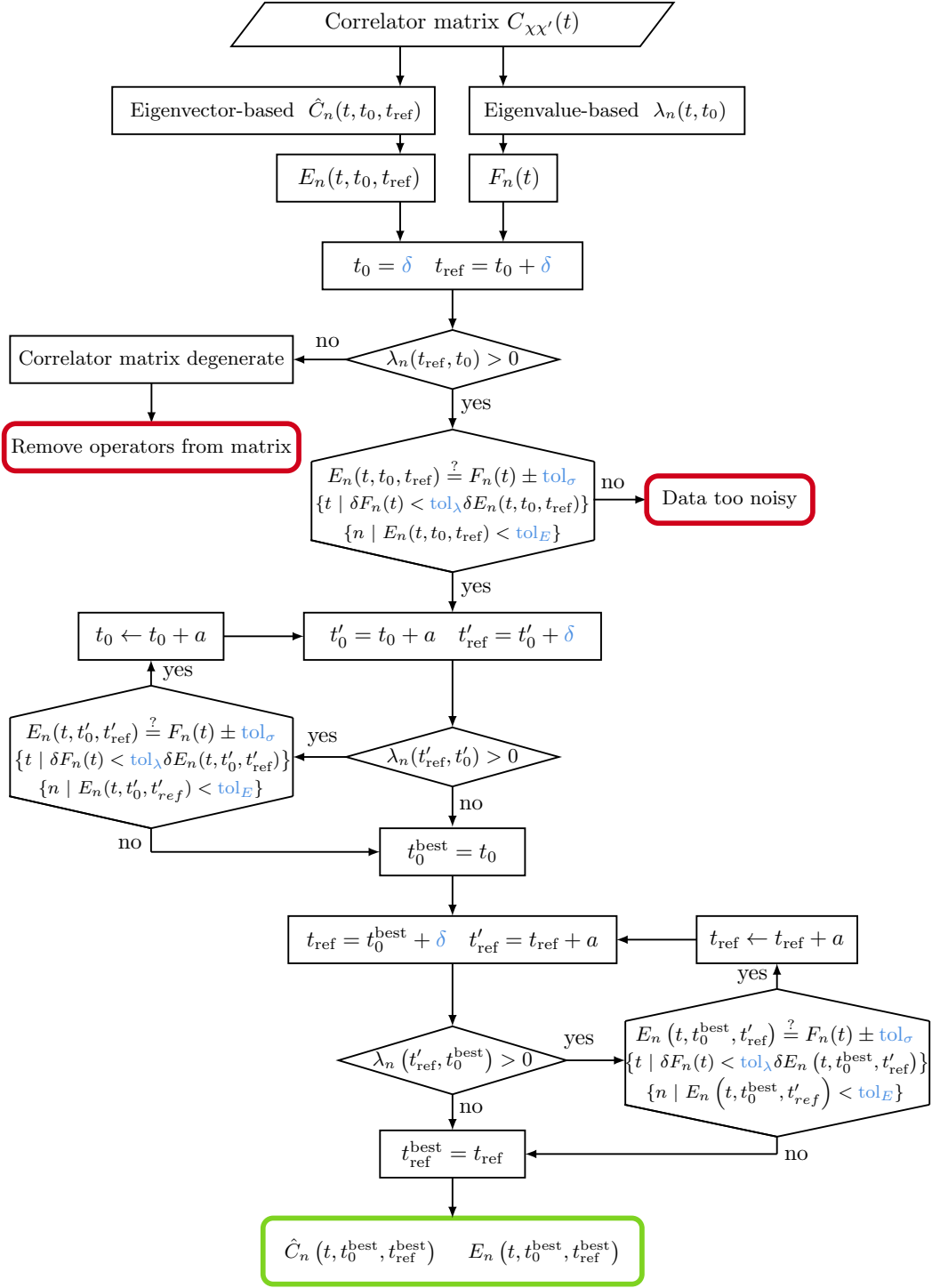
\begin{figure}[!h]
\centering
\resizebox{0.85\linewidth}{!}{\begin{tikzpicture}[x=0.75pt,y=0.75pt,yscale=-1,xscale=1]

\draw   (239.9,16) -- (482,16) -- (462.1,42) -- (220,42) -- cycle ;
\draw[->]    (320,42) -- (320,66) ;
\draw   (247,106) -- (340.5,106) -- (340.5,134) -- (247,134) -- cycle ;
\draw   (149.5,66) -- (340.5,66) -- (340.5,94) -- (149.5,94) -- cycle ;
\draw   (274,158) -- (428,158) -- (428,186) -- (274,186) -- cycle ;
\draw  [color={rgb, 255:red, 126; green, 211; blue, 33 }  ,draw opacity=1 ][line width=2.25]  (238.79,742.25) .. controls (238.79,738.94) and (241.48,736.25) .. (244.79,736.25) -- (457.21,736.25) .. controls (460.52,736.25) and (463.21,738.94) .. (463.21,742.25) -- (463.21,769.35) .. controls (463.21,772.67) and (460.52,775.35) .. (457.21,775.35) -- (244.79,775.35) .. controls (241.48,775.35) and (238.79,772.67) .. (238.79,769.35) -- cycle ;
\draw[->]    (288,227.5) -- (256.5,227.5) ;
\draw[->]    (381,42) -- (381,66) ;
\draw   (361,66) -- (524,66) -- (524,94) -- (361,94) -- cycle ;
\draw[->]    (320,94) -- (320,106) ;
\draw[->]    (381,94) -- (381,106) ;
\draw   (361,106) -- (413,106) -- (413,134) -- (361,134) -- cycle ;
\draw[->]    (320,134) -- (320,158) ;
\draw[->]    (381,134) -- (381,158) ;
\draw[->]    (351,186) -- (351,210) ;
\draw   (351,210) -- (414,227.5) -- (351,245) -- (288,227.5) -- cycle ;
\draw[->]    (351,245) -- (351,269) ;
\draw   (89,213.5) -- (256.5,213.5) -- (256.5,241.5) -- (89,241.5) -- cycle ;
\draw[->]    (172.75,241.5) -- (172.75,265.5) ;
\draw  [color={rgb, 255:red, 208; green, 2; blue, 27 }  ,draw opacity=1 ][line width=2.25]  (85.5,272.5) .. controls (85.5,268.63) and (88.63,265.5) .. (92.5,265.5) -- (253,265.5) .. controls (256.87,265.5) and (260,268.63) .. (260,272.5) -- (260,286.5) .. controls (260,290.37) and (256.87,293.5) .. (253,293.5) -- (92.5,293.5) .. controls (88.63,293.5) and (85.5,290.37) .. (85.5,286.5) -- cycle ;
\draw    (267,334.5) -- (351,353.5) ;
\draw    (351,269) -- (267.5,288) ;
\draw    (351,353.5) -- (435,334.5) ;
\draw    (435,334.5) -- (435,288) ;
\draw    (435,288) -- (351,269) ;
\draw    (267,334.5) -- (351,353.5) ;

\draw    (267.5,288) -- (267.5,334.5) ;

\draw[<-]    (466,310.5) -- (434.5,310.5) ;
\draw[->]   (351,353.5) -- (351,377.5) ;
\draw   (263,377.5) -- (439,377.5) -- (439,405.5) -- (263,405.5) -- cycle ;
\draw[->]    (351,405.5) -- (351,445.5) ;
\draw   (351,445.5) -- (414,463) -- (351,480.5) -- (288,463) -- cycle ;
\draw[->]    (351,480.5) -- (351,504.5) ;
\draw[->]    (288,463) -- (256.5,463) ;
\draw    (88.5,486.25) -- (172.5,505.25) ;
\draw    (172.5,420.75) -- (89,439.75) ;
\draw    (172.5,505.25) -- (256.5,486.25) ;
\draw    (256.5,486.25) -- (256.5,439.75) ;
\draw    (256.5,439.75) -- (172.5,420.75) ;
\draw    (88.5,486.25) -- (172.5,505.25) ;
\draw    (89,439.75) -- (89,486.25) ;

\draw   (129.56,377.5) -- (215.44,377.5) -- (215.44,405.5) -- (129.56,405.5) -- cycle ;
\draw[<-]    (172.5,405.5) -- (172.5,420.75) ;
\draw[<-]    (263,391.5) -- (215.44,391.5) ;
\draw    (172.5,505.25) -- (172.5,519.25) ;
\draw   (308.06,505.25) -- (393.94,505.25) -- (393.94,533.25) -- (308.06,533.25) -- cycle ;
\draw[->]    (172.5,519.25) -- (308.06,519.25) ;
\draw[->]    (351,533.25) -- (351,557.25) ;
\draw   (253.8,557.25) -- (448.2,557.25) -- (448.2,585.25) -- (253.8,585.25) -- cycle ;
\draw    (452.8,669) -- (543.14,688) ;
\draw    (543.14,600.5) -- (453.34,619.5) ;
\draw    (543.14,688) -- (633.47,669) ;
\draw    (633.47,669) -- (633.47,619.5) ;
\draw    (633.47,619.5) -- (543.14,600.5) ;
\draw    (452.8,669) -- (543.14,688) ;
\draw    (453.34,619.5) -- (453.34,669) ;

\draw   (495.56,557.25) -- (590.71,557.25) -- (590.71,585.25) -- (495.56,585.25) -- cycle ;
\draw[->]    (495.56,571.25) -- (448,571.25) ;
\draw[->]    (351,662.89) -- (351,684.25) ;
\draw   (308.06,684.25) -- (393.94,684.25) -- (393.94,712.25) -- (308.06,712.25) -- cycle ;
\draw[->]    (351,712.25) -- (351,736.25) ;
\draw  [color={rgb, 255:red, 208; green, 2; blue, 27 }  ,draw opacity=1 ][line width=2.25]  (466,303.5) .. controls (466,299.63) and (469.13,296.5) .. (473,296.5) -- (557.33,296.5) .. controls (561.2,296.5) and (564.33,299.63) .. (564.33,303.5) -- (564.33,317.5) .. controls (564.33,321.37) and (561.2,324.5) .. (557.33,324.5) -- (473,324.5) .. controls (469.13,324.5) and (466,321.37) .. (466,317.5) -- cycle ;
\draw[<-]    (543.14,585.25) -- (543.14,600.5) ;
\draw[->]    (351,585.25) -- (351,622.61) ;
\draw   (351,622.61) -- (418.5,642.75) -- (351,662.89) -- (283.5,642.75) -- cycle ;
\draw[<-]    (453.34,642.75) -- (418.5,642.75) ;
\draw[<-]    (393.94,698.25) -- (543.14,698.25) ;
\draw    (543.14,688) -- (543.14,698.25) ;

\draw (351,28.5) node  [font=\small]  {$\text{Correlator matrix} \ C{\displaystyle _{\chi \chi '}( t)}$};
\draw (295.75,120) node  [font=\small]  {$E_{n}( t,t_{0} ,t_{\text{ref}})$};
\draw (246.5,79.5) node  [font=\footnotesize]  {$\text{Eigenvector-based} \ \ \hat{C}_{n}( t,t_{0} ,t_{\text{ref}})$};
\draw (277,217) node  [font=\small] [align=left] {no};
\draw (441.5,80.5) node  [font=\footnotesize]  {$\text{Eigenvalue-based} \ \ \lambda _{n}( t,t_{0})$};
\draw (387.5,120) node  [font=\small]  {$F_{n}( t)$};
\draw (351,172) node  [font=\small]  {$t_{0} =\textcolor[rgb]{0.29,0.56,0.89}{\delta } \ \ \ t_{\text{ref}} =t_{0} +\textcolor[rgb]{0.29,0.56,0.89}{\delta }$};
\draw (351,227.5) node  [font=\footnotesize]  {$\lambda _{n}( t_{\mathrm{ref}} ,t_{0})  >0$};
\draw (365,255.5) node  [font=\small] [align=left] {yes};
\draw (172.75,227.5) node  [font=\footnotesize]  {$\text{Correlator matrix degenerate}$};
\draw (172.75,279.5) node  [font=\footnotesize]  {$\text{Remove operators from matrix}$};
\draw (351,295) node  [font=\footnotesize]  {$E_{n}( t,t_{0} ,t_{\text{ref}}) \stackrel{?}{=}F_{n}( t) \pm \textcolor[rgb]{0.29,0.56,0.89}{\text{tol}}\textcolor[rgb]{0.29,0.56,0.89}{_{\sigma }}$};
\draw (351,313.5) node  [font=\scriptsize]  {$\left\{t\ |\ \delta F_{n}( t) < \textcolor[rgb]{0.29,0.56,0.89}{\text{tol}}\textcolor[rgb]{0.29,0.56,0.89}{_{\lambda }} \delta E_{n}( t,t_{0} ,t_{\text{ref}})\right\}$};
\draw (351,330.5) node  [font=\scriptsize]  {$\{n\ |\ E_{n}( t,t_{0} ,t_{\text{ref}}) < \textcolor[rgb]{0.29,0.56,0.89}{\text{tol}}\textcolor[rgb]{0.29,0.56,0.89}{_{E}}\}$};
\draw (446,300.75) node  [font=\small] [align=left] {no};
\draw (365,364) node  [font=\small] [align=left] {yes};
\draw (351,391.5) node  [font=\small]  {$t'_{0} =t_{0} +a\ \ \ t'_{\text{ref}} =t'_{0} +\textcolor[rgb]{0.29,0.56,0.89}{\delta }$};
\draw (351,463) node  [font=\footnotesize]  {$\lambda _{n}( t'_{\text{ref}} ,t'_{0})  >0$};
\draw (274,452.5) node  [font=\small] [align=left] {yes};
\draw (362,491) node  [font=\small] [align=left] {no};
\draw (172.5,483.25) node  [font=\scriptsize]  {$\{n\ |\ E_{n}( t,t'_{0} ,t'_{ref}) < \textcolor[rgb]{0.29,0.56,0.89}{\text{tol}}\textcolor[rgb]{0.29,0.56,0.89}{_{E}}\}$};
\draw (173.5,465.25) node  [font=\scriptsize]  {$\left\{t\ |\ \delta F_{n}( t) < \textcolor[rgb]{0.29,0.56,0.89}{\text{tol}}\textcolor[rgb]{0.29,0.56,0.89}{_{\lambda }} \delta E_{n}( t,t'_{0} ,t'_{\text{ref}})\right\}$};
\draw (172.5,446.75) node  [font=\footnotesize]  {$E_{n}( t,t'_{0} ,t'_{\text{ref}}) \stackrel{?}{=}F_{n}( t) \pm \textcolor[rgb]{0.29,0.56,0.89}{\text{tol}}\textcolor[rgb]{0.29,0.56,0.89}{_{\sigma }}$};
\draw (172.5,391.5) node  [font=\small]  {$t_{0}\leftarrow t_{0} +a$};
\draw (188.5,413.25) node  [font=\small] [align=left] {yes};
\draw (351,519.25) node  [font=\small]  {$t_{0}^{\text{best}} =t_{0}$};
\draw (160,514.25) node  [font=\small] [align=left] {no};
\draw (351,571.25) node  [font=\small]  {$t_{\text{ref}} =t_{0}^{\text{best}} +\textcolor[rgb]{0.29,0.56,0.89}{\delta } \ \ \ t'_{\text{ref}} =t_{\text{ref}} +a$};
\draw (543.14,571.25) node  [font=\small]  {$t_{\text{ref}}\leftarrow t_{\text{ref}} +a$};
\draw (351,698.25) node  [font=\small]  {$t_{\text{ref}}^{\text{best}} =t_{\text{ref}}$};
\draw (362,670.75) node  [font=\small] [align=left] {no};
\draw (351,756.25) node  [font=\footnotesize]  {$\hat{C}_{n}\left( t,t_{0}^{\text{best}} ,t_{\text{ref}}^{\text{best}}\right) \ \ \ \ E_{n}\left( t,t_{0}^{\text{best}} ,t_{\text{ref}}^{\text{best}}\right)$};
\draw (515.17,310.5) node  [font=\footnotesize]  {$\text{Data too noisy}$};
\draw (542,626.25) node  [font=\footnotesize]  {$E_{n}\left( t,t_{0}^{\text{best}} ,t'_{\text{ref}}\right) \stackrel{?}{=}F_{n}( t) \pm \textcolor[rgb]{0.29,0.56,0.89}{\text{tol}}\textcolor[rgb]{0.29,0.56,0.89}{_{\sigma }}$};
\draw (543,644.25) node  [font=\scriptsize]  {$\left\{t\ |\ \delta F_{n}( t) < \textcolor[rgb]{0.29,0.56,0.89}{\text{tol}}\textcolor[rgb]{0.29,0.56,0.89}{_{\lambda }} \delta E_{n}\left( t,t_{0}^{\text{best}} ,t'_{\text{ref}}\right)\right\}$};
\draw (542,662.25) node  [font=\scriptsize]  {$\{n\ |\ E_{n}\left( t,t_{0}^{\text{best}} ,t'_{ref}\right) < \textcolor[rgb]{0.29,0.56,0.89}{\text{tol}}\textcolor[rgb]{0.29,0.56,0.89}{_{E}}\}$};
\draw (528.64,593.25) node  [font=\small] [align=left] {yes};
\draw (351,642.75) node  [font=\footnotesize]  {$\lambda _{n}\left( t'_{\text{ref}} ,t_{0}^{\text{best}}\right)  >0$};
\draw (426,631.75) node  [font=\small] [align=left] {yes};
\draw (554.64,699) node  [font=\small] [align=left] {no};

\end{tikzpicture}}
\caption{Flow chart representing the steps of the algorithm to find the best $t_0$ and $t_{\rm ref}$.}
\label{fig:flowchart}
\end{figure}

As discussed in the main text, $t_{\text{ref}}$ and $t_0$ should be chosen so that the effective masses obtained using the eigenvector- and eigenvalue-based definitions of the principal correlation functions in Eqs.~\eqref{eq:emPC} and~\eqref{eq:emEV}, respectively, agree within uncertainties. This is achieved using the following algorithm, shown in Fig.~\ref{fig:flowchart} with  the algorithm parameters highlighted in blue.
\begin{itemize}
  \item Set $t_0 = \delta$ and $t_{\text{ref}} = t_0 + \delta$, where $\delta$ is the minimum distance over which a sum-of-exponential spectral representation is expected to be valid. 
  
  \item Verify that $\lambda_{\mathsf{n}}(t_{\text{ref}},t_0) > 0$  are positive. If not, the correlation-function matrix is approximately degenerate at this statistical precision, and either an operator must be removed to decrease the size of the correlation-function matrix or statistical precision must be increased. 
  
  \item Check whether $E_{\mathsf{n}}(t,t_0,t_{\text{ref}}) \approx F_{\mathsf{n}}(t)$ for all $t$ where $\delta F_{\mathsf{n}}(t) < \text{tol}_{\lambda} \delta E_{\mathsf{n}}(t,t_0,t_{\text{ref}})$ and $\approx$ denotes equality to within a combined statistical uncertainty of $\text{tol}_{\sigma}$. Here $\text{tol}_{\lambda}$ and $\text{tol}_{\sigma}$ are hyperparameters that are set to $\text{tol}_{\lambda} = 1.5$ and $\text{tol}_{\sigma} = 2 \sigma$ for the calculations in this work.\footnote{
      Note that the introduction of $\text{tol}_{\lambda}$ is needed because $E_{\mathsf{n}}(t,t_{\text{ref}},t_0)$ and $F_{\mathsf{n}}(t)$ become decorrelated for $t \gg t_{\text{ref}}$ and in this region $\approx$ would need to be defined with a more sophisticated statistical measure of similarity.}
  If this check is passed for all $n$ with $E_{\mathsf{n}}(t,t_0,t_{\text{ref}}) < \text{tol}_E$,\footnote{The introduction of $\text{tol}_E$ is required when some eigenvectors describe relatively noisy energy levels much higher in the spectrum than a region of physical interest.}  then a ``plateau region" of $t_0$ and $t_{\text{ref}}$ where the eigenvector- and eigenvalue-based principal correlation functions are consistent has been identified.
  
  \item Define $t_0' = t_0 + 1$ and $t_{\text{ref}}' = t_0' + \delta$.
  
  \item Verify that $\lambda_{\mathsf{n}}(t_{\text{ref}}',t_0' ) > 0$  are positive. If there is a negative eigenvalue, then if a plateau region has been found, $t_0$ is the largest acceptable value before noise sets in and $t_0^{\text{best}} \equiv t_0$ (if a plateau has not been found, then there is insufficient statistical precision to analyze the correlation-function matrix). 
  
  \item Check whether $E_{\mathsf{n}}(t,t_0',t_{\text{ref}}') \approx F_{\mathsf{n}}(t)$ to with $\text{tol}_{\sigma}$ for all $t$ where $\delta F_{\mathsf{n}}(t) < \text{tol}_{\lambda} \delta E_{\mathsf{n}}(t,t_0',t_{\text{ref}}')$ and all $n$ with $E_{\mathsf{n}}(t,t_0',t_{\text{ref}}') < \text{tol}_E$. If so, a plateau region has been found. If they do not agree, then if a plateau region has been found, $t_0$ is the largest acceptable value before noise sets in and $t_0^{\text{best}} \equiv t_0$ (if a plateau has not been found, then there is insufficient statistical precision to analyze the correlation-function matrix).
  
  \item Take $t_0 \rightarrow t_0 + 1$.
  
  \item Repeat the previous four steps until $t_0^{\text{best}}$ has been identified.
  
  \item Define $t_{\text{ref}} \rightarrow t_{\text{ref}}' = t_{\text{ref}} + 1$.
  
  \item Verify that $\lambda_{\mathsf{n}}(t_{\text{ref}}',t_0^{\text{best}}) > 0$  are positive. If there is a negative eigenvalue, then $t_{\text{ref}}^{\text{best}} \equiv t_{\text{ref}}$.
  
  \item Check whether $E_{\mathsf{n}}(t,t_0^{\text{best}},t_{\text{ref}}') \approx F_{\mathsf{n}}(t)$ to with $\text{tol}_{\sigma}$ for all $t$ where $\delta F_{\mathsf{n}}(t) < \text{tol}_{\lambda} \delta E_{\mathsf{n}}(t,t_{\text{ref}}',t_0')$ and all $n$ with $E_{\mathsf{n}}(t,t_0',t_{\text{ref}}') < \text{tol}_E$. If they do not agree, then $t_{\text{ref}}^{\text{best}} \equiv t_{\text{ref}}$.
  
  \item Take $t_{\text{ref}} \rightarrow t_{\text{ref}} + 1$.
  
  \item Repeat the previous four steps until $t_{\text{ref}}^{\text{best}}$ has been identified.
\end{itemize}
The desired principal correlation functions are then defined as $\hat{C}_{\mathsf{n}}(t) \equiv \hat{C}_{\mathsf{n}}(t,t_{0}^{\text{best}},t_{\text{ref}}^{\text{best}})$, and the associated effective mass functions are denoted $E_{\mathsf{n}}(t) \equiv E_{\mathsf{n}}(t,t_{0}^{\text{best}},t_{\text{ref}}^{\text{best}})$.

\section{Variational bounds}\label{app:tabs}

The strongest variational bounds for each of the one- and two-nucleon channels that are studied in this work are presented in Tables~\ref{tab:B1}-\ref{tab:B2I0}.
Fits are performed using methods adapted from Refs.~\cite{NPLQCD:2020lxg,Amarasinghe:2021lqa} and the $t_0$, $t_{\text{ref}}$ selection criteria described in the main text and Appendix~\ref{app:t0tref}.
The uncertainties shown for $a E_{\mathsf{n}}^Q$ and $a \Delta E_{\mathsf{n}}^Q$ include systematic uncertainties associated with the variation in fit results obtained with different choices of $t_{\rm min}$ (the minimum temporal extent used in the fit) added in quadrature to statistical uncertainties calculated using bootstrap methods.
Results for energies in physical units include uncertainties in the determination of $a = 0.1453(16)$ fm~\cite{NPLQCD:2012mex} added in quadrature. 
Ambiguities in defining the lattice spacing away from the physical values of the quark masses are not quantified.

\begin{table}[h!]
\input{figures/GEVPspectrumI0p5G1p} \\\vspace{20pt}
\input{figures/GEVPspectrumI0p5G1m}
\caption{Single-nucleon positive- and negative-parity variational bounds in lattice and physical units obtained from weighted averages of  multi-exponential fits to GEVP correlation functions as described in the main text. \label{tab:B1}}
\end{table}

\begin{table}[h!]
\input{figures/GEVPspectrumI1A1} \\\vspace{20pt}
\input{figures/GEVPspectrumI1E} \\\vspace{20pt}
\input{figures/GEVPspectrumI1T2} 
  \caption{Two-nucleon $I=1$ variational bounds in positive-parity total-angular-momentum cubic irreps $\Gamma_J \in \{A_1^+,E^+,T_2^+\}$ analogous to the results in Table~\ref{tab:B1}. The interpolating-operator set $\mathbb{S}_{15D}^{nn}\cup \mathbb{S}_{16H}^{nn}$ is used.
  \label{tab:B2I1}}
\end{table}

\begin{table}[h!]
\input{figures/GEVPspectrumI0T1} \\\vspace{20pt}
\input{figures/GEVPspectrumI0T2} \\\vspace{20pt}
\input{figures/GEVPspectrumI0E} \\\vspace{20pt}
\input{figures/GEVPspectrumI0A2}
  \caption{Two-nucleon $I=0$ variational bounds in positive-parity total-angular-momentum cubic irreps $\Gamma_J \in \{T_1^+,T_2^+,E^+,A_2^+\}$ analogous to the results in Table~\ref{tab:B1}. The interpolating-operator set $\mathbb{S}_{30D}^d\cup \mathbb{S}_{16H}^d$ is used.
  \label{tab:B2I0}}
\end{table}

\section{Effective masses and overlaps}\label{app:energies}

\begin{figure}[!t]
	\includegraphics[width=0.47\columnwidth]{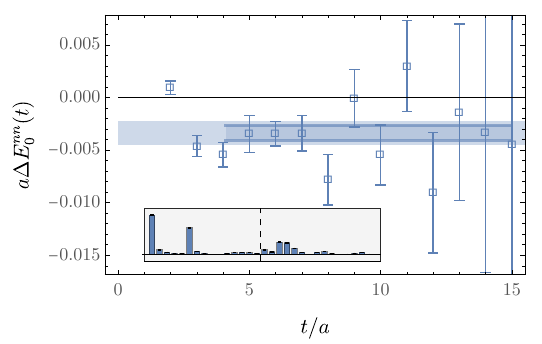}
	\includegraphics[width=0.47\columnwidth]{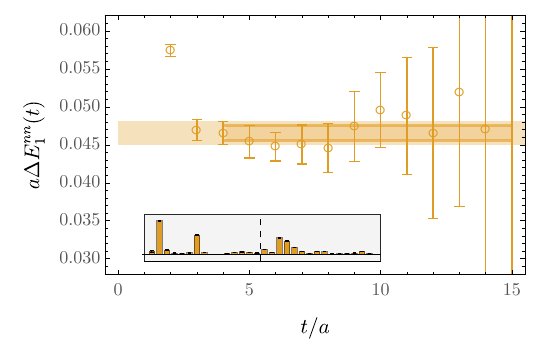}
 
	\includegraphics[width=0.47\columnwidth]{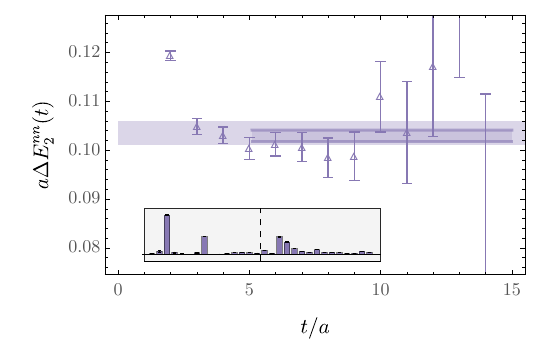}
	\includegraphics[width=0.47\columnwidth]{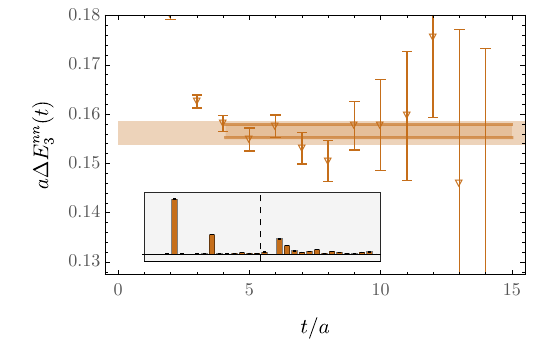}
 
	\includegraphics[width=0.47\columnwidth]{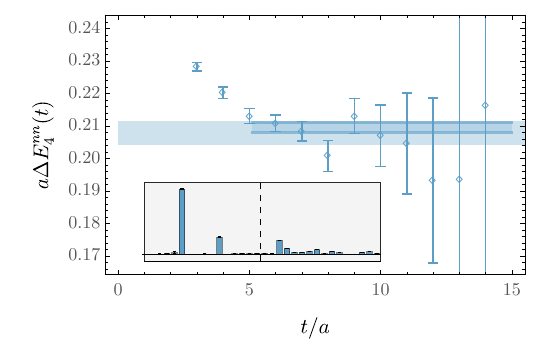}
	\includegraphics[width=0.47\columnwidth]{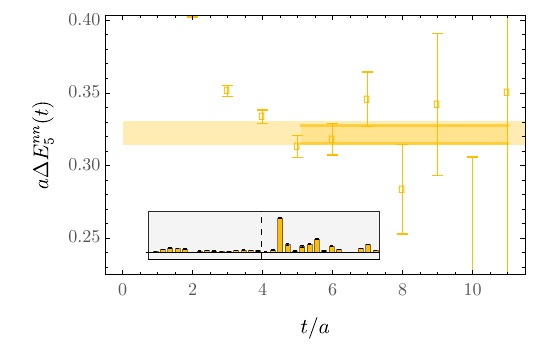}
 
   \caption{\label{fig:B2I1A1_FV_fits} Results for two-nucleon GEVP effective FV energy shifts for the dineutron channel from the $\mathbb{S}_{15D}^{nn}\cup \mathbb{S}_{16H}^{nn}$ operator set. Histograms show the overlaps with interpolating operators corresponding to $s \in \{0,\ldots,4\}$ shell dibaryons with upper/lower/negative-parity spin components (15 bars left of dashed line) and a complete basis of hexaquarks (16 bars right of dashed line) ordered as in Fig.~\ref{fig:B2I1A1_Zplot}.}
\end{figure}

\begin{figure}[!t]
	\includegraphics[width=0.47\columnwidth]{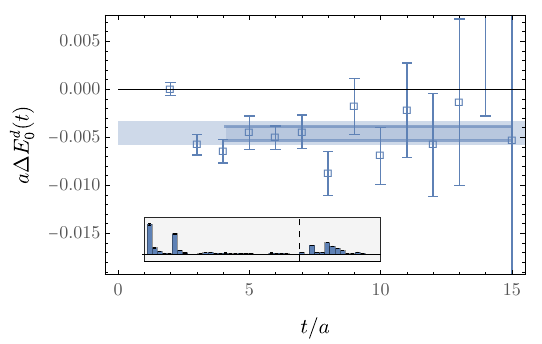}
	\includegraphics[width=0.47\columnwidth]{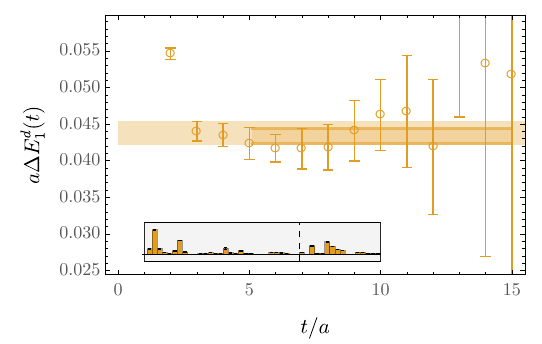}
 
	\includegraphics[width=0.47\columnwidth]{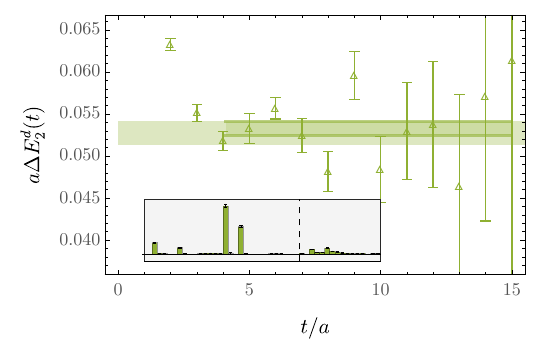}
	\includegraphics[width=0.47\columnwidth]{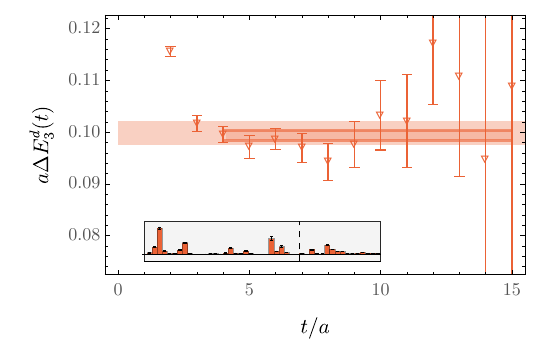}
 
	\includegraphics[width=0.47\columnwidth]{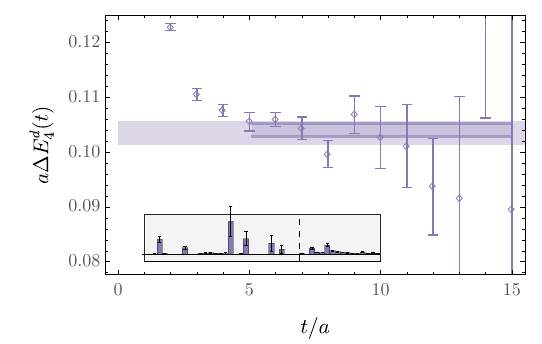}
	\includegraphics[width=0.47\columnwidth]{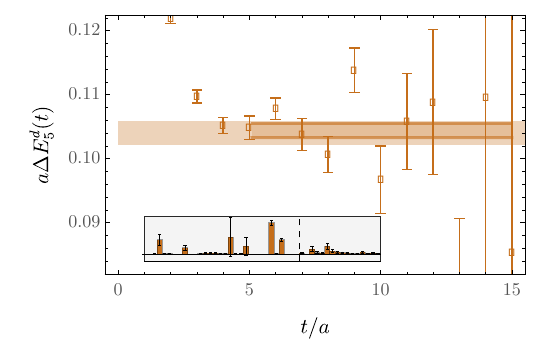}
   \caption{\label{fig:B2I0T1_FV_fits} Results for two-nucleon GEVP effective FV energy shifts for the deuteron channel from the $\mathbb{S}_{30D}^d\cup \mathbb{S}_{16H}^d$ operator set. Histograms show the overlaps with interpolating operators corresponding to $s \in \{0,\ldots,4\}$ shell dibaryons with upper/lower/negative-parity spin components (30 bars left of dashed line) and a complete basis of hexaquarks (16 bars right of dashed line) ordered as in Fig.~\ref{fig:B2I0T1_Zplot}.}
\end{figure}

\begin{figure}[!t]
	\includegraphics[width=0.47\columnwidth]{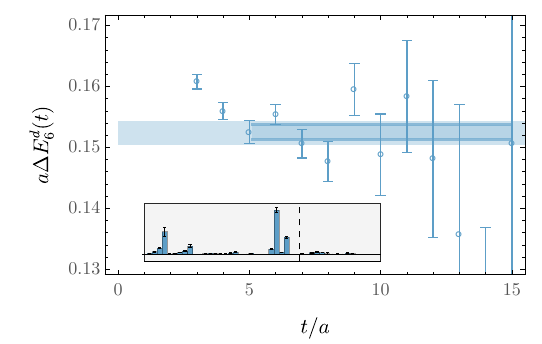}
	\includegraphics[width=0.47\columnwidth]{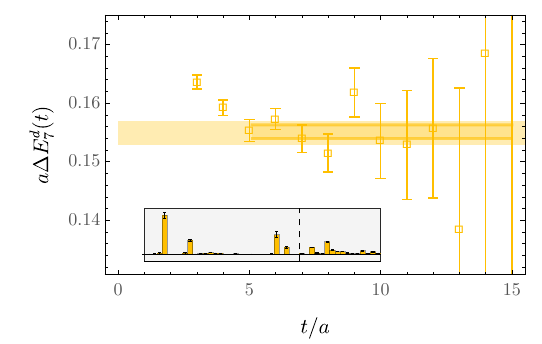}
 
	\includegraphics[width=0.47\columnwidth]{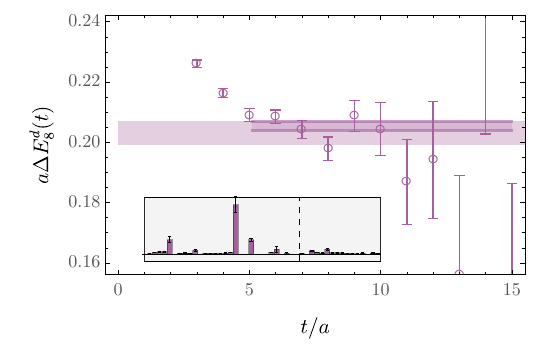}
	\includegraphics[width=0.47\columnwidth]{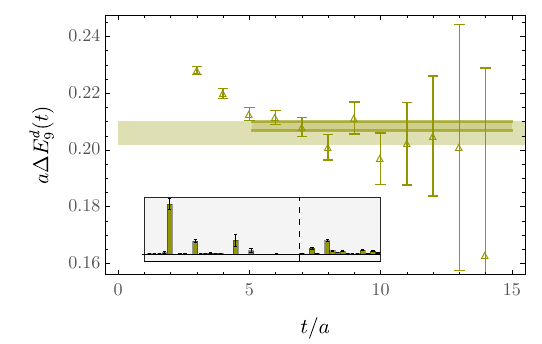}

	\includegraphics[width=0.47\columnwidth]{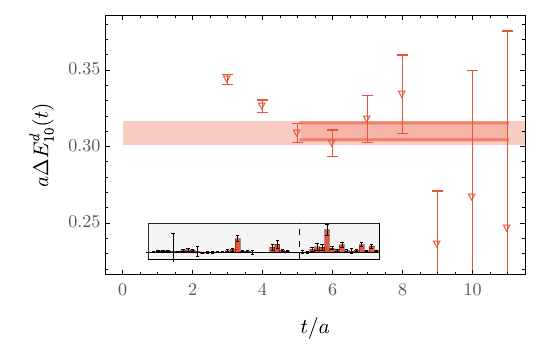}
   \caption{\label{fig:B2I0T1_FV_fits_2} Additional results for two-nucleon GEVP effective FV energy shifts for the deuteron channel. Details are as in Fig.~\ref{fig:B2I0T1_FV_fits}.}
\end{figure}

This appendix presents more details on the effective mass functions and overlaps associated with variational bounds below $\DLE$.
Effective masses for the dineutron channel, which are shown collectively in Fig.~\ref{fig:dineutron_effmass},
are shown individually in more detail in Fig.~\ref{fig:B2I1A1_FV_fits}.
The corresponding overlap factors, whose central values are indicated in Fig.~\ref{fig:B2I1A1_Zplot}, are also shown here as histograms with error bars.
Analogous results for the deuteron channel are shown in Figs.~\ref{fig:B2I0T1_FV_fits} and \ref{fig:B2I0T1_FV_fits_2}.

\clearpage

\bibliography{bib}
\end{document}

%% file: figures/GEVPspectrumI0p5G1p.tex
\begin{tabular}{|c|c|c|} \hline
$n$ & $a E_{\mathsf{n}}^{(2,\frac{1}{2},G_1^+)}$ & $E_{\mathsf{n}}^{(2,\frac{1}{2},G_1^+)}\text{ [GeV]}$ \\\hline
0 & $1.2021(22)$ & $1.636(23)$ \\\hline
1 & $1.9115(285)$ & $2.601(53)$ \\\hline
\end{tabular}

%% file: figures/GEVPspectrumI0p5G1m.tex
\begin{tabular}{|c|c|c|} \hline
$n$ & $a E_{\mathsf{n}}^{(2,\frac{1}{2},G_1^-)}$ & $E_{\mathsf{n}}^{(2,\frac{1}{2},G_1^-)}\text{ [GeV]}$ \\\hline
0 & $1.6362(85)$ & $2.227(33)$ \\\hline
\end{tabular}

%% file: figures/GEVPspectrumI1A1.tex
\begin{tabular}{|c|c|c|c|} \hline
$n$ & $a E_{\mathsf{n}}^{(2,1,A_1^+)}$ & $a \Delta E_{\mathsf{n}}^{(2,1,A_1^+)}$ & $\Delta E_{\mathsf{n}}^{(2,1,A_1^+)}\text{ [MeV]}$ \\\hline
0 & $2.4024(37)$ & $-0.0034(11)$ & $-4.6(1.5)$ \\\hline
1 & $2.4524(40)$ & $0.0465(16)$ & $63.3(2.3)$ \\\hline
2 & $2.5101(39)$ & $0.1035(24)$ & $140.8(3.8)$ \\\hline
3 & $2.5624(40)$ & $0.1561(24)$ & $212.5(4.4)$ \\\hline
4 & $2.6135(55)$ & $0.2078(36)$ & $282.8(6.3)$ \\\hline
5 & $2.7287(95)$ & $0.3222(81)$ & $438.5(12.6)$ \\\hline
\end{tabular}

%% file: figures/GEVPspectrumI1E.tex
\begin{tabular}{|c|c|c|c|} \hline
$n$ & $a E_{\mathsf{n}}^{(2,1,E^+)}$ & $a \Delta E_{\mathsf{n}}^{(2,1,E^+)}$ & $\Delta E_{\mathsf{n}}^{(2,1,E^+)}\text{ [MeV]}$ \\\hline
0 & $2.4574(39)$ & $0.0519(16)$ & $70.6(2.3)$ \\\hline
1 & $2.5088(42)$ & $0.1029(23)$ & $140.1(3.7)$ \\\hline
2 & $2.6127(51)$ & $0.2063(29)$ & $280.7(5.5)$ \\\hline
\end{tabular}

%% file: figures/GEVPspectrumI1T2.tex
\begin{tabular}{|c|c|c|c|} \hline
$n$ & $a E_{\mathsf{n}}^{(2,1,T_2^+)}$ & $a \Delta E_{\mathsf{n}}^{(2,1,T_2^+)}$ & $\Delta E_{\mathsf{n}}^{(2,1,T_2^+)}\text{ [MeV]}$ \\\hline
0 & $2.5096(42)$ & $0.1039(18)$ & $141.4(3.1)$ \\\hline
1 & $2.5610(41)$ & $0.1548(20)$ & $210.7(3.9)$ \\\hline
\end{tabular}

%% file: figures/GEVPspectrumI0T1.tex
\begin{tabular}{|c|c|c|c|} \hline
$n$ & $a E_{\mathsf{n}}^{(2,0,T_1^+)}$ & $a \Delta E_{\mathsf{n}}^{(2,0,T_1^+)}$ & $\Delta E_{\mathsf{n}}^{(2,0,T_1^+)}\text{ [MeV]}$ \\\hline
0 & $2.4017(36)$ & $-0.0046(12)$ & $-6.2(1.7)$ \\\hline
1 & $2.4497(40)$ & $0.0438(16)$ & $59.6(2.3)$ \\\hline
2 & $2.4584(36)$ & $0.0527(14)$ & $71.8(2.2)$ \\\hline
3 & $2.5055(40)$ & $0.0998(23)$ & $135.8(3.6)$ \\\hline
4 & $2.5090(46)$ & $0.1035(22)$ & $140.8(3.5)$ \\\hline
5 & $2.5097(41)$ & $0.1040(18)$ & $141.5(3.2)$ \\\hline
6 & $2.5572(42)$ & $0.1523(19)$ & $207.3(3.9)$ \\\hline
7 & $2.5604(43)$ & $0.1549(20)$ & $210.8(4.0)$ \\\hline
8 & $2.6080(63)$ & $0.2031(39)$ & $276.3(6.6)$ \\\hline
9 & $2.6099(61)$ & $0.2059(42)$ & $280.3(6.9)$ \\\hline
10 & $2.7114(95)$ & $0.3087(77)$ & $420.1(12.0)$ \\\hline
\end{tabular}

%% file: figures/GEVPspectrumI0T2.tex
\begin{tabular}{|c|c|c|c|} \hline
$n$ & $a E_{\mathsf{n}}^{(2,0,T_2^+)}$ & $a \Delta E_{\mathsf{n}}^{(2,0,T_2^+)}$ & $\Delta E_{\mathsf{n}}^{(2,0,T_2^+)}\text{ [MeV]}$ \\\hline
0 & $2.4572(39)$ & $0.0510(16)$ & $69.4(2.4)$ \\\hline
1 & $2.5075(44)$ & $0.1019(18)$ & $138.7(3.1)$ \\\hline
2 & $2.5082(44)$ & $0.1028(21)$ & $139.9(3.5)$ \\\hline
3 & $2.5594(45)$ & $0.1534(20)$ & $208.7(3.9)$ \\\hline
4 & $2.6110(52)$ & $0.2052(31)$ & $279.2(5.7)$ \\\hline
\end{tabular}

%% file: figures/GEVPspectrumI0E.tex
\begin{tabular}{|c|c|c|c|} \hline
$n$ & $a E_{\mathsf{n}}^{(2,0,E^+)}$ & $a \Delta E_{\mathsf{n}}^{(2,0,E^+)}$ & $\Delta E_{\mathsf{n}}^{(2,0,E^+)}\text{ [MeV]}$ \\\hline
0 & $2.5069(40)$ & $0.1009(18)$ & $137.3(3.1)$ \\\hline
1 & $2.5591(41)$ & $0.1525(21)$ & $207.6(4.0)$ \\\hline
\end{tabular}

%% file: figures/GEVPspectrumI0A2.tex
\begin{tabular}{|c|c|c|c|} \hline
$n$ & $a E_{\mathsf{n}}^{(2,0,A_2^+)}$ & $a \Delta E_{\mathsf{n}}^{(2,0,A_2^+)}$ & $\Delta E_{\mathsf{n}}^{(2,0,A_2^+)}\text{ [MeV]}$ \\\hline
0 & $2.5071(53)$ & $0.1023(26)$ & $139.2(4.0)$ \\\hline
1 & $2.5577(54)$ & $0.1512(33)$ & $205.8(5.4)$ \\\hline
\end{tabular}

%% file: main.bbl
\begin{thebibliography}{124}%
\makeatletter
\providecommand \@ifxundefined [1]{%
 \@ifx{#1\undefined}
}%
\providecommand \@ifnum [1]{%
 \ifnum #1\expandafter \@firstoftwo
 \else \expandafter \@secondoftwo
 \fi
}%
\providecommand \@ifx [1]{%
 \ifx #1\expandafter \@firstoftwo
 \else \expandafter \@secondoftwo
 \fi
}%
\providecommand \natexlab [1]{#1}%
\providecommand \enquote  [1]{``#1''}%
\providecommand \bibnamefont  [1]{#1}%
\providecommand \bibfnamefont [1]{#1}%
\providecommand \citenamefont [1]{#1}%
\providecommand \href@noop [0]{\@secondoftwo}%
\providecommand \href [0]{\begingroup \@sanitize@url \@href}%
\providecommand \@href[1]{\@@startlink{#1}\@@href}%
\providecommand \@@href[1]{\endgroup#1\@@endlink}%
\providecommand \@sanitize@url [0]{\catcode `\\12\catcode `\$12\catcode
  `\&12\catcode `\#12\catcode `\^12\catcode `\_12\catcode `\%12\relax}%
\providecommand \@@startlink[1]{}%
\providecommand \@@endlink[0]{}%
\providecommand \url  [0]{\begingroup\@sanitize@url \@url }%
\providecommand \@url [1]{\endgroup\@href {#1}{\urlprefix }}%
\providecommand \urlprefix  [0]{URL }%
\providecommand \Eprint [0]{\href }%
\providecommand \doibase [0]{https://doi.org/}%
\providecommand \selectlanguage [0]{\@gobble}%
\providecommand \bibinfo  [0]{\@secondoftwo}%
\providecommand \bibfield  [0]{\@secondoftwo}%
\providecommand \translation [1]{[#1]}%
\providecommand \BibitemOpen [0]{}%
\providecommand \bibitemStop [0]{}%
\providecommand \bibitemNoStop [0]{.\EOS\space}%
\providecommand \EOS [0]{\spacefactor3000\relax}%
\providecommand \BibitemShut  [1]{\csname bibitem#1\endcsname}%
\let\auto@bib@innerbib\@empty
\bibitem [{\citenamefont {Detmold}\ \emph {et~al.}(2019)\citenamefont
  {Detmold}, \citenamefont {Edwards}, \citenamefont {Dudek}, \citenamefont
  {Engelhardt}, \citenamefont {Lin}, \citenamefont {Meinel}, \citenamefont
  {Orginos},\ and\ \citenamefont {Shanahan}}]{Detmold:2019ghl}%
  \BibitemOpen
  \bibfield  {author} {\bibinfo {author} {\bibfnamefont {W.}~\bibnamefont
  {Detmold}}, \bibinfo {author} {\bibfnamefont {R.~G.}\ \bibnamefont
  {Edwards}}, \bibinfo {author} {\bibfnamefont {J.~J.}\ \bibnamefont {Dudek}},
  \bibinfo {author} {\bibfnamefont {M.}~\bibnamefont {Engelhardt}}, \bibinfo
  {author} {\bibfnamefont {H.-W.}\ \bibnamefont {Lin}}, \bibinfo {author}
  {\bibfnamefont {S.}~\bibnamefont {Meinel}}, \bibinfo {author} {\bibfnamefont
  {K.}~\bibnamefont {Orginos}},\ and\ \bibinfo {author} {\bibfnamefont
  {P.}~\bibnamefont {Shanahan}} (\bibinfo {collaboration} {USQCD}),\ }\bibfield
   {title} {\bibinfo {title} {{Hadrons and Nuclei}},\ }\href
  {https://doi.org/10.1140/epja/i2019-12902-4} {\bibfield  {journal} {\bibinfo
  {journal} {Eur. Phys. J. A}\ }\textbf {\bibinfo {volume} {55}},\ \bibinfo
  {pages} {193} (\bibinfo {year} {2019})},\ \Eprint
  {https://arxiv.org/abs/1904.09512} {arXiv:1904.09512 [hep-lat]} \BibitemShut
  {NoStop}%
\bibitem [{\citenamefont {Drischler}\ \emph {et~al.}(2021)\citenamefont
  {Drischler}, \citenamefont {Haxton}, \citenamefont {McElvain}, \citenamefont
  {Mereghetti}, \citenamefont {Nicholson}, \citenamefont {Vranas},\ and\
  \citenamefont {Walker-Loud}}]{Drischler:2019xuo}%
  \BibitemOpen
  \bibfield  {author} {\bibinfo {author} {\bibfnamefont {C.}~\bibnamefont
  {Drischler}}, \bibinfo {author} {\bibfnamefont {W.}~\bibnamefont {Haxton}},
  \bibinfo {author} {\bibfnamefont {K.}~\bibnamefont {McElvain}}, \bibinfo
  {author} {\bibfnamefont {E.}~\bibnamefont {Mereghetti}}, \bibinfo {author}
  {\bibfnamefont {A.}~\bibnamefont {Nicholson}}, \bibinfo {author}
  {\bibfnamefont {P.}~\bibnamefont {Vranas}},\ and\ \bibinfo {author}
  {\bibfnamefont {A.}~\bibnamefont {Walker-Loud}},\ }\bibfield  {title}
  {\bibinfo {title} {{Towards grounding nuclear physics in QCD}},\ }\href
  {https://doi.org/10.1016/j.ppnp.2021.103888} {\bibfield  {journal} {\bibinfo
  {journal} {Prog. Part. Nucl. Phys.}\ }\textbf {\bibinfo {volume} {121}},\
  \bibinfo {pages} {103888} (\bibinfo {year} {2021})},\ \Eprint
  {https://arxiv.org/abs/1910.07961} {arXiv:1910.07961 [nucl-th]} \BibitemShut
  {NoStop}%
\bibitem [{\citenamefont {Davoudi}\ \emph {et~al.}(2021)\citenamefont
  {Davoudi}, \citenamefont {Detmold}, \citenamefont {Orginos}, \citenamefont
  {Parre\~no}, \citenamefont {Savage}, \citenamefont {Shanahan},\ and\
  \citenamefont {Wagman}}]{Davoudi:2020ngi}%
  \BibitemOpen
  \bibfield  {author} {\bibinfo {author} {\bibfnamefont {Z.}~\bibnamefont
  {Davoudi}}, \bibinfo {author} {\bibfnamefont {W.}~\bibnamefont {Detmold}},
  \bibinfo {author} {\bibfnamefont {K.}~\bibnamefont {Orginos}}, \bibinfo
  {author} {\bibfnamefont {A.}~\bibnamefont {Parre\~no}}, \bibinfo {author}
  {\bibfnamefont {M.~J.}\ \bibnamefont {Savage}}, \bibinfo {author}
  {\bibfnamefont {P.}~\bibnamefont {Shanahan}},\ and\ \bibinfo {author}
  {\bibfnamefont {M.~L.}\ \bibnamefont {Wagman}},\ }\bibfield  {title}
  {\bibinfo {title} {{Nuclear matrix elements from lattice QCD for electroweak
  and beyond-Standard-Model processes}},\ }\href
  {https://doi.org/10.1016/j.physrep.2020.10.004} {\bibfield  {journal}
  {\bibinfo  {journal} {Phys. Rept.}\ }\textbf {\bibinfo {volume} {900}},\
  \bibinfo {pages} {1} (\bibinfo {year} {2021})},\ \Eprint
  {https://arxiv.org/abs/2008.11160} {arXiv:2008.11160 [hep-lat]} \BibitemShut
  {NoStop}%
\bibitem [{\citenamefont {Tews}\ \emph {et~al.}(2022)\citenamefont {Tews} \emph
  {et~al.}}]{Tews:2022yfb}%
  \BibitemOpen
  \bibfield  {author} {\bibinfo {author} {\bibfnamefont {I.}~\bibnamefont
  {Tews}} \emph {et~al.},\ }\bibfield  {title} {\bibinfo {title} {{Nuclear
  Forces for Precision Nuclear Physics: A Collection of Perspectives}},\ }\href
  {https://doi.org/10.1007/s00601-022-01749-x} {\bibfield  {journal} {\bibinfo
  {journal} {Few Body Syst.}\ }\textbf {\bibinfo {volume} {63}},\ \bibinfo
  {pages} {67} (\bibinfo {year} {2022})},\ \Eprint
  {https://arxiv.org/abs/2202.01105} {arXiv:2202.01105 [nucl-th]} \BibitemShut
  {NoStop}%
\bibitem [{\citenamefont {Kronfeld}\ \emph {et~al.}(2022)\citenamefont
  {Kronfeld} \emph {et~al.}}]{USQCD:2022mmc}%
  \BibitemOpen
  \bibfield  {author} {\bibinfo {author} {\bibfnamefont {A.~S.}\ \bibnamefont
  {Kronfeld}} \emph {et~al.} (\bibinfo {collaboration} {USQCD}),\ }\href@noop
  {} {\bibinfo {title} {{Lattice QCD and Particle Physics}}} (\bibinfo {year}
  {2022}),\ \Eprint {https://arxiv.org/abs/2207.07641} {arXiv:2207.07641
  [hep-lat]} \BibitemShut {NoStop}%
\bibitem [{\citenamefont {Prezeau}\ \emph
  {et~al.}(2003{\natexlab{a}})\citenamefont {Prezeau}, \citenamefont {Kurylov},
  \citenamefont {Kamionkowski},\ and\ \citenamefont {Vogel}}]{Prezeau:2003sv}%
  \BibitemOpen
  \bibfield  {author} {\bibinfo {author} {\bibfnamefont {G.}~\bibnamefont
  {Prezeau}}, \bibinfo {author} {\bibfnamefont {A.}~\bibnamefont {Kurylov}},
  \bibinfo {author} {\bibfnamefont {M.}~\bibnamefont {Kamionkowski}},\ and\
  \bibinfo {author} {\bibfnamefont {P.}~\bibnamefont {Vogel}},\ }\bibfield
  {title} {\bibinfo {title} {{New contribution to wimp-nucleus scattering}},\
  }\href {https://doi.org/10.1103/PhysRevLett.91.231301} {\bibfield  {journal}
  {\bibinfo  {journal} {Phys. Rev. Lett.}\ }\textbf {\bibinfo {volume} {91}},\
  \bibinfo {pages} {231301} (\bibinfo {year} {2003}{\natexlab{a}})},\ \Eprint
  {https://arxiv.org/abs/astro-ph/0309115} {arXiv:astro-ph/0309115}
  \BibitemShut {NoStop}%
\bibitem [{\citenamefont {Chang}\ \emph {et~al.}(2018)\citenamefont {Chang},
  \citenamefont {Davoudi}, \citenamefont {Detmold}, \citenamefont {Gambhir},
  \citenamefont {Orginos}, \citenamefont {Savage}, \citenamefont {Shanahan},
  \citenamefont {Wagman},\ and\ \citenamefont {Winter}}]{Chang:2017eiq}%
  \BibitemOpen
  \bibfield  {author} {\bibinfo {author} {\bibfnamefont {E.}~\bibnamefont
  {Chang}}, \bibinfo {author} {\bibfnamefont {Z.}~\bibnamefont {Davoudi}},
  \bibinfo {author} {\bibfnamefont {W.}~\bibnamefont {Detmold}}, \bibinfo
  {author} {\bibfnamefont {A.~S.}\ \bibnamefont {Gambhir}}, \bibinfo {author}
  {\bibfnamefont {K.}~\bibnamefont {Orginos}}, \bibinfo {author} {\bibfnamefont
  {M.~J.}\ \bibnamefont {Savage}}, \bibinfo {author} {\bibfnamefont {P.~E.}\
  \bibnamefont {Shanahan}}, \bibinfo {author} {\bibfnamefont {M.~L.}\
  \bibnamefont {Wagman}},\ and\ \bibinfo {author} {\bibfnamefont
  {F.}~\bibnamefont {Winter}} (\bibinfo {collaboration} {NPLQCD}),\ }\bibfield
  {title} {\bibinfo {title} {{Scalar, Axial, and Tensor Interactions of Light
  Nuclei from Lattice QCD}},\ }\href
  {https://doi.org/10.1103/PhysRevLett.120.152002} {\bibfield  {journal}
  {\bibinfo  {journal} {Phys. Rev. Lett.}\ }\textbf {\bibinfo {volume} {120}},\
  \bibinfo {pages} {152002} (\bibinfo {year} {2018})},\ \Eprint
  {https://arxiv.org/abs/1712.03221} {arXiv:1712.03221 [hep-lat]} \BibitemShut
  {NoStop}%
\bibitem [{\citenamefont {Krebs}\ \emph {et~al.}(2020)\citenamefont {Krebs},
  \citenamefont {Epelbaum},\ and\ \citenamefont {Mei\ss{}ner}}]{Krebs:2020plh}%
  \BibitemOpen
  \bibfield  {author} {\bibinfo {author} {\bibfnamefont {H.}~\bibnamefont
  {Krebs}}, \bibinfo {author} {\bibfnamefont {E.}~\bibnamefont {Epelbaum}},\
  and\ \bibinfo {author} {\bibfnamefont {U.-G.}\ \bibnamefont {Mei\ss{}ner}},\
  }\bibfield  {title} {\bibinfo {title} {{Subleading contributions to the
  nuclear scalar isoscalar current}},\ }\href
  {https://doi.org/10.1140/epja/s10050-020-00249-y} {\bibfield  {journal}
  {\bibinfo  {journal} {Eur. Phys. J. A}\ }\textbf {\bibinfo {volume} {56}},\
  \bibinfo {pages} {240} (\bibinfo {year} {2020})},\ \Eprint
  {https://arxiv.org/abs/2005.07433} {arXiv:2005.07433 [nucl-th]} \BibitemShut
  {NoStop}%
\bibitem [{\citenamefont {de~Vries}\ \emph {et~al.}(2023)\citenamefont
  {de~Vries}, \citenamefont {K{\"o}rber}, \citenamefont {Nogga},\ and\
  \citenamefont {Shain}}]{deVries:2023hin}%
  \BibitemOpen
  \bibfield  {author} {\bibinfo {author} {\bibfnamefont {J.}~\bibnamefont
  {de~Vries}}, \bibinfo {author} {\bibfnamefont {C.}~\bibnamefont
  {K{\"o}rber}}, \bibinfo {author} {\bibfnamefont {A.}~\bibnamefont {Nogga}},\
  and\ \bibinfo {author} {\bibfnamefont {S.}~\bibnamefont {Shain}},\
  }\href@noop {} {\bibinfo {title} {{Dark matter scattering off ${}^4$He in
  chiral effective field theory}}} (\bibinfo {year} {2023}),\ \Eprint
  {https://arxiv.org/abs/2310.11343} {arXiv:2310.11343 [hep-ph]} \BibitemShut
  {NoStop}%
\bibitem [{\citenamefont {Prezeau}\ \emph
  {et~al.}(2003{\natexlab{b}})\citenamefont {Prezeau}, \citenamefont
  {Ramsey-Musolf},\ and\ \citenamefont {Vogel}}]{Prezeau:2003xn}%
  \BibitemOpen
  \bibfield  {author} {\bibinfo {author} {\bibfnamefont {G.}~\bibnamefont
  {Prezeau}}, \bibinfo {author} {\bibfnamefont {M.}~\bibnamefont
  {Ramsey-Musolf}},\ and\ \bibinfo {author} {\bibfnamefont {P.}~\bibnamefont
  {Vogel}},\ }\bibfield  {title} {\bibinfo {title} {{Neutrinoless double beta
  decay and effective field theory}},\ }\href
  {https://doi.org/10.1103/PhysRevD.68.034016} {\bibfield  {journal} {\bibinfo
  {journal} {Phys. Rev. D}\ }\textbf {\bibinfo {volume} {68}},\ \bibinfo
  {pages} {034016} (\bibinfo {year} {2003}{\natexlab{b}})},\ \Eprint
  {https://arxiv.org/abs/hep-ph/0303205} {arXiv:hep-ph/0303205} \BibitemShut
  {NoStop}%
\bibitem [{\citenamefont {Cirigliano}\ \emph {et~al.}(2021)\citenamefont
  {Cirigliano}, \citenamefont {Dekens}, \citenamefont {de~Vries}, \citenamefont
  {Hoferichter},\ and\ \citenamefont {Mereghetti}}]{Cirigliano:2020dmx}%
  \BibitemOpen
  \bibfield  {author} {\bibinfo {author} {\bibfnamefont {V.}~\bibnamefont
  {Cirigliano}}, \bibinfo {author} {\bibfnamefont {W.}~\bibnamefont {Dekens}},
  \bibinfo {author} {\bibfnamefont {J.}~\bibnamefont {de~Vries}}, \bibinfo
  {author} {\bibfnamefont {M.}~\bibnamefont {Hoferichter}},\ and\ \bibinfo
  {author} {\bibfnamefont {E.}~\bibnamefont {Mereghetti}},\ }\bibfield  {title}
  {\bibinfo {title} {{Toward Complete Leading-Order Predictions for
  Neutrinoless Double $\beta$ Decay}},\ }\href
  {https://doi.org/10.1103/PhysRevLett.126.172002} {\bibfield  {journal}
  {\bibinfo  {journal} {Phys. Rev. Lett.}\ }\textbf {\bibinfo {volume} {126}},\
  \bibinfo {pages} {172002} (\bibinfo {year} {2021})},\ \Eprint
  {https://arxiv.org/abs/2012.11602} {arXiv:2012.11602 [nucl-th]} \BibitemShut
  {NoStop}%
\bibitem [{\citenamefont {Cirigliano}\ \emph {et~al.}(2022)\citenamefont
  {Cirigliano} \emph {et~al.}}]{Cirigliano:2022rmf}%
  \BibitemOpen
  \bibfield  {author} {\bibinfo {author} {\bibfnamefont {V.}~\bibnamefont
  {Cirigliano}} \emph {et~al.},\ }\bibfield  {title} {\bibinfo {title}
  {{Towards precise and accurate calculations of neutrinoless double-beta
  decay}},\ }\href {https://doi.org/10.1088/1361-6471/aca03e} {\bibfield
  {journal} {\bibinfo  {journal} {J. Phys. G}\ }\textbf {\bibinfo {volume}
  {49}},\ \bibinfo {pages} {120502} (\bibinfo {year} {2022})},\ \Eprint
  {https://arxiv.org/abs/2207.01085} {arXiv:2207.01085 [nucl-th]} \BibitemShut
  {NoStop}%
\bibitem [{\citenamefont {Davoudi}\ \emph {et~al.}(2024)\citenamefont
  {Davoudi}, \citenamefont {Detmold}, \citenamefont {Fu}, \citenamefont
  {Grebe}, \citenamefont {Jay}, \citenamefont {Murphy}, \citenamefont {Oare},
  \citenamefont {Shanahan},\ and\ \citenamefont {Wagman}}]{Davoudi:2024ukx}%
  \BibitemOpen
  \bibfield  {author} {\bibinfo {author} {\bibfnamefont {Z.}~\bibnamefont
  {Davoudi}}, \bibinfo {author} {\bibfnamefont {W.}~\bibnamefont {Detmold}},
  \bibinfo {author} {\bibfnamefont {Z.}~\bibnamefont {Fu}}, \bibinfo {author}
  {\bibfnamefont {A.~V.}\ \bibnamefont {Grebe}}, \bibinfo {author}
  {\bibfnamefont {W.}~\bibnamefont {Jay}}, \bibinfo {author} {\bibfnamefont
  {D.}~\bibnamefont {Murphy}}, \bibinfo {author} {\bibfnamefont
  {P.}~\bibnamefont {Oare}}, \bibinfo {author} {\bibfnamefont {P.~E.}\
  \bibnamefont {Shanahan}},\ and\ \bibinfo {author} {\bibfnamefont {M.~L.}\
  \bibnamefont {Wagman}} (\bibinfo {collaboration} {NPLQCD}),\ }\bibfield
  {title} {\bibinfo {title} {{Long-distance nuclear matrix elements for
  neutrinoless double-beta decay from lattice QCD}},\ }\href
  {https://doi.org/10.1103/PhysRevD.109.114514} {\bibfield  {journal} {\bibinfo
   {journal} {Phys. Rev. D}\ }\textbf {\bibinfo {volume} {109}},\ \bibinfo
  {pages} {114514} (\bibinfo {year} {2024})},\ \Eprint
  {https://arxiv.org/abs/2402.09362} {arXiv:2402.09362 [hep-lat]} \BibitemShut
  {NoStop}%
\bibitem [{\citenamefont {Alvarez-Ruso}\ \emph {et~al.}(2018)\citenamefont
  {Alvarez-Ruso} \emph {et~al.}}]{NuSTEC:2017hzk}%
  \BibitemOpen
  \bibfield  {author} {\bibinfo {author} {\bibfnamefont {L.}~\bibnamefont
  {Alvarez-Ruso}} \emph {et~al.} (\bibinfo {collaboration} {NuSTEC}),\
  }\bibfield  {title} {\bibinfo {title} {{NuSTEC White Paper: Status and
  challenges of neutrino\textendash{}nucleus scattering}},\ }\href
  {https://doi.org/10.1016/j.ppnp.2018.01.006} {\bibfield  {journal} {\bibinfo
  {journal} {Prog. Part. Nucl. Phys.}\ }\textbf {\bibinfo {volume} {100}},\
  \bibinfo {pages} {1} (\bibinfo {year} {2018})},\ \Eprint
  {https://arxiv.org/abs/1706.03621} {arXiv:1706.03621 [hep-ph]} \BibitemShut
  {NoStop}%
\bibitem [{\citenamefont {Kronfeld}\ \emph {et~al.}(2019)\citenamefont
  {Kronfeld}, \citenamefont {Richards}, \citenamefont {Detmold}, \citenamefont
  {Gupta}, \citenamefont {Lin}, \citenamefont {Liu}, \citenamefont {Meyer},
  \citenamefont {Sufian},\ and\ \citenamefont {Syritsyn}}]{Kronfeld:2019nfb}%
  \BibitemOpen
  \bibfield  {author} {\bibinfo {author} {\bibfnamefont {A.~S.}\ \bibnamefont
  {Kronfeld}}, \bibinfo {author} {\bibfnamefont {D.~G.}\ \bibnamefont
  {Richards}}, \bibinfo {author} {\bibfnamefont {W.}~\bibnamefont {Detmold}},
  \bibinfo {author} {\bibfnamefont {R.}~\bibnamefont {Gupta}}, \bibinfo
  {author} {\bibfnamefont {H.-W.}\ \bibnamefont {Lin}}, \bibinfo {author}
  {\bibfnamefont {K.-F.}\ \bibnamefont {Liu}}, \bibinfo {author} {\bibfnamefont
  {A.~S.}\ \bibnamefont {Meyer}}, \bibinfo {author} {\bibfnamefont
  {R.}~\bibnamefont {Sufian}},\ and\ \bibinfo {author} {\bibfnamefont
  {S.}~\bibnamefont {Syritsyn}} (\bibinfo {collaboration} {USQCD}),\ }\bibfield
   {title} {\bibinfo {title} {{Lattice QCD and Neutrino-Nucleus Scattering}},\
  }\href {https://doi.org/10.1140/epja/i2019-12916-x} {\bibfield  {journal}
  {\bibinfo  {journal} {Eur. Phys. J. A}\ }\textbf {\bibinfo {volume} {55}},\
  \bibinfo {pages} {196} (\bibinfo {year} {2019})},\ \Eprint
  {https://arxiv.org/abs/1904.09931} {arXiv:1904.09931 [hep-lat]} \BibitemShut
  {NoStop}%
\bibitem [{\citenamefont {Ruso}\ \emph {et~al.}(2022)\citenamefont {Ruso} \emph
  {et~al.}}]{Ruso:2022qes}%
  \BibitemOpen
  \bibfield  {author} {\bibinfo {author} {\bibfnamefont {L.~A.}\ \bibnamefont
  {Ruso}} \emph {et~al.},\ }\href@noop {} {\bibinfo {title} {{Theoretical tools
  for neutrino scattering: interplay between lattice QCD, EFTs, nuclear
  physics, phenomenology, and neutrino event generators}}} (\bibinfo {year}
  {2022}),\ \Eprint {https://arxiv.org/abs/2203.09030} {arXiv:2203.09030
  [hep-ph]} \BibitemShut {NoStop}%
\bibitem [{\citenamefont {Kronfeld}(2012)}]{Kronfeld:2012uk}%
  \BibitemOpen
  \bibfield  {author} {\bibinfo {author} {\bibfnamefont {A.~S.}\ \bibnamefont
  {Kronfeld}},\ }\bibfield  {title} {\bibinfo {title} {{Twenty-first Century
  Lattice Gauge Theory: Results from the QCD Lagrangian}},\ }\href
  {https://doi.org/10.1146/annurev-nucl-102711-094942} {\bibfield  {journal}
  {\bibinfo  {journal} {Ann. Rev. Nucl. Part. Sci.}\ }\textbf {\bibinfo
  {volume} {62}},\ \bibinfo {pages} {265} (\bibinfo {year} {2012})},\ \Eprint
  {https://arxiv.org/abs/1203.1204} {arXiv:1203.1204 [hep-lat]} \BibitemShut
  {NoStop}%
\bibitem [{\citenamefont {Beane}\ \emph {et~al.}(2011)\citenamefont {Beane},
  \citenamefont {Detmold}, \citenamefont {Orginos},\ and\ \citenamefont
  {Savage}}]{Beane:2010em}%
  \BibitemOpen
  \bibfield  {author} {\bibinfo {author} {\bibfnamefont {S.~R.}\ \bibnamefont
  {Beane}}, \bibinfo {author} {\bibfnamefont {W.}~\bibnamefont {Detmold}},
  \bibinfo {author} {\bibfnamefont {K.}~\bibnamefont {Orginos}},\ and\ \bibinfo
  {author} {\bibfnamefont {M.~J.}\ \bibnamefont {Savage}},\ }\bibfield  {title}
  {\bibinfo {title} {{Nuclear Physics from Lattice QCD}},\ }\href
  {https://doi.org/10.1016/j.ppnp.2010.08.002} {\bibfield  {journal} {\bibinfo
  {journal} {Prog. Part. Nucl. Phys.}\ }\textbf {\bibinfo {volume} {66}},\
  \bibinfo {pages} {1} (\bibinfo {year} {2011})},\ \Eprint
  {https://arxiv.org/abs/1004.2935} {arXiv:1004.2935 [hep-lat]} \BibitemShut
  {NoStop}%
\bibitem [{\citenamefont {Brice\~no}\ \emph {et~al.}(2018)\citenamefont
  {Brice\~no}, \citenamefont {Dudek},\ and\ \citenamefont
  {Young}}]{Briceno:2017max}%
  \BibitemOpen
  \bibfield  {author} {\bibinfo {author} {\bibfnamefont {R.~A.}\ \bibnamefont
  {Brice\~no}}, \bibinfo {author} {\bibfnamefont {J.~J.}\ \bibnamefont
  {Dudek}},\ and\ \bibinfo {author} {\bibfnamefont {R.~D.}\ \bibnamefont
  {Young}},\ }\bibfield  {title} {\bibinfo {title} {{Scattering processes and
  resonances from lattice QCD}},\ }\href
  {https://doi.org/10.1103/RevModPhys.90.025001} {\bibfield  {journal}
  {\bibinfo  {journal} {Rev. Mod. Phys.}\ }\textbf {\bibinfo {volume} {90}},\
  \bibinfo {pages} {025001} (\bibinfo {year} {2018})},\ \Eprint
  {https://arxiv.org/abs/1706.06223} {arXiv:1706.06223 [hep-lat]} \BibitemShut
  {NoStop}%
\bibitem [{\citenamefont {Hansen}\ and\ \citenamefont
  {Sharpe}(2019)}]{Hansen:2019nir}%
  \BibitemOpen
  \bibfield  {author} {\bibinfo {author} {\bibfnamefont {M.~T.}\ \bibnamefont
  {Hansen}}\ and\ \bibinfo {author} {\bibfnamefont {S.~R.}\ \bibnamefont
  {Sharpe}},\ }\bibfield  {title} {\bibinfo {title} {{Lattice QCD and
  Three-particle Decays of Resonances}},\ }\href
  {https://doi.org/10.1146/annurev-nucl-101918-023723} {\bibfield  {journal}
  {\bibinfo  {journal} {Ann. Rev. Nucl. Part. Sci.}\ }\textbf {\bibinfo
  {volume} {69}},\ \bibinfo {pages} {65} (\bibinfo {year} {2019})},\ \Eprint
  {https://arxiv.org/abs/1901.00483} {arXiv:1901.00483 [hep-lat]} \BibitemShut
  {NoStop}%
\bibitem [{\citenamefont {H\"orz}(2022)}]{Horz:2022glt}%
  \BibitemOpen
  \bibfield  {author} {\bibinfo {author} {\bibfnamefont {B.}~\bibnamefont
  {H\"orz}},\ }\bibfield  {title} {\bibinfo {title} {{Spectroscopy and Hadron
  Interactions}},\ }\href {https://doi.org/10.22323/1.396.0006} {\bibfield
  {journal} {\bibinfo  {journal} {PoS}\ }\textbf {\bibinfo {volume}
  {LATTICE2021}},\ \bibinfo {pages} {006} (\bibinfo {year} {2022})}\BibitemShut
  {NoStop}%
\bibitem [{\citenamefont {Romero-L\'opez}(2023)}]{Romero-Lopez:2022usb}%
  \BibitemOpen
  \bibfield  {author} {\bibinfo {author} {\bibfnamefont {F.}~\bibnamefont
  {Romero-L\'opez}},\ }\bibfield  {title} {\bibinfo {title} {{Multi-hadron
  interactions from lattice QCD}},\ }\href
  {https://doi.org/10.22323/1.430.0235} {\bibfield  {journal} {\bibinfo
  {journal} {PoS}\ }\textbf {\bibinfo {volume} {LATTICE2022}},\ \bibinfo
  {pages} {235} (\bibinfo {year} {2023})},\ \Eprint
  {https://arxiv.org/abs/2212.13793} {arXiv:2212.13793 [hep-lat]} \BibitemShut
  {NoStop}%
\bibitem [{\citenamefont {Hanlon}(2024)}]{Hanlon:2024fjd}%
  \BibitemOpen
  \bibfield  {author} {\bibinfo {author} {\bibfnamefont {A.~D.}\ \bibnamefont
  {Hanlon}},\ }\bibfield  {title} {\bibinfo {title} {{Hadron spectroscopy and
  few-body dynamics from lattice QCD}},\ }in\ \href@noop {} {\emph {\bibinfo
  {booktitle} {{40th International Symposium on Lattice Field Theory}}}}\
  (\bibinfo {year} {2024})\ \Eprint {https://arxiv.org/abs/2402.05185}
  {arXiv:2402.05185 [hep-lat]} \BibitemShut {NoStop}%
\bibitem [{\citenamefont {Parisi}(1984)}]{Parisi:1983ae}%
  \BibitemOpen
  \bibfield  {author} {\bibinfo {author} {\bibfnamefont {G.}~\bibnamefont
  {Parisi}},\ }\bibfield  {title} {\bibinfo {title} {{The Strategy for
  Computing the Hadronic Mass Spectrum}},\ }\href
  {https://doi.org/10.1016/0370-1573(84)90081-4} {\bibfield  {journal}
  {\bibinfo  {journal} {Phys. Rept.}\ }\textbf {\bibinfo {volume} {103}},\
  \bibinfo {pages} {203} (\bibinfo {year} {1984})}\BibitemShut {NoStop}%
\bibitem [{\citenamefont {Lepage}(1989)}]{Lepage:1989hd}%
  \BibitemOpen
  \bibfield  {author} {\bibinfo {author} {\bibfnamefont {G.~P.}\ \bibnamefont
  {Lepage}},\ }\bibfield  {title} {\bibinfo {title} {{The Analysis of
  Algorithms for Lattice Field Theory}},\ }in\ \href@noop {} {\emph {\bibinfo
  {booktitle} {{Theoretical Advanced Study Institute in Elementary Particle
  Physics}}}}\ (\bibinfo {year} {1989})\BibitemShut {NoStop}%
\bibitem [{\citenamefont {Beane}\ \emph
  {et~al.}(2009{\natexlab{a}})\citenamefont {Beane}, \citenamefont {Detmold},
  \citenamefont {Luu}, \citenamefont {Orginos}, \citenamefont {Parre\~{n}o},
  \citenamefont {Savage}, \citenamefont {Torok},\ and\ \citenamefont
  {Walker-Loud}}]{Beane:2009gs}%
  \BibitemOpen
  \bibfield  {author} {\bibinfo {author} {\bibfnamefont {S.~R.}\ \bibnamefont
  {Beane}}, \bibinfo {author} {\bibfnamefont {W.}~\bibnamefont {Detmold}},
  \bibinfo {author} {\bibfnamefont {T.~C.}\ \bibnamefont {Luu}}, \bibinfo
  {author} {\bibfnamefont {K.}~\bibnamefont {Orginos}}, \bibinfo {author}
  {\bibfnamefont {A.}~\bibnamefont {Parre\~{n}o}}, \bibinfo {author}
  {\bibfnamefont {M.~J.}\ \bibnamefont {Savage}}, \bibinfo {author}
  {\bibfnamefont {A.}~\bibnamefont {Torok}},\ and\ \bibinfo {author}
  {\bibfnamefont {A.}~\bibnamefont {Walker-Loud}} (\bibinfo {collaboration}
  {NPLQCD}),\ }\bibfield  {title} {\bibinfo {title} {{High Statistics Analysis
  using Anisotropic Clover Lattices: (II) Three-Baryon Systems}},\ }\href
  {https://doi.org/10.1103/PhysRevD.80.074501} {\bibfield  {journal} {\bibinfo
  {journal} {Phys. Rev. D}\ }\textbf {\bibinfo {volume} {80}},\ \bibinfo
  {pages} {074501} (\bibinfo {year} {2009}{\natexlab{a}})},\ \Eprint
  {https://arxiv.org/abs/0905.0466} {arXiv:0905.0466 [hep-lat]} \BibitemShut
  {NoStop}%
\bibitem [{\citenamefont {Beane}\ and\ \citenamefont
  {Savage}(2003{\natexlab{a}})}]{Beane:2002vs}%
  \BibitemOpen
  \bibfield  {author} {\bibinfo {author} {\bibfnamefont {S.~R.}\ \bibnamefont
  {Beane}}\ and\ \bibinfo {author} {\bibfnamefont {M.~J.}\ \bibnamefont
  {Savage}},\ }\bibfield  {title} {\bibinfo {title} {{Variation of fundamental
  couplings and nuclear forces}},\ }\href
  {https://doi.org/10.1016/S0375-9474(02)01268-X} {\bibfield  {journal}
  {\bibinfo  {journal} {Nucl. Phys. A}\ }\textbf {\bibinfo {volume} {713}},\
  \bibinfo {pages} {148} (\bibinfo {year} {2003}{\natexlab{a}})},\ \Eprint
  {https://arxiv.org/abs/hep-ph/0206113} {arXiv:hep-ph/0206113} \BibitemShut
  {NoStop}%
\bibitem [{\citenamefont {Epelbaum}\ \emph {et~al.}(2003)\citenamefont
  {Epelbaum}, \citenamefont {Mei{\ss}ner},\ and\ \citenamefont
  {Gloeckle}}]{Epelbaum:2002gb}%
  \BibitemOpen
  \bibfield  {author} {\bibinfo {author} {\bibfnamefont {E.}~\bibnamefont
  {Epelbaum}}, \bibinfo {author} {\bibfnamefont {U.-G.}\ \bibnamefont
  {Mei{\ss}ner}},\ and\ \bibinfo {author} {\bibfnamefont {W.}~\bibnamefont
  {Gloeckle}},\ }\bibfield  {title} {\bibinfo {title} {{Nuclear forces in the
  chiral limit}},\ }\href {https://doi.org/10.1016/S0375-9474(02)01393-3}
  {\bibfield  {journal} {\bibinfo  {journal} {Nucl. Phys. A}\ }\textbf
  {\bibinfo {volume} {714}},\ \bibinfo {pages} {535} (\bibinfo {year}
  {2003})},\ \Eprint {https://arxiv.org/abs/nucl-th/0207089}
  {arXiv:nucl-th/0207089} \BibitemShut {NoStop}%
\bibitem [{\citenamefont {Beane}\ and\ \citenamefont
  {Savage}(2003{\natexlab{b}})}]{Beane:2002xf}%
  \BibitemOpen
  \bibfield  {author} {\bibinfo {author} {\bibfnamefont {S.~R.}\ \bibnamefont
  {Beane}}\ and\ \bibinfo {author} {\bibfnamefont {M.~J.}\ \bibnamefont
  {Savage}},\ }\bibfield  {title} {\bibinfo {title} {{The Quark mass dependence
  of two nucleon systems}},\ }\href
  {https://doi.org/10.1016/S0375-9474(02)01586-5} {\bibfield  {journal}
  {\bibinfo  {journal} {Nucl. Phys. A}\ }\textbf {\bibinfo {volume} {717}},\
  \bibinfo {pages} {91} (\bibinfo {year} {2003}{\natexlab{b}})},\ \Eprint
  {https://arxiv.org/abs/nucl-th/0208021} {arXiv:nucl-th/0208021} \BibitemShut
  {NoStop}%
\bibitem [{\citenamefont {Braaten}\ and\ \citenamefont
  {Hammer}(2003)}]{Braaten:2003eu}%
  \BibitemOpen
  \bibfield  {author} {\bibinfo {author} {\bibfnamefont {E.}~\bibnamefont
  {Braaten}}\ and\ \bibinfo {author} {\bibfnamefont {H.~W.}\ \bibnamefont
  {Hammer}},\ }\bibfield  {title} {\bibinfo {title} {{An Infrared
  renormalization group limit cycle in QCD}},\ }\href
  {https://doi.org/10.1103/PhysRevLett.91.102002} {\bibfield  {journal}
  {\bibinfo  {journal} {Phys. Rev. Lett.}\ }\textbf {\bibinfo {volume} {91}},\
  \bibinfo {pages} {102002} (\bibinfo {year} {2003})},\ \Eprint
  {https://arxiv.org/abs/nucl-th/0303038} {arXiv:nucl-th/0303038} \BibitemShut
  {NoStop}%
\bibitem [{\citenamefont {Flambaum}\ and\ \citenamefont
  {Wiringa}(2007)}]{Flambaum:2007mj}%
  \BibitemOpen
  \bibfield  {author} {\bibinfo {author} {\bibfnamefont {V.~V.}\ \bibnamefont
  {Flambaum}}\ and\ \bibinfo {author} {\bibfnamefont {R.~B.}\ \bibnamefont
  {Wiringa}},\ }\bibfield  {title} {\bibinfo {title} {{Dependence of nuclear
  binding on hadronic mass variation}},\ }\href
  {https://doi.org/10.1103/PhysRevC.76.054002} {\bibfield  {journal} {\bibinfo
  {journal} {Phys. Rev. C}\ }\textbf {\bibinfo {volume} {76}},\ \bibinfo
  {pages} {054002} (\bibinfo {year} {2007})},\ \Eprint
  {https://arxiv.org/abs/0709.0077} {arXiv:0709.0077 [nucl-th]} \BibitemShut
  {NoStop}%
\bibitem [{\citenamefont {Bedaque}\ \emph {et~al.}(2011)\citenamefont
  {Bedaque}, \citenamefont {Luu},\ and\ \citenamefont
  {Platter}}]{Bedaque:2010hr}%
  \BibitemOpen
  \bibfield  {author} {\bibinfo {author} {\bibfnamefont {P.~F.}\ \bibnamefont
  {Bedaque}}, \bibinfo {author} {\bibfnamefont {T.}~\bibnamefont {Luu}},\ and\
  \bibinfo {author} {\bibfnamefont {L.}~\bibnamefont {Platter}},\ }\bibfield
  {title} {\bibinfo {title} {{Quark mass variation constraints from Big Bang
  nucleosynthesis}},\ }\href {https://doi.org/10.1103/PhysRevC.83.045803}
  {\bibfield  {journal} {\bibinfo  {journal} {Phys. Rev. C}\ }\textbf {\bibinfo
  {volume} {83}},\ \bibinfo {pages} {045803} (\bibinfo {year} {2011})},\
  \Eprint {https://arxiv.org/abs/1012.3840} {arXiv:1012.3840 [nucl-th]}
  \BibitemShut {NoStop}%
\bibitem [{\citenamefont {Chen}\ \emph {et~al.}(2012)\citenamefont {Chen},
  \citenamefont {Lee}, \citenamefont {Liu},\ and\ \citenamefont
  {Liu}}]{Chen:2010yt}%
  \BibitemOpen
  \bibfield  {author} {\bibinfo {author} {\bibfnamefont {J.-W.}\ \bibnamefont
  {Chen}}, \bibinfo {author} {\bibfnamefont {T.-K.}\ \bibnamefont {Lee}},
  \bibinfo {author} {\bibfnamefont {C.~P.}\ \bibnamefont {Liu}},\ and\ \bibinfo
  {author} {\bibfnamefont {Y.-S.}\ \bibnamefont {Liu}},\ }\bibfield  {title}
  {\bibinfo {title} {{On the Quark Mass Dependence of Two Nucleon
  Observables}},\ }\href {https://doi.org/10.1103/PhysRevC.86.054001}
  {\bibfield  {journal} {\bibinfo  {journal} {Phys. Rev. C}\ }\textbf {\bibinfo
  {volume} {86}},\ \bibinfo {pages} {054001} (\bibinfo {year} {2012})},\
  \Eprint {https://arxiv.org/abs/1012.0453} {arXiv:1012.0453 [nucl-th]}
  \BibitemShut {NoStop}%
\bibitem [{\citenamefont {Soto}\ and\ \citenamefont
  {Tarrus}(2012)}]{Soto:2011tb}%
  \BibitemOpen
  \bibfield  {author} {\bibinfo {author} {\bibfnamefont {J.}~\bibnamefont
  {Soto}}\ and\ \bibinfo {author} {\bibfnamefont {J.}~\bibnamefont {Tarrus}},\
  }\bibfield  {title} {\bibinfo {title} {{On the quark mass dependence of
  nucleon-nucleon S-wave scattering lengths}},\ }\href
  {https://doi.org/10.1103/PhysRevC.85.044001} {\bibfield  {journal} {\bibinfo
  {journal} {Phys. Rev. C}\ }\textbf {\bibinfo {volume} {85}},\ \bibinfo
  {pages} {044001} (\bibinfo {year} {2012})},\ \Eprint
  {https://arxiv.org/abs/1112.4426} {arXiv:1112.4426 [nucl-th]} \BibitemShut
  {NoStop}%
\bibitem [{\citenamefont {Epelbaum}\ \emph {et~al.}(2013)\citenamefont
  {Epelbaum}, \citenamefont {Krebs}, \citenamefont {L\"ahde}, \citenamefont
  {Lee},\ and\ \citenamefont {Mei\ss{}ner}}]{Epelbaum:2012iu}%
  \BibitemOpen
  \bibfield  {author} {\bibinfo {author} {\bibfnamefont {E.}~\bibnamefont
  {Epelbaum}}, \bibinfo {author} {\bibfnamefont {H.}~\bibnamefont {Krebs}},
  \bibinfo {author} {\bibfnamefont {T.~A.}\ \bibnamefont {L\"ahde}}, \bibinfo
  {author} {\bibfnamefont {D.}~\bibnamefont {Lee}},\ and\ \bibinfo {author}
  {\bibfnamefont {U.-G.}\ \bibnamefont {Mei\ss{}ner}},\ }\bibfield  {title}
  {\bibinfo {title} {{Viability of Carbon-Based Life as a Function of the Light
  Quark Mass}},\ }\href {https://doi.org/10.1103/PhysRevLett.110.112502}
  {\bibfield  {journal} {\bibinfo  {journal} {Phys. Rev. Lett.}\ }\textbf
  {\bibinfo {volume} {110}},\ \bibinfo {pages} {112502} (\bibinfo {year}
  {2013})},\ \Eprint {https://arxiv.org/abs/1212.4181} {arXiv:1212.4181
  [nucl-th]} \BibitemShut {NoStop}%
\bibitem [{\citenamefont {Detmold}\ \emph
  {et~al.}(2014{\natexlab{a}})\citenamefont {Detmold}, \citenamefont
  {McCullough},\ and\ \citenamefont {Pochinsky}}]{Detmold:2014qqa}%
  \BibitemOpen
  \bibfield  {author} {\bibinfo {author} {\bibfnamefont {W.}~\bibnamefont
  {Detmold}}, \bibinfo {author} {\bibfnamefont {M.}~\bibnamefont
  {McCullough}},\ and\ \bibinfo {author} {\bibfnamefont {A.}~\bibnamefont
  {Pochinsky}},\ }\bibfield  {title} {\bibinfo {title} {{Dark Nuclei I:
  Cosmology and Indirect Detection}},\ }\href
  {https://doi.org/10.1103/PhysRevD.90.115013} {\bibfield  {journal} {\bibinfo
  {journal} {Phys. Rev. D}\ }\textbf {\bibinfo {volume} {90}},\ \bibinfo
  {pages} {115013} (\bibinfo {year} {2014}{\natexlab{a}})},\ \Eprint
  {https://arxiv.org/abs/1406.2276} {arXiv:1406.2276 [hep-ph]} \BibitemShut
  {NoStop}%
\bibitem [{\citenamefont {Detmold}\ \emph
  {et~al.}(2014{\natexlab{b}})\citenamefont {Detmold}, \citenamefont
  {McCullough},\ and\ \citenamefont {Pochinsky}}]{Detmold:2014kba}%
  \BibitemOpen
  \bibfield  {author} {\bibinfo {author} {\bibfnamefont {W.}~\bibnamefont
  {Detmold}}, \bibinfo {author} {\bibfnamefont {M.}~\bibnamefont
  {McCullough}},\ and\ \bibinfo {author} {\bibfnamefont {A.}~\bibnamefont
  {Pochinsky}},\ }\bibfield  {title} {\bibinfo {title} {{Dark nuclei. II.
  Nuclear spectroscopy in two-color QCD}},\ }\href
  {https://doi.org/10.1103/PhysRevD.90.114506} {\bibfield  {journal} {\bibinfo
  {journal} {Phys. Rev. D}\ }\textbf {\bibinfo {volume} {90}},\ \bibinfo
  {pages} {114506} (\bibinfo {year} {2014}{\natexlab{b}})},\ \Eprint
  {https://arxiv.org/abs/1406.4116} {arXiv:1406.4116 [hep-lat]} \BibitemShut
  {NoStop}%
\bibitem [{\citenamefont {DeGrand}\ and\ \citenamefont
  {Neil}(2020)}]{DeGrand:2019vbx}%
  \BibitemOpen
  \bibfield  {author} {\bibinfo {author} {\bibfnamefont {T.}~\bibnamefont
  {DeGrand}}\ and\ \bibinfo {author} {\bibfnamefont {E.~T.}\ \bibnamefont
  {Neil}},\ }\bibfield  {title} {\bibinfo {title} {{Repurposing lattice QCD
  results for composite phenomenology}},\ }\href
  {https://doi.org/10.1103/PhysRevD.101.034504} {\bibfield  {journal} {\bibinfo
   {journal} {Phys. Rev. D}\ }\textbf {\bibinfo {volume} {101}},\ \bibinfo
  {pages} {034504} (\bibinfo {year} {2020})},\ \Eprint
  {https://arxiv.org/abs/1910.08561} {arXiv:1910.08561 [hep-ph]} \BibitemShut
  {NoStop}%
\bibitem [{\citenamefont {Fukugita}\ \emph {et~al.}(1994)\citenamefont
  {Fukugita}, \citenamefont {Kuramashi}, \citenamefont {Mino}, \citenamefont
  {Okawa},\ and\ \citenamefont {Ukawa}}]{Fukugita:1994na}%
  \BibitemOpen
  \bibfield  {author} {\bibinfo {author} {\bibfnamefont {M.}~\bibnamefont
  {Fukugita}}, \bibinfo {author} {\bibfnamefont {Y.}~\bibnamefont {Kuramashi}},
  \bibinfo {author} {\bibfnamefont {H.}~\bibnamefont {Mino}}, \bibinfo {author}
  {\bibfnamefont {M.}~\bibnamefont {Okawa}},\ and\ \bibinfo {author}
  {\bibfnamefont {A.}~\bibnamefont {Ukawa}},\ }\bibfield  {title} {\bibinfo
  {title} {{An Exploratory study of nucleon-nucleon scattering lengths in
  lattice QCD}},\ }\href {https://doi.org/10.1103/PhysRevLett.73.2176}
  {\bibfield  {journal} {\bibinfo  {journal} {Phys. Rev. Lett.}\ }\textbf
  {\bibinfo {volume} {73}},\ \bibinfo {pages} {2176} (\bibinfo {year}
  {1994})},\ \Eprint {https://arxiv.org/abs/hep-lat/9407012}
  {arXiv:hep-lat/9407012} \BibitemShut {NoStop}%
\bibitem [{\citenamefont {Fukugita}\ \emph {et~al.}(1995)\citenamefont
  {Fukugita}, \citenamefont {Kuramashi}, \citenamefont {Okawa}, \citenamefont
  {Mino},\ and\ \citenamefont {Ukawa}}]{Fukugita:1994ve}%
  \BibitemOpen
  \bibfield  {author} {\bibinfo {author} {\bibfnamefont {M.}~\bibnamefont
  {Fukugita}}, \bibinfo {author} {\bibfnamefont {Y.}~\bibnamefont {Kuramashi}},
  \bibinfo {author} {\bibfnamefont {M.}~\bibnamefont {Okawa}}, \bibinfo
  {author} {\bibfnamefont {H.}~\bibnamefont {Mino}},\ and\ \bibinfo {author}
  {\bibfnamefont {A.}~\bibnamefont {Ukawa}},\ }\bibfield  {title} {\bibinfo
  {title} {{Hadron scattering lengths in lattice QCD}},\ }\href
  {https://doi.org/10.1103/PhysRevD.52.3003} {\bibfield  {journal} {\bibinfo
  {journal} {Phys. Rev. D}\ }\textbf {\bibinfo {volume} {52}},\ \bibinfo
  {pages} {3003} (\bibinfo {year} {1995})},\ \Eprint
  {https://arxiv.org/abs/hep-lat/9501024} {arXiv:hep-lat/9501024} \BibitemShut
  {NoStop}%
\bibitem [{\citenamefont {Beane}\ \emph {et~al.}(2006)\citenamefont {Beane},
  \citenamefont {Bedaque}, \citenamefont {Orginos},\ and\ \citenamefont
  {Savage}}]{Beane:2006mx}%
  \BibitemOpen
  \bibfield  {author} {\bibinfo {author} {\bibfnamefont {S.~R.}\ \bibnamefont
  {Beane}}, \bibinfo {author} {\bibfnamefont {P.~F.}\ \bibnamefont {Bedaque}},
  \bibinfo {author} {\bibfnamefont {K.}~\bibnamefont {Orginos}},\ and\ \bibinfo
  {author} {\bibfnamefont {M.~J.}\ \bibnamefont {Savage}},\ }\bibfield  {title}
  {\bibinfo {title} {{Nucleon-nucleon scattering from fully-dynamical lattice
  QCD}},\ }\href {https://doi.org/10.1103/PhysRevLett.97.012001} {\bibfield
  {journal} {\bibinfo  {journal} {Phys. Rev. Lett.}\ }\textbf {\bibinfo
  {volume} {97}},\ \bibinfo {pages} {012001} (\bibinfo {year} {2006})},\
  \Eprint {https://arxiv.org/abs/hep-lat/0602010} {arXiv:hep-lat/0602010}
  \BibitemShut {NoStop}%
\bibitem [{\citenamefont {Beane}\ \emph {et~al.}(2007)\citenamefont {Beane},
  \citenamefont {Bedaque}, \citenamefont {Luu}, \citenamefont {Orginos},
  \citenamefont {Pallante}, \citenamefont {Parre\~{n}o},\ and\ \citenamefont
  {Savage}}]{Beane:2006gf}%
  \BibitemOpen
  \bibfield  {author} {\bibinfo {author} {\bibfnamefont {S.~R.}\ \bibnamefont
  {Beane}}, \bibinfo {author} {\bibfnamefont {P.~F.}\ \bibnamefont {Bedaque}},
  \bibinfo {author} {\bibfnamefont {T.~C.}\ \bibnamefont {Luu}}, \bibinfo
  {author} {\bibfnamefont {K.}~\bibnamefont {Orginos}}, \bibinfo {author}
  {\bibfnamefont {E.}~\bibnamefont {Pallante}}, \bibinfo {author}
  {\bibfnamefont {A.}~\bibnamefont {Parre\~{n}o}},\ and\ \bibinfo {author}
  {\bibfnamefont {M.~J.}\ \bibnamefont {Savage}} (\bibinfo {collaboration}
  {NPLQCD}),\ }\bibfield  {title} {\bibinfo {title} {{Hyperon-Nucleon
  Scattering from Fully-Dynamical Lattice QCD}},\ }\href
  {https://doi.org/10.1016/j.nuclphysa.2007.07.006} {\bibfield  {journal}
  {\bibinfo  {journal} {Nucl. Phys. A}\ }\textbf {\bibinfo {volume} {794}},\
  \bibinfo {pages} {62} (\bibinfo {year} {2007})},\ \Eprint
  {https://arxiv.org/abs/hep-lat/0612026} {arXiv:hep-lat/0612026} \BibitemShut
  {NoStop}%
\bibitem [{\citenamefont {Beane}\ \emph {et~al.}(2010)\citenamefont {Beane},
  \citenamefont {Detmold}, \citenamefont {Lin}, \citenamefont {Luu},
  \citenamefont {Orginos}, \citenamefont {Savage}, \citenamefont {Torok},\ and\
  \citenamefont {Walker-Loud}}]{Beane:2009py}%
  \BibitemOpen
  \bibfield  {author} {\bibinfo {author} {\bibfnamefont {S.~R.}\ \bibnamefont
  {Beane}}, \bibinfo {author} {\bibfnamefont {W.}~\bibnamefont {Detmold}},
  \bibinfo {author} {\bibfnamefont {H.-W.}\ \bibnamefont {Lin}}, \bibinfo
  {author} {\bibfnamefont {T.~C.}\ \bibnamefont {Luu}}, \bibinfo {author}
  {\bibfnamefont {K.}~\bibnamefont {Orginos}}, \bibinfo {author} {\bibfnamefont
  {M.~J.}\ \bibnamefont {Savage}}, \bibinfo {author} {\bibfnamefont
  {A.}~\bibnamefont {Torok}},\ and\ \bibinfo {author} {\bibfnamefont
  {A.}~\bibnamefont {Walker-Loud}} (\bibinfo {collaboration} {NPLQCD}),\
  }\bibfield  {title} {\bibinfo {title} {{High Statistics Analysis using
  Anisotropic Clover Lattices: (III) Baryon-Baryon Interactions}},\ }\href
  {https://doi.org/10.1103/PhysRevD.81.054505} {\bibfield  {journal} {\bibinfo
  {journal} {Phys. Rev. D}\ }\textbf {\bibinfo {volume} {81}},\ \bibinfo
  {pages} {054505} (\bibinfo {year} {2010})},\ \Eprint
  {https://arxiv.org/abs/0912.4243} {arXiv:0912.4243 [hep-lat]} \BibitemShut
  {NoStop}%
\bibitem [{\citenamefont {Yamazaki}\ \emph {et~al.}(2010)\citenamefont
  {Yamazaki}, \citenamefont {Kuramashi},\ and\ \citenamefont
  {Ukawa}}]{Yamazaki:2009ua}%
  \BibitemOpen
  \bibfield  {author} {\bibinfo {author} {\bibfnamefont {T.}~\bibnamefont
  {Yamazaki}}, \bibinfo {author} {\bibfnamefont {Y.}~\bibnamefont
  {Kuramashi}},\ and\ \bibinfo {author} {\bibfnamefont {A.}~\bibnamefont
  {Ukawa}} (\bibinfo {collaboration} {PACS-CS}),\ }\bibfield  {title} {\bibinfo
  {title} {{Helium Nuclei in Quenched Lattice QCD}},\ }\href
  {https://doi.org/10.1103/PhysRevD.81.111504} {\bibfield  {journal} {\bibinfo
  {journal} {Phys. Rev. D}\ }\textbf {\bibinfo {volume} {81}},\ \bibinfo
  {pages} {111504} (\bibinfo {year} {2010})},\ \Eprint
  {https://arxiv.org/abs/0912.1383} {arXiv:0912.1383 [hep-lat]} \BibitemShut
  {NoStop}%
\bibitem [{\citenamefont {Beane}\ \emph {et~al.}(2012)\citenamefont {Beane},
  \citenamefont {Chang}, \citenamefont {Detmold}, \citenamefont {Lin},
  \citenamefont {Luu}, \citenamefont {Orginos}, \citenamefont {Parre\~{n}o},
  \citenamefont {Savage}, \citenamefont {Torok},\ and\ \citenamefont
  {Walker-Loud}}]{NPLQCD:2011naw}%
  \BibitemOpen
  \bibfield  {author} {\bibinfo {author} {\bibfnamefont {S.~R.}\ \bibnamefont
  {Beane}}, \bibinfo {author} {\bibfnamefont {E.}~\bibnamefont {Chang}},
  \bibinfo {author} {\bibfnamefont {W.}~\bibnamefont {Detmold}}, \bibinfo
  {author} {\bibfnamefont {H.~W.}\ \bibnamefont {Lin}}, \bibinfo {author}
  {\bibfnamefont {T.~C.}\ \bibnamefont {Luu}}, \bibinfo {author} {\bibfnamefont
  {K.}~\bibnamefont {Orginos}}, \bibinfo {author} {\bibfnamefont
  {A.}~\bibnamefont {Parre\~{n}o}}, \bibinfo {author} {\bibfnamefont {M.~J.}\
  \bibnamefont {Savage}}, \bibinfo {author} {\bibfnamefont {A.}~\bibnamefont
  {Torok}},\ and\ \bibinfo {author} {\bibfnamefont {A.}~\bibnamefont
  {Walker-Loud}} (\bibinfo {collaboration} {NPLQCD}),\ }\bibfield  {title}
  {\bibinfo {title} {{The Deuteron and Exotic Two-Body Bound States from
  Lattice QCD}},\ }\href {https://doi.org/10.1103/PhysRevD.85.054511}
  {\bibfield  {journal} {\bibinfo  {journal} {Phys. Rev. D}\ }\textbf {\bibinfo
  {volume} {85}},\ \bibinfo {pages} {054511} (\bibinfo {year} {2012})},\
  \Eprint {https://arxiv.org/abs/1109.2889} {arXiv:1109.2889 [hep-lat]}
  \BibitemShut {NoStop}%
\bibitem [{\citenamefont {Beane}\ \emph
  {et~al.}(2013{\natexlab{a}})\citenamefont {Beane}, \citenamefont {Chang},
  \citenamefont {Cohen}, \citenamefont {Detmold}, \citenamefont {Lin},
  \citenamefont {Luu}, \citenamefont {Orginos}, \citenamefont {Parre\~{n}o},
  \citenamefont {Savage},\ and\ \citenamefont {Walker-Loud}}]{NPLQCD:2012mex}%
  \BibitemOpen
  \bibfield  {author} {\bibinfo {author} {\bibfnamefont {S.~R.}\ \bibnamefont
  {Beane}}, \bibinfo {author} {\bibfnamefont {E.}~\bibnamefont {Chang}},
  \bibinfo {author} {\bibfnamefont {S.~D.}\ \bibnamefont {Cohen}}, \bibinfo
  {author} {\bibfnamefont {W.}~\bibnamefont {Detmold}}, \bibinfo {author}
  {\bibfnamefont {H.~W.}\ \bibnamefont {Lin}}, \bibinfo {author} {\bibfnamefont
  {T.~C.}\ \bibnamefont {Luu}}, \bibinfo {author} {\bibfnamefont
  {K.}~\bibnamefont {Orginos}}, \bibinfo {author} {\bibfnamefont
  {A.}~\bibnamefont {Parre\~{n}o}}, \bibinfo {author} {\bibfnamefont {M.~J.}\
  \bibnamefont {Savage}},\ and\ \bibinfo {author} {\bibfnamefont
  {A.}~\bibnamefont {Walker-Loud}} (\bibinfo {collaboration} {NPLQCD}),\
  }\bibfield  {title} {\bibinfo {title} {{Light Nuclei and Hypernuclei from
  Quantum Chromodynamics in the Limit of SU(3) Flavor Symmetry}},\ }\href
  {https://doi.org/10.1103/PhysRevD.87.034506} {\bibfield  {journal} {\bibinfo
  {journal} {Phys. Rev. D}\ }\textbf {\bibinfo {volume} {87}},\ \bibinfo
  {pages} {034506} (\bibinfo {year} {2013}{\natexlab{a}})},\ \Eprint
  {https://arxiv.org/abs/1206.5219} {arXiv:1206.5219 [hep-lat]} \BibitemShut
  {NoStop}%
\bibitem [{\citenamefont {Beane}\ \emph
  {et~al.}(2013{\natexlab{b}})\citenamefont {Beane} \emph
  {et~al.}}]{NPLQCD:2013bqy}%
  \BibitemOpen
  \bibfield  {author} {\bibinfo {author} {\bibfnamefont {S.~R.}\ \bibnamefont
  {Beane}} \emph {et~al.} (\bibinfo {collaboration} {NPLQCD}),\ }\bibfield
  {title} {\bibinfo {title} {{Nucleon-Nucleon Scattering Parameters in the
  Limit of SU(3) Flavor Symmetry}},\ }\href
  {https://doi.org/10.1103/PhysRevC.88.024003} {\bibfield  {journal} {\bibinfo
  {journal} {Phys. Rev. C}\ }\textbf {\bibinfo {volume} {88}},\ \bibinfo
  {pages} {024003} (\bibinfo {year} {2013}{\natexlab{b}})},\ \Eprint
  {https://arxiv.org/abs/1301.5790} {arXiv:1301.5790 [hep-lat]} \BibitemShut
  {NoStop}%
\bibitem [{\citenamefont {Wagman}\ \emph {et~al.}(2017)\citenamefont {Wagman},
  \citenamefont {Winter}, \citenamefont {Chang}, \citenamefont {Davoudi},
  \citenamefont {Detmold}, \citenamefont {Orginos}, \citenamefont {Savage},\
  and\ \citenamefont {Shanahan}}]{Wagman:2017tmp}%
  \BibitemOpen
  \bibfield  {author} {\bibinfo {author} {\bibfnamefont {M.~L.}\ \bibnamefont
  {Wagman}}, \bibinfo {author} {\bibfnamefont {F.}~\bibnamefont {Winter}},
  \bibinfo {author} {\bibfnamefont {E.}~\bibnamefont {Chang}}, \bibinfo
  {author} {\bibfnamefont {Z.}~\bibnamefont {Davoudi}}, \bibinfo {author}
  {\bibfnamefont {W.}~\bibnamefont {Detmold}}, \bibinfo {author} {\bibfnamefont
  {K.}~\bibnamefont {Orginos}}, \bibinfo {author} {\bibfnamefont {M.~J.}\
  \bibnamefont {Savage}},\ and\ \bibinfo {author} {\bibfnamefont {P.~E.}\
  \bibnamefont {Shanahan}},\ }\bibfield  {title} {\bibinfo {title}
  {{Baryon-Baryon Interactions and Spin-Flavor Symmetry from Lattice Quantum
  Chromodynamics}},\ }\href {https://doi.org/10.1103/PhysRevD.96.114510}
  {\bibfield  {journal} {\bibinfo  {journal} {Phys. Rev. D}\ }\textbf {\bibinfo
  {volume} {96}},\ \bibinfo {pages} {114510} (\bibinfo {year} {2017})},\
  \Eprint {https://arxiv.org/abs/1706.06550} {arXiv:1706.06550 [hep-lat]}
  \BibitemShut {NoStop}%
\bibitem [{\citenamefont {Orginos}\ \emph {et~al.}(2015)\citenamefont
  {Orginos}, \citenamefont {Parre\~{n}o}, \citenamefont {Savage}, \citenamefont
  {Beane}, \citenamefont {Chang},\ and\ \citenamefont
  {Detmold}}]{Orginos:2015aya}%
  \BibitemOpen
  \bibfield  {author} {\bibinfo {author} {\bibfnamefont {K.}~\bibnamefont
  {Orginos}}, \bibinfo {author} {\bibfnamefont {A.}~\bibnamefont
  {Parre\~{n}o}}, \bibinfo {author} {\bibfnamefont {M.~J.}\ \bibnamefont
  {Savage}}, \bibinfo {author} {\bibfnamefont {S.~R.}\ \bibnamefont {Beane}},
  \bibinfo {author} {\bibfnamefont {E.}~\bibnamefont {Chang}},\ and\ \bibinfo
  {author} {\bibfnamefont {W.}~\bibnamefont {Detmold}},\ }\bibfield  {title}
  {\bibinfo {title} {{Two nucleon systems at $m_\pi\sim 450~{\rm MeV}$ from
  lattice QCD}},\ }\href {https://doi.org/10.1103/PhysRevD.92.114512}
  {\bibfield  {journal} {\bibinfo  {journal} {Phys. Rev. D}\ }\textbf {\bibinfo
  {volume} {92}},\ \bibinfo {pages} {114512} (\bibinfo {year} {2015})},\
  \bibinfo {note} {[Erratum: Phys.Rev.D 102, 039903 (2020)]},\ \Eprint
  {https://arxiv.org/abs/1508.07583} {arXiv:1508.07583 [hep-lat]} \BibitemShut
  {NoStop}%
\bibitem [{\citenamefont {Beane}\ \emph {et~al.}(2021)\citenamefont {Beane}
  \emph {et~al.}}]{NPLQCD:2020ozd}%
  \BibitemOpen
  \bibfield  {author} {\bibinfo {author} {\bibfnamefont {S.~R.}\ \bibnamefont
  {Beane}} \emph {et~al.} (\bibinfo {collaboration} {NPLQCD, QCDSF}),\
  }\bibfield  {title} {\bibinfo {title} {{Charged multihadron systems in
  lattice QCD+QED}},\ }\href {https://doi.org/10.1103/PhysRevD.103.054504}
  {\bibfield  {journal} {\bibinfo  {journal} {Phys. Rev. D}\ }\textbf {\bibinfo
  {volume} {103}},\ \bibinfo {pages} {054504} (\bibinfo {year} {2021})},\
  \Eprint {https://arxiv.org/abs/2003.12130} {arXiv:2003.12130 [hep-lat]}
  \BibitemShut {NoStop}%
\bibitem [{\citenamefont {Illa}\ \emph {et~al.}(2021)\citenamefont {Illa} \emph
  {et~al.}}]{NPLQCD:2020lxg}%
  \BibitemOpen
  \bibfield  {author} {\bibinfo {author} {\bibfnamefont {M.}~\bibnamefont
  {Illa}} \emph {et~al.} (\bibinfo {collaboration} {NPLQCD}),\ }\bibfield
  {title} {\bibinfo {title} {{Low-energy scattering and effective interactions
  of two baryons at $m_{\pi}\sim 450$ MeV from lattice quantum
  chromodynamics}},\ }\href {https://doi.org/10.1103/PhysRevD.103.054508}
  {\bibfield  {journal} {\bibinfo  {journal} {Phys. Rev. D}\ }\textbf {\bibinfo
  {volume} {103}},\ \bibinfo {pages} {054508} (\bibinfo {year} {2021})},\
  \Eprint {https://arxiv.org/abs/2009.12357} {arXiv:2009.12357 [hep-lat]}
  \BibitemShut {NoStop}%
\bibitem [{\citenamefont {Yamazaki}\ \emph {et~al.}(2012)\citenamefont
  {Yamazaki}, \citenamefont {Ishikawa}, \citenamefont {Kuramashi},\ and\
  \citenamefont {Ukawa}}]{Yamazaki:2012hi}%
  \BibitemOpen
  \bibfield  {author} {\bibinfo {author} {\bibfnamefont {T.}~\bibnamefont
  {Yamazaki}}, \bibinfo {author} {\bibfnamefont {K.-i.}\ \bibnamefont
  {Ishikawa}}, \bibinfo {author} {\bibfnamefont {Y.}~\bibnamefont
  {Kuramashi}},\ and\ \bibinfo {author} {\bibfnamefont {A.}~\bibnamefont
  {Ukawa}},\ }\bibfield  {title} {\bibinfo {title} {{Helium nuclei, deuteron
  and dineutron in 2+1 flavor lattice QCD}},\ }\href
  {https://doi.org/10.1103/PhysRevD.86.074514} {\bibfield  {journal} {\bibinfo
  {journal} {Phys. Rev. D}\ }\textbf {\bibinfo {volume} {86}},\ \bibinfo
  {pages} {074514} (\bibinfo {year} {2012})},\ \Eprint
  {https://arxiv.org/abs/1207.4277} {arXiv:1207.4277 [hep-lat]} \BibitemShut
  {NoStop}%
\bibitem [{\citenamefont {Yamazaki}\ \emph {et~al.}(2015)\citenamefont
  {Yamazaki}, \citenamefont {Ishikawa}, \citenamefont {Kuramashi},\ and\
  \citenamefont {Ukawa}}]{Yamazaki:2015asa}%
  \BibitemOpen
  \bibfield  {author} {\bibinfo {author} {\bibfnamefont {T.}~\bibnamefont
  {Yamazaki}}, \bibinfo {author} {\bibfnamefont {K.-i.}\ \bibnamefont
  {Ishikawa}}, \bibinfo {author} {\bibfnamefont {Y.}~\bibnamefont
  {Kuramashi}},\ and\ \bibinfo {author} {\bibfnamefont {A.}~\bibnamefont
  {Ukawa}},\ }\bibfield  {title} {\bibinfo {title} {{Study of quark mass
  dependence of binding energy for light nuclei in 2+1 flavor lattice QCD}},\
  }\href {https://doi.org/10.1103/PhysRevD.92.014501} {\bibfield  {journal}
  {\bibinfo  {journal} {Phys. Rev. D}\ }\textbf {\bibinfo {volume} {92}},\
  \bibinfo {pages} {014501} (\bibinfo {year} {2015})},\ \Eprint
  {https://arxiv.org/abs/1502.04182} {arXiv:1502.04182 [hep-lat]} \BibitemShut
  {NoStop}%
\bibitem [{\citenamefont {Berkowitz}\ \emph {et~al.}(2017)\citenamefont
  {Berkowitz}, \citenamefont {Kurth}, \citenamefont {Nicholson}, \citenamefont
  {Jo\'o}, \citenamefont {Rinaldi}, \citenamefont {Strother}, \citenamefont
  {Vranas},\ and\ \citenamefont {Walker-Loud}}]{Berkowitz:2015eaa}%
  \BibitemOpen
  \bibfield  {author} {\bibinfo {author} {\bibfnamefont {E.}~\bibnamefont
  {Berkowitz}}, \bibinfo {author} {\bibfnamefont {T.}~\bibnamefont {Kurth}},
  \bibinfo {author} {\bibfnamefont {A.}~\bibnamefont {Nicholson}}, \bibinfo
  {author} {\bibfnamefont {B.}~\bibnamefont {Jo\'o}}, \bibinfo {author}
  {\bibfnamefont {E.}~\bibnamefont {Rinaldi}}, \bibinfo {author} {\bibfnamefont
  {M.}~\bibnamefont {Strother}}, \bibinfo {author} {\bibfnamefont {P.~M.}\
  \bibnamefont {Vranas}},\ and\ \bibinfo {author} {\bibfnamefont
  {A.}~\bibnamefont {Walker-Loud}},\ }\bibfield  {title} {\bibinfo {title}
  {{Two-Nucleon Higher Partial-Wave Scattering from Lattice QCD}},\ }\href
  {https://doi.org/10.1016/j.physletb.2016.12.024} {\bibfield  {journal}
  {\bibinfo  {journal} {Phys. Lett. B}\ }\textbf {\bibinfo {volume} {765}},\
  \bibinfo {pages} {285} (\bibinfo {year} {2017})},\ \Eprint
  {https://arxiv.org/abs/1508.00886} {arXiv:1508.00886 [hep-lat]} \BibitemShut
  {NoStop}%
\bibitem [{\citenamefont {Francis}\ \emph {et~al.}(2019)\citenamefont
  {Francis}, \citenamefont {Green}, \citenamefont {Junnarkar}, \citenamefont
  {Miao}, \citenamefont {Rae},\ and\ \citenamefont {Wittig}}]{Francis:2018qch}%
  \BibitemOpen
  \bibfield  {author} {\bibinfo {author} {\bibfnamefont {A.}~\bibnamefont
  {Francis}}, \bibinfo {author} {\bibfnamefont {J.~R.}\ \bibnamefont {Green}},
  \bibinfo {author} {\bibfnamefont {P.~M.}\ \bibnamefont {Junnarkar}}, \bibinfo
  {author} {\bibfnamefont {C.}~\bibnamefont {Miao}}, \bibinfo {author}
  {\bibfnamefont {T.~D.}\ \bibnamefont {Rae}},\ and\ \bibinfo {author}
  {\bibfnamefont {H.}~\bibnamefont {Wittig}},\ }\bibfield  {title} {\bibinfo
  {title} {{Lattice QCD study of the $H$ dibaryon using hexaquark and
  two-baryon interpolators}},\ }\href
  {https://doi.org/10.1103/PhysRevD.99.074505} {\bibfield  {journal} {\bibinfo
  {journal} {Phys. Rev. D}\ }\textbf {\bibinfo {volume} {99}},\ \bibinfo
  {pages} {074505} (\bibinfo {year} {2019})},\ \Eprint
  {https://arxiv.org/abs/1805.03966} {arXiv:1805.03966 [hep-lat]} \BibitemShut
  {NoStop}%
\bibitem [{\citenamefont {Amarasinghe}\ \emph {et~al.}(2023)\citenamefont
  {Amarasinghe}, \citenamefont {Baghdadi}, \citenamefont {Davoudi},
  \citenamefont {Detmold}, \citenamefont {Illa}, \citenamefont {Parre\~{n}o},
  \citenamefont {Pochinsky}, \citenamefont {Shanahan},\ and\ \citenamefont
  {Wagman}}]{Amarasinghe:2021lqa}%
  \BibitemOpen
  \bibfield  {author} {\bibinfo {author} {\bibfnamefont {S.}~\bibnamefont
  {Amarasinghe}}, \bibinfo {author} {\bibfnamefont {R.}~\bibnamefont
  {Baghdadi}}, \bibinfo {author} {\bibfnamefont {Z.}~\bibnamefont {Davoudi}},
  \bibinfo {author} {\bibfnamefont {W.}~\bibnamefont {Detmold}}, \bibinfo
  {author} {\bibfnamefont {M.}~\bibnamefont {Illa}}, \bibinfo {author}
  {\bibfnamefont {A.}~\bibnamefont {Parre\~{n}o}}, \bibinfo {author}
  {\bibfnamefont {A.~V.}\ \bibnamefont {Pochinsky}}, \bibinfo {author}
  {\bibfnamefont {P.~E.}\ \bibnamefont {Shanahan}},\ and\ \bibinfo {author}
  {\bibfnamefont {M.~L.}\ \bibnamefont {Wagman}},\ }\bibfield  {title}
  {\bibinfo {title} {{Variational study of two-nucleon systems with lattice
  QCD}},\ }\href {https://doi.org/10.1103/PhysRevD.107.094508} {\bibfield
  {journal} {\bibinfo  {journal} {Phys. Rev. D}\ }\textbf {\bibinfo {volume}
  {107}},\ \bibinfo {pages} {094508} (\bibinfo {year} {2023})},\ \Eprint
  {https://arxiv.org/abs/2108.10835} {arXiv:2108.10835 [hep-lat]} \BibitemShut
  {NoStop}%
\bibitem [{\citenamefont {H\"orz}\ \emph {et~al.}(2021)\citenamefont {H\"orz}
  \emph {et~al.}}]{Horz:2020zvv}%
  \BibitemOpen
  \bibfield  {author} {\bibinfo {author} {\bibfnamefont {B.}~\bibnamefont
  {H\"orz}} \emph {et~al.},\ }\bibfield  {title} {\bibinfo {title}
  {{Two-nucleon S-wave interactions at the $SU(3)$ flavor-symmetric point with
  $m_{ud}\simeq m_s^{\rm phys}$: A first lattice QCD calculation with the
  stochastic Laplacian Heaviside method}},\ }\href
  {https://doi.org/10.1103/PhysRevC.103.014003} {\bibfield  {journal} {\bibinfo
   {journal} {Phys. Rev. C}\ }\textbf {\bibinfo {volume} {103}},\ \bibinfo
  {pages} {014003} (\bibinfo {year} {2021})},\ \Eprint
  {https://arxiv.org/abs/2009.11825} {arXiv:2009.11825 [hep-lat]} \BibitemShut
  {NoStop}%
\bibitem [{\citenamefont {L{\"u}scher}(1986)}]{Luscher:1986pf}%
  \BibitemOpen
  \bibfield  {author} {\bibinfo {author} {\bibfnamefont {M.}~\bibnamefont
  {L{\"u}scher}},\ }\bibfield  {title} {\bibinfo {title} {{Volume Dependence of
  the Energy Spectrum in Massive Quantum Field Theories. 2. Scattering
  States}},\ }\href {https://doi.org/10.1007/BF01211097} {\bibfield  {journal}
  {\bibinfo  {journal} {Commun. Math. Phys.}\ }\textbf {\bibinfo {volume}
  {105}},\ \bibinfo {pages} {153} (\bibinfo {year} {1986})}\BibitemShut
  {NoStop}%
\bibitem [{\citenamefont {Lin}\ \emph {et~al.}(2001)\citenamefont {Lin},
  \citenamefont {Martinelli}, \citenamefont {Sachrajda},\ and\ \citenamefont
  {Testa}}]{Lin:2001ek}%
  \BibitemOpen
  \bibfield  {author} {\bibinfo {author} {\bibfnamefont {C.~J.~D.}\
  \bibnamefont {Lin}}, \bibinfo {author} {\bibfnamefont {G.}~\bibnamefont
  {Martinelli}}, \bibinfo {author} {\bibfnamefont {C.~T.}\ \bibnamefont
  {Sachrajda}},\ and\ \bibinfo {author} {\bibfnamefont {M.}~\bibnamefont
  {Testa}},\ }\bibfield  {title} {\bibinfo {title} {{K --\ensuremath{>} pi pi
  decays in a finite volume}},\ }\href
  {https://doi.org/10.1016/S0550-3213(01)00495-3} {\bibfield  {journal}
  {\bibinfo  {journal} {Nucl. Phys. B}\ }\textbf {\bibinfo {volume} {619}},\
  \bibinfo {pages} {467} (\bibinfo {year} {2001})},\ \Eprint
  {https://arxiv.org/abs/hep-lat/0104006} {arXiv:hep-lat/0104006} \BibitemShut
  {NoStop}%
\bibitem [{\citenamefont {Aoki}\ \emph {et~al.}(2005)\citenamefont {Aoki} \emph
  {et~al.}}]{CP-PACS:2005gzm}%
  \BibitemOpen
  \bibfield  {author} {\bibinfo {author} {\bibfnamefont {S.}~\bibnamefont
  {Aoki}} \emph {et~al.} (\bibinfo {collaboration} {CP-PACS}),\ }\bibfield
  {title} {\bibinfo {title} {{I=2 pion scattering length from two-pion wave
  functions}},\ }\href {https://doi.org/10.1103/PhysRevD.71.094504} {\bibfield
  {journal} {\bibinfo  {journal} {Phys. Rev. D}\ }\textbf {\bibinfo {volume}
  {71}},\ \bibinfo {pages} {094504} (\bibinfo {year} {2005})},\ \Eprint
  {https://arxiv.org/abs/hep-lat/0503025} {arXiv:hep-lat/0503025} \BibitemShut
  {NoStop}%
\bibitem [{\citenamefont {Ishii}\ \emph {et~al.}(2007)\citenamefont {Ishii},
  \citenamefont {Aoki},\ and\ \citenamefont {Hatsuda}}]{Ishii:2006ec}%
  \BibitemOpen
  \bibfield  {author} {\bibinfo {author} {\bibfnamefont {N.}~\bibnamefont
  {Ishii}}, \bibinfo {author} {\bibfnamefont {S.}~\bibnamefont {Aoki}},\ and\
  \bibinfo {author} {\bibfnamefont {T.}~\bibnamefont {Hatsuda}},\ }\bibfield
  {title} {\bibinfo {title} {{The Nuclear Force from Lattice QCD}},\ }\href
  {https://doi.org/10.1103/PhysRevLett.99.022001} {\bibfield  {journal}
  {\bibinfo  {journal} {Phys. Rev. Lett.}\ }\textbf {\bibinfo {volume} {99}},\
  \bibinfo {pages} {022001} (\bibinfo {year} {2007})},\ \Eprint
  {https://arxiv.org/abs/nucl-th/0611096} {arXiv:nucl-th/0611096} \BibitemShut
  {NoStop}%
\bibitem [{\citenamefont {Murano}\ \emph {et~al.}(2011)\citenamefont {Murano},
  \citenamefont {Ishii}, \citenamefont {Aoki},\ and\ \citenamefont
  {Hatsuda}}]{Murano:2011nz}%
  \BibitemOpen
  \bibfield  {author} {\bibinfo {author} {\bibfnamefont {K.}~\bibnamefont
  {Murano}}, \bibinfo {author} {\bibfnamefont {N.}~\bibnamefont {Ishii}},
  \bibinfo {author} {\bibfnamefont {S.}~\bibnamefont {Aoki}},\ and\ \bibinfo
  {author} {\bibfnamefont {T.}~\bibnamefont {Hatsuda}},\ }\bibfield  {title}
  {\bibinfo {title} {{Nucleon-Nucleon Potential and its Non-locality in Lattice
  QCD}},\ }\href {https://doi.org/10.1143/PTP.125.1225} {\bibfield  {journal}
  {\bibinfo  {journal} {Prog. Theor. Phys.}\ }\textbf {\bibinfo {volume}
  {125}},\ \bibinfo {pages} {1225} (\bibinfo {year} {2011})},\ \Eprint
  {https://arxiv.org/abs/1103.0619} {arXiv:1103.0619 [hep-lat]} \BibitemShut
  {NoStop}%
\bibitem [{\citenamefont {Aoki}\ \emph {et~al.}(2011)\citenamefont {Aoki},
  \citenamefont {Ishii}, \citenamefont {Doi}, \citenamefont {Hatsuda},
  \citenamefont {Ikeda}, \citenamefont {Inoue}, \citenamefont {Murano},
  \citenamefont {Nemura},\ and\ \citenamefont {Sasaki}}]{Aoki:2011gt}%
  \BibitemOpen
  \bibfield  {author} {\bibinfo {author} {\bibfnamefont {S.}~\bibnamefont
  {Aoki}}, \bibinfo {author} {\bibfnamefont {N.}~\bibnamefont {Ishii}},
  \bibinfo {author} {\bibfnamefont {T.}~\bibnamefont {Doi}}, \bibinfo {author}
  {\bibfnamefont {T.}~\bibnamefont {Hatsuda}}, \bibinfo {author} {\bibfnamefont
  {Y.}~\bibnamefont {Ikeda}}, \bibinfo {author} {\bibfnamefont
  {T.}~\bibnamefont {Inoue}}, \bibinfo {author} {\bibfnamefont
  {K.}~\bibnamefont {Murano}}, \bibinfo {author} {\bibfnamefont
  {H.}~\bibnamefont {Nemura}},\ and\ \bibinfo {author} {\bibfnamefont
  {K.}~\bibnamefont {Sasaki}} (\bibinfo {collaboration} {HAL QCD}),\ }\bibfield
   {title} {\bibinfo {title} {{Extraction of Hadron Interactions above
  Inelastic Threshold in Lattice QCD}},\ }\href
  {https://doi.org/10.2183/pjab.87.509} {\bibfield  {journal} {\bibinfo
  {journal} {Proc. Japan Acad. B}\ }\textbf {\bibinfo {volume} {87}},\ \bibinfo
  {pages} {509} (\bibinfo {year} {2011})},\ \Eprint
  {https://arxiv.org/abs/1106.2281} {arXiv:1106.2281 [hep-lat]} \BibitemShut
  {NoStop}%
\bibitem [{\citenamefont {Ishii}\ \emph {et~al.}(2012)\citenamefont {Ishii},
  \citenamefont {Aoki}, \citenamefont {Doi}, \citenamefont {Hatsuda},
  \citenamefont {Ikeda}, \citenamefont {Inoue}, \citenamefont {Murano},
  \citenamefont {Nemura},\ and\ \citenamefont {Sasaki}}]{Ishii:2012ssm}%
  \BibitemOpen
  \bibfield  {author} {\bibinfo {author} {\bibfnamefont {N.}~\bibnamefont
  {Ishii}}, \bibinfo {author} {\bibfnamefont {S.}~\bibnamefont {Aoki}},
  \bibinfo {author} {\bibfnamefont {T.}~\bibnamefont {Doi}}, \bibinfo {author}
  {\bibfnamefont {T.}~\bibnamefont {Hatsuda}}, \bibinfo {author} {\bibfnamefont
  {Y.}~\bibnamefont {Ikeda}}, \bibinfo {author} {\bibfnamefont
  {T.}~\bibnamefont {Inoue}}, \bibinfo {author} {\bibfnamefont
  {K.}~\bibnamefont {Murano}}, \bibinfo {author} {\bibfnamefont
  {H.}~\bibnamefont {Nemura}},\ and\ \bibinfo {author} {\bibfnamefont
  {K.}~\bibnamefont {Sasaki}} (\bibinfo {collaboration} {HAL QCD}),\ }\bibfield
   {title} {\bibinfo {title} {{Hadron\textendash{}hadron interactions from
  imaginary-time Nambu\textendash{}Bethe\textendash{}Salpeter wave function on
  the lattice}},\ }\href {https://doi.org/10.1016/j.physletb.2012.04.076}
  {\bibfield  {journal} {\bibinfo  {journal} {Phys. Lett. B}\ }\textbf
  {\bibinfo {volume} {712}},\ \bibinfo {pages} {437} (\bibinfo {year}
  {2012})},\ \Eprint {https://arxiv.org/abs/1203.3642} {arXiv:1203.3642
  [hep-lat]} \BibitemShut {NoStop}%
\bibitem [{\citenamefont {Sasaki}\ \emph {et~al.}(2020)\citenamefont {Sasaki}
  \emph {et~al.}}]{HALQCD:2019wsz}%
  \BibitemOpen
  \bibfield  {author} {\bibinfo {author} {\bibfnamefont {K.}~\bibnamefont
  {Sasaki}} \emph {et~al.} (\bibinfo {collaboration} {HAL QCD}),\ }\bibfield
  {title} {\bibinfo {title} {{$\Lambda\Lambda$ and N$\Xi$ interactions from
  lattice QCD near the physical point}},\ }\href
  {https://doi.org/10.1016/j.nuclphysa.2020.121737} {\bibfield  {journal}
  {\bibinfo  {journal} {Nucl. Phys. A}\ }\textbf {\bibinfo {volume} {998}},\
  \bibinfo {pages} {121737} (\bibinfo {year} {2020})},\ \Eprint
  {https://arxiv.org/abs/1912.08630} {arXiv:1912.08630 [hep-lat]} \BibitemShut
  {NoStop}%
\bibitem [{\citenamefont {Fox}\ \emph {et~al.}(1982)\citenamefont {Fox},
  \citenamefont {Gupta}, \citenamefont {Martin},\ and\ \citenamefont
  {Otto}}]{Fox:1981xz}%
  \BibitemOpen
  \bibfield  {author} {\bibinfo {author} {\bibfnamefont {G.}~\bibnamefont
  {Fox}}, \bibinfo {author} {\bibfnamefont {R.}~\bibnamefont {Gupta}}, \bibinfo
  {author} {\bibfnamefont {O.}~\bibnamefont {Martin}},\ and\ \bibinfo {author}
  {\bibfnamefont {S.}~\bibnamefont {Otto}},\ }\bibfield  {title} {\bibinfo
  {title} {{Monte Carlo Estimates of the Mass Gap of the O(2) and O(3) Spin
  Models in (1+1)-dimensions}},\ }\href
  {https://doi.org/10.1016/0550-3213(82)90384-4} {\bibfield  {journal}
  {\bibinfo  {journal} {Nucl. Phys. B}\ }\textbf {\bibinfo {volume} {205}},\
  \bibinfo {pages} {188} (\bibinfo {year} {1982})}\BibitemShut {NoStop}%
\bibitem [{\citenamefont {Michael}\ and\ \citenamefont
  {Teasdale}(1983)}]{Michael:1982gb}%
  \BibitemOpen
  \bibfield  {author} {\bibinfo {author} {\bibfnamefont {C.}~\bibnamefont
  {Michael}}\ and\ \bibinfo {author} {\bibfnamefont {I.}~\bibnamefont
  {Teasdale}},\ }\bibfield  {title} {\bibinfo {title} {{Extracting Glueball
  Masses From Lattice {QCD}}},\ }\href
  {https://doi.org/10.1016/0550-3213(83)90674-0} {\bibfield  {journal}
  {\bibinfo  {journal} {Nucl. Phys. B}\ }\textbf {\bibinfo {volume} {215}},\
  \bibinfo {pages} {433} (\bibinfo {year} {1983})}\BibitemShut {NoStop}%
\bibitem [{\citenamefont {L{\"u}scher}\ and\ \citenamefont
  {Wolff}(1990)}]{Luscher:1990ck}%
  \BibitemOpen
  \bibfield  {author} {\bibinfo {author} {\bibfnamefont {M.}~\bibnamefont
  {L{\"u}scher}}\ and\ \bibinfo {author} {\bibfnamefont {U.}~\bibnamefont
  {Wolff}},\ }\bibfield  {title} {\bibinfo {title} {{How to Calculate the
  Elastic Scattering Matrix in Two-dimensional Quantum Field Theories by
  Numerical Simulation}},\ }\href
  {https://doi.org/10.1016/0550-3213(90)90540-T} {\bibfield  {journal}
  {\bibinfo  {journal} {Nucl. Phys. B}\ }\textbf {\bibinfo {volume} {339}},\
  \bibinfo {pages} {222} (\bibinfo {year} {1990})}\BibitemShut {NoStop}%
\bibitem [{\citenamefont {Brodsky}\ \emph {et~al.}(1983)\citenamefont
  {Brodsky}, \citenamefont {Ji},\ and\ \citenamefont
  {Lepage}}]{Brodsky:1983vf}%
  \BibitemOpen
  \bibfield  {author} {\bibinfo {author} {\bibfnamefont {S.~J.}\ \bibnamefont
  {Brodsky}}, \bibinfo {author} {\bibfnamefont {C.-R.}\ \bibnamefont {Ji}},\
  and\ \bibinfo {author} {\bibfnamefont {G.~P.}\ \bibnamefont {Lepage}},\
  }\bibfield  {title} {\bibinfo {title} {{Quantum Chromodynamic Predictions for
  the Deuteron Form-Factor}},\ }\href
  {https://doi.org/10.1103/PhysRevLett.51.83} {\bibfield  {journal} {\bibinfo
  {journal} {Phys. Rev. Lett.}\ }\textbf {\bibinfo {volume} {51}},\ \bibinfo
  {pages} {83} (\bibinfo {year} {1983})}\BibitemShut {NoStop}%
\bibitem [{\citenamefont {Miller}(1984)}]{Miller:1984twp}%
  \BibitemOpen
  \bibfield  {author} {\bibinfo {author} {\bibfnamefont {G.~A.}\ \bibnamefont
  {Miller}},\ }\bibfield  {title} {\bibinfo {title} {{Six Quark Cluster
  Components of Nuclear Wave Functions and the Pion Nucleus Double Charge
  Exchange Reaction}},\ }\href {https://doi.org/10.1103/PhysRevLett.53.2008}
  {\bibfield  {journal} {\bibinfo  {journal} {Phys. Rev. Lett.}\ }\textbf
  {\bibinfo {volume} {53}},\ \bibinfo {pages} {2008} (\bibinfo {year}
  {1984})}\BibitemShut {NoStop}%
\bibitem [{\citenamefont {Bashkanov}\ \emph {et~al.}(2013)\citenamefont
  {Bashkanov}, \citenamefont {Brodsky},\ and\ \citenamefont
  {Clement}}]{Bashkanov:2013cla}%
  \BibitemOpen
  \bibfield  {author} {\bibinfo {author} {\bibfnamefont {M.}~\bibnamefont
  {Bashkanov}}, \bibinfo {author} {\bibfnamefont {S.~J.}\ \bibnamefont
  {Brodsky}},\ and\ \bibinfo {author} {\bibfnamefont {H.}~\bibnamefont
  {Clement}},\ }\bibfield  {title} {\bibinfo {title} {{Novel Six-Quark
  Hidden-Color Dibaryon States in QCD}},\ }\href
  {https://doi.org/10.1016/j.physletb.2013.10.059} {\bibfield  {journal}
  {\bibinfo  {journal} {Phys. Lett. B}\ }\textbf {\bibinfo {volume} {727}},\
  \bibinfo {pages} {438} (\bibinfo {year} {2013})},\ \Eprint
  {https://arxiv.org/abs/1308.6404} {arXiv:1308.6404 [hep-ph]} \BibitemShut
  {NoStop}%
\bibitem [{\citenamefont {Miller}(2014)}]{Miller:2013hla}%
  \BibitemOpen
  \bibfield  {author} {\bibinfo {author} {\bibfnamefont {G.~A.}\ \bibnamefont
  {Miller}},\ }\bibfield  {title} {\bibinfo {title} {{Pionic and Hidden-Color,
  Six-Quark Contributions to the Deuteron b1 Structure Function}},\ }\href
  {https://doi.org/10.1103/PhysRevC.89.045203} {\bibfield  {journal} {\bibinfo
  {journal} {Phys. Rev. C}\ }\textbf {\bibinfo {volume} {89}},\ \bibinfo
  {pages} {045203} (\bibinfo {year} {2014})},\ \Eprint
  {https://arxiv.org/abs/1311.4561} {arXiv:1311.4561 [nucl-th]} \BibitemShut
  {NoStop}%
\bibitem [{\citenamefont {Schiel}(2015)}]{Schiel:2015kwa}%
  \BibitemOpen
  \bibfield  {author} {\bibinfo {author} {\bibfnamefont {R.~W.}\ \bibnamefont
  {Schiel}},\ }\bibfield  {title} {\bibinfo {title} {{Expanding the
  Interpolator Basis in the Variational Method to Explicitly Account for
  Backward Running States}},\ }\href
  {https://doi.org/10.1103/PhysRevD.92.034512} {\bibfield  {journal} {\bibinfo
  {journal} {Phys. Rev. D}\ }\textbf {\bibinfo {volume} {92}},\ \bibinfo
  {pages} {034512} (\bibinfo {year} {2015})},\ \Eprint
  {https://arxiv.org/abs/1503.02588} {arXiv:1503.02588 [hep-lat]} \BibitemShut
  {NoStop}%
\bibitem [{\citenamefont {Beane}\ \emph
  {et~al.}(2009{\natexlab{b}})\citenamefont {Beane}, \citenamefont {Detmold},
  \citenamefont {Luu}, \citenamefont {Orginos}, \citenamefont {Parre\~{n}o},
  \citenamefont {Savage}, \citenamefont {Torok},\ and\ \citenamefont
  {Walker-Loud}}]{Beane:2009kya}%
  \BibitemOpen
  \bibfield  {author} {\bibinfo {author} {\bibfnamefont {S.~R.}\ \bibnamefont
  {Beane}}, \bibinfo {author} {\bibfnamefont {W.}~\bibnamefont {Detmold}},
  \bibinfo {author} {\bibfnamefont {T.~C.}\ \bibnamefont {Luu}}, \bibinfo
  {author} {\bibfnamefont {K.}~\bibnamefont {Orginos}}, \bibinfo {author}
  {\bibfnamefont {A.}~\bibnamefont {Parre\~{n}o}}, \bibinfo {author}
  {\bibfnamefont {M.~J.}\ \bibnamefont {Savage}}, \bibinfo {author}
  {\bibfnamefont {A.}~\bibnamefont {Torok}},\ and\ \bibinfo {author}
  {\bibfnamefont {A.}~\bibnamefont {Walker-Loud}} (\bibinfo {collaboration}
  {NPLQCD}),\ }\bibfield  {title} {\bibinfo {title} {{High Statistics Analysis
  using Anisotropic Clover Lattices: (I) Single Hadron Correlation
  Functions}},\ }\href {https://doi.org/10.1103/PhysRevD.79.114502} {\bibfield
  {journal} {\bibinfo  {journal} {Phys. Rev. D}\ }\textbf {\bibinfo {volume}
  {79}},\ \bibinfo {pages} {114502} (\bibinfo {year} {2009}{\natexlab{b}})},\
  \Eprint {https://arxiv.org/abs/0903.2990} {arXiv:0903.2990 [hep-lat]}
  \BibitemShut {NoStop}%
\bibitem [{\citenamefont {Endres}\ \emph {et~al.}(2011)\citenamefont {Endres},
  \citenamefont {Kaplan}, \citenamefont {Lee},\ and\ \citenamefont
  {Nicholson}}]{Endres:2011jm}%
  \BibitemOpen
  \bibfield  {author} {\bibinfo {author} {\bibfnamefont {M.~G.}\ \bibnamefont
  {Endres}}, \bibinfo {author} {\bibfnamefont {D.~B.}\ \bibnamefont {Kaplan}},
  \bibinfo {author} {\bibfnamefont {J.-W.}\ \bibnamefont {Lee}},\ and\ \bibinfo
  {author} {\bibfnamefont {A.~N.}\ \bibnamefont {Nicholson}},\ }\bibfield
  {title} {\bibinfo {title} {{Noise, sign problems, and statistics}},\ }\href
  {https://doi.org/10.1103/PhysRevLett.107.201601} {\bibfield  {journal}
  {\bibinfo  {journal} {Phys. Rev. Lett.}\ }\textbf {\bibinfo {volume} {107}},\
  \bibinfo {pages} {201601} (\bibinfo {year} {2011})},\ \Eprint
  {https://arxiv.org/abs/1106.0073} {arXiv:1106.0073 [hep-lat]} \BibitemShut
  {NoStop}%
\bibitem [{\citenamefont {Grabowska}\ \emph {et~al.}(2013)\citenamefont
  {Grabowska}, \citenamefont {Kaplan},\ and\ \citenamefont
  {Nicholson}}]{Grabowska:2012ik}%
  \BibitemOpen
  \bibfield  {author} {\bibinfo {author} {\bibfnamefont {D.}~\bibnamefont
  {Grabowska}}, \bibinfo {author} {\bibfnamefont {D.~B.}\ \bibnamefont
  {Kaplan}},\ and\ \bibinfo {author} {\bibfnamefont {A.~N.}\ \bibnamefont
  {Nicholson}},\ }\bibfield  {title} {\bibinfo {title} {{Sign problems, noise,
  and chiral symmetry breaking in a QCD-like theory}},\ }\href
  {https://doi.org/10.1103/PhysRevD.87.014504} {\bibfield  {journal} {\bibinfo
  {journal} {Phys. Rev. D}\ }\textbf {\bibinfo {volume} {87}},\ \bibinfo
  {pages} {014504} (\bibinfo {year} {2013})},\ \Eprint
  {https://arxiv.org/abs/1208.5760} {arXiv:1208.5760 [hep-lat]} \BibitemShut
  {NoStop}%
\bibitem [{\citenamefont {Beane}\ \emph {et~al.}(2015)\citenamefont {Beane},
  \citenamefont {Detmold}, \citenamefont {Orginos},\ and\ \citenamefont
  {Savage}}]{Beane:2014oea}%
  \BibitemOpen
  \bibfield  {author} {\bibinfo {author} {\bibfnamefont {S.~R.}\ \bibnamefont
  {Beane}}, \bibinfo {author} {\bibfnamefont {W.}~\bibnamefont {Detmold}},
  \bibinfo {author} {\bibfnamefont {K.}~\bibnamefont {Orginos}},\ and\ \bibinfo
  {author} {\bibfnamefont {M.~J.}\ \bibnamefont {Savage}},\ }\bibfield  {title}
  {\bibinfo {title} {{Uncertainty Quantification in Lattice QCD Calculations
  for Nuclear Physics}},\ }\href
  {https://doi.org/10.1088/0954-3899/42/3/034022} {\bibfield  {journal}
  {\bibinfo  {journal} {J. Phys. G}\ }\textbf {\bibinfo {volume} {42}},\
  \bibinfo {pages} {034022} (\bibinfo {year} {2015})},\ \Eprint
  {https://arxiv.org/abs/1410.2937} {arXiv:1410.2937 [nucl-th]} \BibitemShut
  {NoStop}%
\bibitem [{\citenamefont {Wagman}\ and\ \citenamefont
  {Savage}(2017{\natexlab{a}})}]{Wagman:2016bam}%
  \BibitemOpen
  \bibfield  {author} {\bibinfo {author} {\bibfnamefont {M.~L.}\ \bibnamefont
  {Wagman}}\ and\ \bibinfo {author} {\bibfnamefont {M.~J.}\ \bibnamefont
  {Savage}},\ }\bibfield  {title} {\bibinfo {title} {{Statistics of baryon
  correlation functions in lattice QCD}},\ }\href
  {https://doi.org/10.1103/PhysRevD.96.114508} {\bibfield  {journal} {\bibinfo
  {journal} {Phys. Rev. D}\ }\textbf {\bibinfo {volume} {96}},\ \bibinfo
  {pages} {114508} (\bibinfo {year} {2017}{\natexlab{a}})},\ \Eprint
  {https://arxiv.org/abs/1611.07643} {arXiv:1611.07643 [hep-lat]} \BibitemShut
  {NoStop}%
\bibitem [{\citenamefont {Wagman}\ and\ \citenamefont
  {Savage}(2017{\natexlab{b}})}]{Wagman:2017xfh}%
  \BibitemOpen
  \bibfield  {author} {\bibinfo {author} {\bibfnamefont {M.~L.}\ \bibnamefont
  {Wagman}}\ and\ \bibinfo {author} {\bibfnamefont {M.~J.}\ \bibnamefont
  {Savage}},\ }\href@noop {} {\bibinfo {title} {{Taming the Signal-to-Noise
  Problem in Lattice QCD by Phase Reweighting}}} (\bibinfo {year}
  {2017}{\natexlab{b}}),\ \Eprint {https://arxiv.org/abs/1704.07356}
  {arXiv:1704.07356 [hep-lat]} \BibitemShut {NoStop}%
\bibitem [{\citenamefont {Detmold}\ and\ \citenamefont
  {Endres}(2014)}]{Detmold:2014hla}%
  \BibitemOpen
  \bibfield  {author} {\bibinfo {author} {\bibfnamefont {W.}~\bibnamefont
  {Detmold}}\ and\ \bibinfo {author} {\bibfnamefont {M.~G.}\ \bibnamefont
  {Endres}},\ }\bibfield  {title} {\bibinfo {title} {{Signal/noise enhancement
  strategies for stochastically estimated correlation functions}},\ }\href
  {https://doi.org/10.1103/PhysRevD.90.034503} {\bibfield  {journal} {\bibinfo
  {journal} {Phys. Rev. D}\ }\textbf {\bibinfo {volume} {90}},\ \bibinfo
  {pages} {034503} (\bibinfo {year} {2014})},\ \Eprint
  {https://arxiv.org/abs/1404.6816} {arXiv:1404.6816 [hep-lat]} \BibitemShut
  {NoStop}%
\bibitem [{\citenamefont {Hwang}(2004)}]{Hwang:2024}%
  \BibitemOpen
  \bibfield  {author} {\bibinfo {author} {\bibfnamefont {S.-G.}\ \bibnamefont
  {Hwang}},\ }\bibfield  {title} {\bibinfo {title} {Cauchy's interlace theorem
  for eigenvalues of hermitian matrices},\ }\href
  {http://www.jstor.org/stable/4145217} {\bibfield  {journal} {\bibinfo
  {journal} {The American Mathematical Monthly}\ }\textbf {\bibinfo {volume}
  {111}},\ \bibinfo {pages} {157} (\bibinfo {year} {2004})}\BibitemShut
  {NoStop}%
\bibitem [{\citenamefont {Horn}\ and\ \citenamefont
  {Johnson}(1985)}]{Horn:1985}%
  \BibitemOpen
  \bibfield  {author} {\bibinfo {author} {\bibfnamefont {R.~A.}\ \bibnamefont
  {Horn}}\ and\ \bibinfo {author} {\bibfnamefont {C.~R.}\ \bibnamefont
  {Johnson}},\ }\href
  {https://doi.org/https://doi.org/10.1017/CBO9780511810817} {\emph {\bibinfo
  {title} {Matrix analysis}}},\ \bibinfo {edition} {1st}\ ed.\ (\bibinfo
  {publisher} {Cambridge University Press},\ \bibinfo {year}
  {1985})\BibitemShut {NoStop}%
\bibitem [{\citenamefont {Golub}\ and\ \citenamefont
  {Van~Loan}(2013)}]{doi:10.1137/1.9781421407944}%
  \BibitemOpen
  \bibfield  {author} {\bibinfo {author} {\bibfnamefont {G.~H.}\ \bibnamefont
  {Golub}}\ and\ \bibinfo {author} {\bibfnamefont {C.~F.}\ \bibnamefont
  {Van~Loan}},\ }\href {https://doi.org/10.1137/1.9781421407944} {\emph
  {\bibinfo {title} {Matrix Computations}}}\ (\bibinfo  {publisher} {Johns
  Hopkins University Press},\ \bibinfo {address} {Philadelphia, PA},\ \bibinfo
  {year} {2013})\BibitemShut {NoStop}%
\bibitem [{\citenamefont {Fleming}(2023)}]{Fleming:2023zml}%
  \BibitemOpen
  \bibfield  {author} {\bibinfo {author} {\bibfnamefont {G.~T.}\ \bibnamefont
  {Fleming}},\ }\bibfield  {title} {\bibinfo {title} {{Beyond Generalized
  Eigenvalues in Lattice Quantum Field Theory}},\ }in\ \href@noop {} {\emph
  {\bibinfo {booktitle} {{40th International Symposium on Lattice Field
  Theory}}}}\ (\bibinfo {year} {2023})\ \Eprint
  {https://arxiv.org/abs/2309.05111} {arXiv:2309.05111 [hep-lat]} \BibitemShut
  {NoStop}%
\bibitem [{\citenamefont {Kogut}\ and\ \citenamefont
  {Susskind}(1975)}]{Kogut:1974ag}%
  \BibitemOpen
  \bibfield  {author} {\bibinfo {author} {\bibfnamefont {J.~B.}\ \bibnamefont
  {Kogut}}\ and\ \bibinfo {author} {\bibfnamefont {L.}~\bibnamefont
  {Susskind}},\ }\bibfield  {title} {\bibinfo {title} {{Hamiltonian Formulation
  of Wilson's Lattice Gauge Theories}},\ }\href
  {https://doi.org/10.1103/PhysRevD.11.395} {\bibfield  {journal} {\bibinfo
  {journal} {Phys. Rev. D}\ }\textbf {\bibinfo {volume} {11}},\ \bibinfo
  {pages} {395} (\bibinfo {year} {1975})}\BibitemShut {NoStop}%
\bibitem [{\citenamefont {Bulava}\ \emph {et~al.}(2010)\citenamefont {Bulava},
  \citenamefont {Edwards}, \citenamefont {Engelson}, \citenamefont {Jo\'o},
  \citenamefont {Lin}, \citenamefont {Morningstar}, \citenamefont {Richards},\
  and\ \citenamefont {Wallace}}]{Bulava:2010yg}%
  \BibitemOpen
  \bibfield  {author} {\bibinfo {author} {\bibfnamefont {J.}~\bibnamefont
  {Bulava}}, \bibinfo {author} {\bibfnamefont {R.~G.}\ \bibnamefont {Edwards}},
  \bibinfo {author} {\bibfnamefont {E.}~\bibnamefont {Engelson}}, \bibinfo
  {author} {\bibfnamefont {B.}~\bibnamefont {Jo\'o}}, \bibinfo {author}
  {\bibfnamefont {H.-W.}\ \bibnamefont {Lin}}, \bibinfo {author} {\bibfnamefont
  {C.}~\bibnamefont {Morningstar}}, \bibinfo {author} {\bibfnamefont {D.~G.}\
  \bibnamefont {Richards}},\ and\ \bibinfo {author} {\bibfnamefont {S.~J.}\
  \bibnamefont {Wallace}},\ }\bibfield  {title} {\bibinfo {title} {{Nucleon,
  $\Delta$ and $\Omega$ excited states in $N_f=2+1$ lattice QCD}},\ }\href
  {https://doi.org/10.1103/PhysRevD.82.014507} {\bibfield  {journal} {\bibinfo
  {journal} {Phys. Rev. D}\ }\textbf {\bibinfo {volume} {82}},\ \bibinfo
  {pages} {014507} (\bibinfo {year} {2010})},\ \Eprint
  {https://arxiv.org/abs/1004.5072} {arXiv:1004.5072 [hep-lat]} \BibitemShut
  {NoStop}%
\bibitem [{\citenamefont {Bulava}\ \emph {et~al.}(2016)\citenamefont {Bulava},
  \citenamefont {Fahy}, \citenamefont {H\"orz}, \citenamefont {Juge},
  \citenamefont {Morningstar},\ and\ \citenamefont {Wong}}]{Bulava:2016mks}%
  \BibitemOpen
  \bibfield  {author} {\bibinfo {author} {\bibfnamefont {J.}~\bibnamefont
  {Bulava}}, \bibinfo {author} {\bibfnamefont {B.}~\bibnamefont {Fahy}},
  \bibinfo {author} {\bibfnamefont {B.}~\bibnamefont {H\"orz}}, \bibinfo
  {author} {\bibfnamefont {K.~J.}\ \bibnamefont {Juge}}, \bibinfo {author}
  {\bibfnamefont {C.}~\bibnamefont {Morningstar}},\ and\ \bibinfo {author}
  {\bibfnamefont {C.~H.}\ \bibnamefont {Wong}},\ }\bibfield  {title} {\bibinfo
  {title} {{$I=1$ and $I=2$ $\pi-\pi$ scattering phase shifts from
  $N_{\mathrm{f}} = 2+1$ lattice QCD}},\ }\href
  {https://doi.org/10.1016/j.nuclphysb.2016.07.024} {\bibfield  {journal}
  {\bibinfo  {journal} {Nucl. Phys. B}\ }\textbf {\bibinfo {volume} {910}},\
  \bibinfo {pages} {842} (\bibinfo {year} {2016})},\ \Eprint
  {https://arxiv.org/abs/1604.05593} {arXiv:1604.05593 [hep-lat]} \BibitemShut
  {NoStop}%
\bibitem [{\citenamefont {Blossier}\ \emph {et~al.}(2009)\citenamefont
  {Blossier}, \citenamefont {Della~Morte}, \citenamefont {von Hippel},
  \citenamefont {Mendes},\ and\ \citenamefont {Sommer}}]{Blossier:2009kd}%
  \BibitemOpen
  \bibfield  {author} {\bibinfo {author} {\bibfnamefont {B.}~\bibnamefont
  {Blossier}}, \bibinfo {author} {\bibfnamefont {M.}~\bibnamefont
  {Della~Morte}}, \bibinfo {author} {\bibfnamefont {G.}~\bibnamefont {von
  Hippel}}, \bibinfo {author} {\bibfnamefont {T.}~\bibnamefont {Mendes}},\ and\
  \bibinfo {author} {\bibfnamefont {R.}~\bibnamefont {Sommer}},\ }\bibfield
  {title} {\bibinfo {title} {{On the generalized eigenvalue method for energies
  and matrix elements in lattice field theory}},\ }\href
  {https://doi.org/10.1088/1126-6708/2009/04/094} {\bibfield  {journal}
  {\bibinfo  {journal} {JHEP}\ }\textbf {\bibinfo {volume} {04}},\ \bibinfo
  {pages} {094}},\ \Eprint {https://arxiv.org/abs/0902.1265} {arXiv:0902.1265
  [hep-lat]} \BibitemShut {NoStop}%
\bibitem [{\citenamefont {Basak}\ \emph {et~al.}(2005)\citenamefont {Basak},
  \citenamefont {Edwards}, \citenamefont {Fleming}, \citenamefont {Heller},
  \citenamefont {Morningstar}, \citenamefont {Richards}, \citenamefont {Sato},\
  and\ \citenamefont {Wallace}}]{Basak:2005ir}%
  \BibitemOpen
  \bibfield  {author} {\bibinfo {author} {\bibfnamefont {S.}~\bibnamefont
  {Basak}}, \bibinfo {author} {\bibfnamefont {R.}~\bibnamefont {Edwards}},
  \bibinfo {author} {\bibfnamefont {G.~T.}\ \bibnamefont {Fleming}}, \bibinfo
  {author} {\bibfnamefont {U.~M.}\ \bibnamefont {Heller}}, \bibinfo {author}
  {\bibfnamefont {C.}~\bibnamefont {Morningstar}}, \bibinfo {author}
  {\bibfnamefont {D.}~\bibnamefont {Richards}}, \bibinfo {author}
  {\bibfnamefont {I.}~\bibnamefont {Sato}},\ and\ \bibinfo {author}
  {\bibfnamefont {S.~J.}\ \bibnamefont {Wallace}} (\bibinfo {collaboration}
  {Lattice Hadron Physics (LHPC)}),\ }\bibfield  {title} {\bibinfo {title}
  {{Clebsch-Gordan construction of lattice interpolating fields for excited
  baryons}},\ }\href {https://doi.org/10.1103/PhysRevD.72.074501} {\bibfield
  {journal} {\bibinfo  {journal} {Phys. Rev. D}\ }\textbf {\bibinfo {volume}
  {72}},\ \bibinfo {pages} {074501} (\bibinfo {year} {2005})},\ \Eprint
  {https://arxiv.org/abs/hep-lat/0508018} {arXiv:hep-lat/0508018} \BibitemShut
  {NoStop}%
\bibitem [{\citenamefont {Sasaki}\ \emph {et~al.}(2002)\citenamefont {Sasaki},
  \citenamefont {Blum},\ and\ \citenamefont {Ohta}}]{Sasaki:2001nf}%
  \BibitemOpen
  \bibfield  {author} {\bibinfo {author} {\bibfnamefont {S.}~\bibnamefont
  {Sasaki}}, \bibinfo {author} {\bibfnamefont {T.}~\bibnamefont {Blum}},\ and\
  \bibinfo {author} {\bibfnamefont {S.}~\bibnamefont {Ohta}},\ }\bibfield
  {title} {\bibinfo {title} {{A Lattice study of the nucleon excited states
  with domain wall fermions}},\ }\href
  {https://doi.org/10.1103/PhysRevD.65.074503} {\bibfield  {journal} {\bibinfo
  {journal} {Phys. Rev. D}\ }\textbf {\bibinfo {volume} {65}},\ \bibinfo
  {pages} {074503} (\bibinfo {year} {2002})},\ \Eprint
  {https://arxiv.org/abs/hep-lat/0102010} {arXiv:hep-lat/0102010} \BibitemShut
  {NoStop}%
\bibitem [{\citenamefont {Melnitchouk}\ \emph {et~al.}(2003)\citenamefont
  {Melnitchouk}, \citenamefont {Bilson-Thompson}, \citenamefont {Bonnet},
  \citenamefont {Hedditch}, \citenamefont {Lee}, \citenamefont {Leinweber},
  \citenamefont {Williams}, \citenamefont {Zanotti},\ and\ \citenamefont
  {Zhang}}]{Melnitchouk:2002eg}%
  \BibitemOpen
  \bibfield  {author} {\bibinfo {author} {\bibfnamefont {W.}~\bibnamefont
  {Melnitchouk}}, \bibinfo {author} {\bibfnamefont {S.~O.}\ \bibnamefont
  {Bilson-Thompson}}, \bibinfo {author} {\bibfnamefont {F.~D.~R.}\ \bibnamefont
  {Bonnet}}, \bibinfo {author} {\bibfnamefont {J.~N.}\ \bibnamefont
  {Hedditch}}, \bibinfo {author} {\bibfnamefont {F.~X.}\ \bibnamefont {Lee}},
  \bibinfo {author} {\bibfnamefont {D.~B.}\ \bibnamefont {Leinweber}}, \bibinfo
  {author} {\bibfnamefont {A.~G.}\ \bibnamefont {Williams}}, \bibinfo {author}
  {\bibfnamefont {J.~M.}\ \bibnamefont {Zanotti}},\ and\ \bibinfo {author}
  {\bibfnamefont {J.~B.}\ \bibnamefont {Zhang}},\ }\bibfield  {title} {\bibinfo
  {title} {{Excited baryons in lattice QCD}},\ }\href
  {https://doi.org/10.1103/PhysRevD.67.114506} {\bibfield  {journal} {\bibinfo
  {journal} {Phys. Rev. D}\ }\textbf {\bibinfo {volume} {67}},\ \bibinfo
  {pages} {114506} (\bibinfo {year} {2003})},\ \Eprint
  {https://arxiv.org/abs/hep-lat/0202022} {arXiv:hep-lat/0202022} \BibitemShut
  {NoStop}%
\bibitem [{\citenamefont {Br{\"o}mmel}\ \emph {et~al.}(2004)\citenamefont
  {Br{\"o}mmel}, \citenamefont {Crompton}, \citenamefont {Gattringer},
  \citenamefont {Glozman}, \citenamefont {Lang}, \citenamefont {Schaefer},\
  and\ \citenamefont {Sch{\"a}fer}}]{Brommel:2003jm}%
  \BibitemOpen
  \bibfield  {author} {\bibinfo {author} {\bibfnamefont {D.}~\bibnamefont
  {Br{\"o}mmel}}, \bibinfo {author} {\bibfnamefont {P.}~\bibnamefont
  {Crompton}}, \bibinfo {author} {\bibfnamefont {C.}~\bibnamefont
  {Gattringer}}, \bibinfo {author} {\bibfnamefont {L.~Y.}\ \bibnamefont
  {Glozman}}, \bibinfo {author} {\bibfnamefont {C.~B.}\ \bibnamefont {Lang}},
  \bibinfo {author} {\bibfnamefont {S.}~\bibnamefont {Schaefer}},\ and\
  \bibinfo {author} {\bibfnamefont {A.}~\bibnamefont {Sch{\"a}fer}} (\bibinfo
  {collaboration} {Bern-Graz-Regensburg}),\ }\bibfield  {title} {\bibinfo
  {title} {{Excited nucleons with chirally improved fermions}},\ }\href
  {https://doi.org/10.1103/PhysRevD.69.094513} {\bibfield  {journal} {\bibinfo
  {journal} {Phys. Rev. D}\ }\textbf {\bibinfo {volume} {69}},\ \bibinfo
  {pages} {094513} (\bibinfo {year} {2004})},\ \Eprint
  {https://arxiv.org/abs/hep-ph/0307073} {arXiv:hep-ph/0307073} \BibitemShut
  {NoStop}%
\bibitem [{\citenamefont {Basak}\ \emph {et~al.}(2007)\citenamefont {Basak},
  \citenamefont {Edwards}, \citenamefont {Fleming}, \citenamefont {Juge},
  \citenamefont {Lichtl}, \citenamefont {Morningstar}, \citenamefont
  {Richards}, \citenamefont {Sato},\ and\ \citenamefont
  {Wallace}}]{Basak:2007kj}%
  \BibitemOpen
  \bibfield  {author} {\bibinfo {author} {\bibfnamefont {S.}~\bibnamefont
  {Basak}}, \bibinfo {author} {\bibfnamefont {R.~G.}\ \bibnamefont {Edwards}},
  \bibinfo {author} {\bibfnamefont {G.~T.}\ \bibnamefont {Fleming}}, \bibinfo
  {author} {\bibfnamefont {K.~J.}\ \bibnamefont {Juge}}, \bibinfo {author}
  {\bibfnamefont {A.}~\bibnamefont {Lichtl}}, \bibinfo {author} {\bibfnamefont
  {C.}~\bibnamefont {Morningstar}}, \bibinfo {author} {\bibfnamefont {D.~G.}\
  \bibnamefont {Richards}}, \bibinfo {author} {\bibfnamefont {I.}~\bibnamefont
  {Sato}},\ and\ \bibinfo {author} {\bibfnamefont {S.~J.}\ \bibnamefont
  {Wallace}},\ }\bibfield  {title} {\bibinfo {title} {{Lattice QCD
  determination of patterns of excited baryon states}},\ }\href
  {https://doi.org/10.1103/PhysRevD.76.074504} {\bibfield  {journal} {\bibinfo
  {journal} {Phys. Rev. D}\ }\textbf {\bibinfo {volume} {76}},\ \bibinfo
  {pages} {074504} (\bibinfo {year} {2007})},\ \Eprint
  {https://arxiv.org/abs/0709.0008} {arXiv:0709.0008 [hep-lat]} \BibitemShut
  {NoStop}%
\bibitem [{\citenamefont {Detmold}\ \emph {et~al.}(2021)\citenamefont
  {Detmold}, \citenamefont {Murphy}, \citenamefont {Pochinsky}, \citenamefont
  {Savage}, \citenamefont {Shanahan},\ and\ \citenamefont
  {Wagman}}]{Detmold:2019fbk}%
  \BibitemOpen
  \bibfield  {author} {\bibinfo {author} {\bibfnamefont {W.}~\bibnamefont
  {Detmold}}, \bibinfo {author} {\bibfnamefont {D.~J.}\ \bibnamefont {Murphy}},
  \bibinfo {author} {\bibfnamefont {A.~V.}\ \bibnamefont {Pochinsky}}, \bibinfo
  {author} {\bibfnamefont {M.~J.}\ \bibnamefont {Savage}}, \bibinfo {author}
  {\bibfnamefont {P.~E.}\ \bibnamefont {Shanahan}},\ and\ \bibinfo {author}
  {\bibfnamefont {M.~L.}\ \bibnamefont {Wagman}},\ }\bibfield  {title}
  {\bibinfo {title} {{Sparsening algorithm for multihadron lattice QCD
  correlation functions}},\ }\href
  {https://doi.org/10.1103/PhysRevD.104.034502} {\bibfield  {journal} {\bibinfo
   {journal} {Phys. Rev. D}\ }\textbf {\bibinfo {volume} {104}},\ \bibinfo
  {pages} {034502} (\bibinfo {year} {2021})},\ \Eprint
  {https://arxiv.org/abs/1908.07050} {arXiv:1908.07050 [hep-lat]} \BibitemShut
  {NoStop}%
\bibitem [{\citenamefont {Li}\ \emph {et~al.}(2021)\citenamefont {Li},
  \citenamefont {Xia}, \citenamefont {Feng}, \citenamefont {Jin},\ and\
  \citenamefont {Liu}}]{Li:2020hbj}%
  \BibitemOpen
  \bibfield  {author} {\bibinfo {author} {\bibfnamefont {Y.}~\bibnamefont
  {Li}}, \bibinfo {author} {\bibfnamefont {S.-C.}\ \bibnamefont {Xia}},
  \bibinfo {author} {\bibfnamefont {X.}~\bibnamefont {Feng}}, \bibinfo {author}
  {\bibfnamefont {L.-C.}\ \bibnamefont {Jin}},\ and\ \bibinfo {author}
  {\bibfnamefont {C.}~\bibnamefont {Liu}},\ }\bibfield  {title} {\bibinfo
  {title} {{Field sparsening for the construction of the correlation functions
  in lattice QCD}},\ }\href {https://doi.org/10.1103/PhysRevD.103.014514}
  {\bibfield  {journal} {\bibinfo  {journal} {Phys. Rev. D}\ }\textbf {\bibinfo
  {volume} {103}},\ \bibinfo {pages} {014514} (\bibinfo {year} {2021})},\
  \Eprint {https://arxiv.org/abs/2009.01029} {arXiv:2009.01029 [hep-lat]}
  \BibitemShut {NoStop}%
\bibitem [{\citenamefont {Luu}\ and\ \citenamefont
  {Savage}(2011)}]{Luu:2011ep}%
  \BibitemOpen
  \bibfield  {author} {\bibinfo {author} {\bibfnamefont {T.}~\bibnamefont
  {Luu}}\ and\ \bibinfo {author} {\bibfnamefont {M.~J.}\ \bibnamefont
  {Savage}},\ }\bibfield  {title} {\bibinfo {title} {{Extracting Scattering
  Phase-Shifts in Higher Partial-Waves from Lattice QCD Calculations}},\ }\href
  {https://doi.org/10.1103/PhysRevD.83.114508} {\bibfield  {journal} {\bibinfo
  {journal} {Phys. Rev. D}\ }\textbf {\bibinfo {volume} {83}},\ \bibinfo
  {pages} {114508} (\bibinfo {year} {2011})},\ \Eprint
  {https://arxiv.org/abs/1101.3347} {arXiv:1101.3347 [hep-lat]} \BibitemShut
  {NoStop}%
\bibitem [{\citenamefont {Detmold}\ \emph {et~al.}(2024)\citenamefont
  {Detmold}, \citenamefont {Jay}, \citenamefont {Kanwar}, \citenamefont
  {Shanahan},\ and\ \citenamefont {Wagman}}]{Detmold:2024ifm}%
  \BibitemOpen
  \bibfield  {author} {\bibinfo {author} {\bibfnamefont {W.}~\bibnamefont
  {Detmold}}, \bibinfo {author} {\bibfnamefont {W.~I.}\ \bibnamefont {Jay}},
  \bibinfo {author} {\bibfnamefont {G.}~\bibnamefont {Kanwar}}, \bibinfo
  {author} {\bibfnamefont {P.~E.}\ \bibnamefont {Shanahan}},\ and\ \bibinfo
  {author} {\bibfnamefont {M.~L.}\ \bibnamefont {Wagman}},\ }\bibfield  {title}
  {\bibinfo {title} {{Multiparticle interpolating operators in quantum field
  theories with cubic symmetry}},\ }\href
  {https://doi.org/10.1103/PhysRevD.109.094516} {\bibfield  {journal} {\bibinfo
   {journal} {Phys. Rev. D}\ }\textbf {\bibinfo {volume} {109}},\ \bibinfo
  {pages} {094516} (\bibinfo {year} {2024})},\ \Eprint
  {https://arxiv.org/abs/2403.00672} {arXiv:2403.00672 [hep-lat]} \BibitemShut
  {NoStop}%
\bibitem [{\citenamefont {L{\"u}scher}(1991)}]{Luscher:1990ux}%
  \BibitemOpen
  \bibfield  {author} {\bibinfo {author} {\bibfnamefont {M.}~\bibnamefont
  {L{\"u}scher}},\ }\bibfield  {title} {\bibinfo {title} {{Two particle states
  on a torus and their relation to the scattering matrix}},\ }\href
  {https://doi.org/10.1016/0550-3213(91)90366-6} {\bibfield  {journal}
  {\bibinfo  {journal} {Nucl. Phys. B}\ }\textbf {\bibinfo {volume} {354}},\
  \bibinfo {pages} {531} (\bibinfo {year} {1991})}\BibitemShut {NoStop}%
\bibitem [{\citenamefont {Koenig}\ \emph {et~al.}(2012)\citenamefont {Koenig},
  \citenamefont {Lee},\ and\ \citenamefont {Hammer}}]{Konig:2011ti}%
  \BibitemOpen
  \bibfield  {author} {\bibinfo {author} {\bibfnamefont {S.}~\bibnamefont
  {Koenig}}, \bibinfo {author} {\bibfnamefont {D.}~\bibnamefont {Lee}},\ and\
  \bibinfo {author} {\bibfnamefont {H.~W.}\ \bibnamefont {Hammer}},\ }\bibfield
   {title} {\bibinfo {title} {{Non-relativistic bound states in a finite
  volume}},\ }\href {https://doi.org/10.1016/j.aop.2011.12.015} {\bibfield
  {journal} {\bibinfo  {journal} {Annals Phys.}\ }\textbf {\bibinfo {volume}
  {327}},\ \bibinfo {pages} {1450} (\bibinfo {year} {2012})},\ \Eprint
  {https://arxiv.org/abs/1109.4577} {arXiv:1109.4577 [hep-lat]} \BibitemShut
  {NoStop}%
\bibitem [{\citenamefont {Briceño}\ \emph {et~al.}(2013)\citenamefont
  {Briceño}, \citenamefont {Davoudi}, \citenamefont {Luu},\ and\ \citenamefont
  {Savage}}]{Briceno:2013bda}%
  \BibitemOpen
  \bibfield  {author} {\bibinfo {author} {\bibfnamefont {R.~A.}\ \bibnamefont
  {Briceño}}, \bibinfo {author} {\bibfnamefont {Z.}~\bibnamefont {Davoudi}},
  \bibinfo {author} {\bibfnamefont {T.}~\bibnamefont {Luu}},\ and\ \bibinfo
  {author} {\bibfnamefont {M.~J.}\ \bibnamefont {Savage}},\ }\bibfield  {title}
  {\bibinfo {title} {{Two-nucleon systems in a finite volume. II. $^3S_1-^3D_1$
  coupled channels and the deuteron}},\ }\href
  {https://doi.org/10.1103/PhysRevD.88.114507} {\bibfield  {journal} {\bibinfo
  {journal} {Phys. Rev. D}\ }\textbf {\bibinfo {volume} {88}},\ \bibinfo
  {pages} {114507} (\bibinfo {year} {2013})},\ \Eprint
  {https://arxiv.org/abs/1309.3556} {arXiv:1309.3556 [hep-lat]} \BibitemShut
  {NoStop}%
\bibitem [{\citenamefont {Rao}\ and\ \citenamefont
  {Shrock}(1982)}]{Rao:1982gt}%
  \BibitemOpen
  \bibfield  {author} {\bibinfo {author} {\bibfnamefont {S.}~\bibnamefont
  {Rao}}\ and\ \bibinfo {author} {\bibfnamefont {R.}~\bibnamefont {Shrock}},\
  }\bibfield  {title} {\bibinfo {title} {{$n \leftrightarrow \bar{n}$
  Transition Operators and Their Matrix Elements in the {MIT} Bag Model}},\
  }\href {https://doi.org/10.1016/0370-2693(82)90333-1} {\bibfield  {journal}
  {\bibinfo  {journal} {Phys. Lett. B}\ }\textbf {\bibinfo {volume} {116}},\
  \bibinfo {pages} {238} (\bibinfo {year} {1982})}\BibitemShut {NoStop}%
\bibitem [{\citenamefont {Detmold}\ and\ \citenamefont
  {Orginos}(2013)}]{Detmold:2012eu}%
  \BibitemOpen
  \bibfield  {author} {\bibinfo {author} {\bibfnamefont {W.}~\bibnamefont
  {Detmold}}\ and\ \bibinfo {author} {\bibfnamefont {K.}~\bibnamefont
  {Orginos}},\ }\bibfield  {title} {\bibinfo {title} {{Nuclear correlation
  functions in lattice QCD}},\ }\href
  {https://doi.org/10.1103/PhysRevD.87.114512} {\bibfield  {journal} {\bibinfo
  {journal} {Phys. Rev. D}\ }\textbf {\bibinfo {volume} {87}},\ \bibinfo
  {pages} {114512} (\bibinfo {year} {2013})},\ \Eprint
  {https://arxiv.org/abs/1207.1452} {arXiv:1207.1452 [hep-lat]} \BibitemShut
  {NoStop}%
\bibitem [{\citenamefont {Buchoff}\ and\ \citenamefont
  {Wagman}(2016)}]{Buchoff:2015qwa}%
  \BibitemOpen
  \bibfield  {author} {\bibinfo {author} {\bibfnamefont {M.~I.}\ \bibnamefont
  {Buchoff}}\ and\ \bibinfo {author} {\bibfnamefont {M.}~\bibnamefont
  {Wagman}},\ }\bibfield  {title} {\bibinfo {title} {{Perturbative
  Renormalization of Neutron-Antineutron Operators}},\ }\href
  {https://doi.org/10.1103/PhysRevD.93.016005} {\bibfield  {journal} {\bibinfo
  {journal} {Phys. Rev. D}\ }\textbf {\bibinfo {volume} {93}},\ \bibinfo
  {pages} {016005} (\bibinfo {year} {2016})},\ \bibinfo {note} {[Erratum:
  Phys.Rev.D 98, 079901 (2018)]},\ \Eprint {https://arxiv.org/abs/1506.00647}
  {arXiv:1506.00647 [hep-ph]} \BibitemShut {NoStop}%
\bibitem [{\citenamefont {L{\"u}scher}\ and\ \citenamefont
  {Weisz}(1985)}]{Luscher:1984xn}%
  \BibitemOpen
  \bibfield  {author} {\bibinfo {author} {\bibfnamefont {M.}~\bibnamefont
  {L{\"u}scher}}\ and\ \bibinfo {author} {\bibfnamefont {P.}~\bibnamefont
  {Weisz}},\ }\bibfield  {title} {\bibinfo {title} {{On-shell improved lattice
  gauge theories}},\ }\href {https://doi.org/10.1007/BF01205792} {\bibfield
  {journal} {\bibinfo  {journal} {Commun. Math. Phys.}\ }\textbf {\bibinfo
  {volume} {98}},\ \bibinfo {pages} {433} (\bibinfo {year} {1985})},\ \bibinfo
  {note} {[Erratum: Commun.Math.Phys. 98, 433 (1985)]}\BibitemShut {NoStop}%
\bibitem [{\citenamefont {Morningstar}\ and\ \citenamefont
  {Peardon}(2004)}]{Morningstar:2003gk}%
  \BibitemOpen
  \bibfield  {author} {\bibinfo {author} {\bibfnamefont {C.}~\bibnamefont
  {Morningstar}}\ and\ \bibinfo {author} {\bibfnamefont {M.~J.}\ \bibnamefont
  {Peardon}},\ }\bibfield  {title} {\bibinfo {title} {{Analytic smearing of
  SU(3) link variables in lattice QCD}},\ }\href
  {https://doi.org/10.1103/PhysRevD.69.054501} {\bibfield  {journal} {\bibinfo
  {journal} {Phys. Rev. D}\ }\textbf {\bibinfo {volume} {69}},\ \bibinfo
  {pages} {054501} (\bibinfo {year} {2004})},\ \Eprint
  {https://arxiv.org/abs/hep-lat/0311018} {arXiv:hep-lat/0311018} \BibitemShut
  {NoStop}%
\bibitem [{\citenamefont {Wilson}(1974)}]{Wilson:1974sk}%
  \BibitemOpen
  \bibfield  {author} {\bibinfo {author} {\bibfnamefont {K.~G.}\ \bibnamefont
  {Wilson}},\ }\bibfield  {title} {\bibinfo {title} {{Confinement of Quarks}},\
  }\href {https://doi.org/10.1103/PhysRevD.10.2445} {\bibfield  {journal}
  {\bibinfo  {journal} {Phys. Rev. D}\ }\textbf {\bibinfo {volume} {10}},\
  \bibinfo {pages} {2445} (\bibinfo {year} {1974})}\BibitemShut {NoStop}%
\bibitem [{\citenamefont {Sheikholeslami}\ and\ \citenamefont
  {Wohlert}(1985)}]{Sheikholeslami:1985ij}%
  \BibitemOpen
  \bibfield  {author} {\bibinfo {author} {\bibfnamefont {B.}~\bibnamefont
  {Sheikholeslami}}\ and\ \bibinfo {author} {\bibfnamefont {R.}~\bibnamefont
  {Wohlert}},\ }\bibfield  {title} {\bibinfo {title} {{Improved Continuum Limit
  Lattice Action for QCD with Wilson Fermions}},\ }\href
  {https://doi.org/10.1016/0550-3213(85)90002-1} {\bibfield  {journal}
  {\bibinfo  {journal} {Nucl. Phys. B}\ }\textbf {\bibinfo {volume} {259}},\
  \bibinfo {pages} {572} (\bibinfo {year} {1985})}\BibitemShut {NoStop}%
\bibitem [{\citenamefont {G{\"u}sken}\ \emph {et~al.}(1989)\citenamefont
  {G{\"u}sken}, \citenamefont {L{\"o}w}, \citenamefont {M{\"u}tter},
  \citenamefont {Sommer}, \citenamefont {Patel},\ and\ \citenamefont
  {Schilling}}]{Gusken:1989ad}%
  \BibitemOpen
  \bibfield  {author} {\bibinfo {author} {\bibfnamefont {S.}~\bibnamefont
  {G{\"u}sken}}, \bibinfo {author} {\bibfnamefont {U.}~\bibnamefont {L{\"o}w}},
  \bibinfo {author} {\bibfnamefont {K.}~\bibnamefont {M{\"u}tter}}, \bibinfo
  {author} {\bibfnamefont {R.}~\bibnamefont {Sommer}}, \bibinfo {author}
  {\bibfnamefont {A.}~\bibnamefont {Patel}},\ and\ \bibinfo {author}
  {\bibfnamefont {K.}~\bibnamefont {Schilling}},\ }\bibfield  {title} {\bibinfo
  {title} {{Nonsinglet Axial Vector Couplings of the Baryon Octet in Lattice
  {QCD}}},\ }\href {https://doi.org/10.1016/S0370-2693(89)80034-6} {\bibfield
  {journal} {\bibinfo  {journal} {Phys. Lett. B}\ }\textbf {\bibinfo {volume}
  {227}},\ \bibinfo {pages} {266} (\bibinfo {year} {1989})}\BibitemShut
  {NoStop}%
\bibitem [{\citenamefont {G{\"u}sken}(1990)}]{Gusken:1989qx}%
  \BibitemOpen
  \bibfield  {author} {\bibinfo {author} {\bibfnamefont {S.}~\bibnamefont
  {G{\"u}sken}},\ }\bibfield  {title} {\bibinfo {title} {{A Study of smearing
  techniques for hadron correlation functions}},\ }\href
  {https://doi.org/10.1016/0920-5632(90)90273-W} {\bibfield  {journal}
  {\bibinfo  {journal} {Nucl. Phys. B Proc. Suppl.}\ }\textbf {\bibinfo
  {volume} {17}},\ \bibinfo {pages} {361} (\bibinfo {year} {1990})}\BibitemShut
  {NoStop}%
\bibitem [{\citenamefont {Chang}\ \emph {et~al.}(2015)\citenamefont {Chang},
  \citenamefont {Detmold}, \citenamefont {Orginos}, \citenamefont
  {Parre\~{n}o}, \citenamefont {Savage}, \citenamefont {Tiburzi},\ and\
  \citenamefont {Beane}}]{Chang:2015qxa}%
  \BibitemOpen
  \bibfield  {author} {\bibinfo {author} {\bibfnamefont {E.}~\bibnamefont
  {Chang}}, \bibinfo {author} {\bibfnamefont {W.}~\bibnamefont {Detmold}},
  \bibinfo {author} {\bibfnamefont {K.}~\bibnamefont {Orginos}}, \bibinfo
  {author} {\bibfnamefont {A.}~\bibnamefont {Parre\~{n}o}}, \bibinfo {author}
  {\bibfnamefont {M.~J.}\ \bibnamefont {Savage}}, \bibinfo {author}
  {\bibfnamefont {B.~C.}\ \bibnamefont {Tiburzi}},\ and\ \bibinfo {author}
  {\bibfnamefont {S.~R.}\ \bibnamefont {Beane}} (\bibinfo {collaboration}
  {NPLQCD}),\ }\bibfield  {title} {\bibinfo {title} {{Magnetic structure of
  light nuclei from lattice QCD}},\ }\href
  {https://doi.org/10.1103/PhysRevD.92.114502} {\bibfield  {journal} {\bibinfo
  {journal} {Phys. Rev. D}\ }\textbf {\bibinfo {volume} {92}},\ \bibinfo
  {pages} {114502} (\bibinfo {year} {2015})},\ \Eprint
  {https://arxiv.org/abs/1506.05518} {arXiv:1506.05518 [hep-lat]} \BibitemShut
  {NoStop}%
\bibitem [{\citenamefont {Mandula}\ \emph
  {et~al.}(1983{\natexlab{a}})\citenamefont {Mandula}, \citenamefont {Zweig},\
  and\ \citenamefont {Govaerts}}]{Mandula:1982us}%
  \BibitemOpen
  \bibfield  {author} {\bibinfo {author} {\bibfnamefont {J.~E.}\ \bibnamefont
  {Mandula}}, \bibinfo {author} {\bibfnamefont {G.}~\bibnamefont {Zweig}},\
  and\ \bibinfo {author} {\bibfnamefont {J.}~\bibnamefont {Govaerts}},\
  }\bibfield  {title} {\bibinfo {title} {{Covariant lattice glueball fields}},\
  }\href {https://doi.org/10.1016/0550-3213(83)90400-5} {\bibfield  {journal}
  {\bibinfo  {journal} {Nucl. Phys. B}\ }\textbf {\bibinfo {volume} {228}},\
  \bibinfo {pages} {109} (\bibinfo {year} {1983}{\natexlab{a}})}\BibitemShut
  {NoStop}%
\bibitem [{\citenamefont {Mandula}\ \emph
  {et~al.}(1983{\natexlab{b}})\citenamefont {Mandula}, \citenamefont {Zweig},\
  and\ \citenamefont {Govaerts}}]{Mandula:1983ut}%
  \BibitemOpen
  \bibfield  {author} {\bibinfo {author} {\bibfnamefont {J.~E.}\ \bibnamefont
  {Mandula}}, \bibinfo {author} {\bibfnamefont {G.}~\bibnamefont {Zweig}},\
  and\ \bibinfo {author} {\bibfnamefont {J.}~\bibnamefont {Govaerts}},\
  }\bibfield  {title} {\bibinfo {title} {{Representations of the Rotation
  Reflection Symmetry Group of the Four-dimensional Cubic Lattice}},\ }\href
  {https://doi.org/10.1016/0550-3213(83)90399-1} {\bibfield  {journal}
  {\bibinfo  {journal} {Nucl. Phys. B}\ }\textbf {\bibinfo {volume} {228}},\
  \bibinfo {pages} {91} (\bibinfo {year} {1983}{\natexlab{b}})}\BibitemShut
  {NoStop}%
\bibitem [{\citenamefont {Brice\~{n}o}\ \emph {et~al.}(2013)\citenamefont
  {Brice\~{n}o}, \citenamefont {Davoudi},\ and\ \citenamefont
  {Luu}}]{Briceno:2013lba}%
  \BibitemOpen
  \bibfield  {author} {\bibinfo {author} {\bibfnamefont {R.~A.}\ \bibnamefont
  {Brice\~{n}o}}, \bibinfo {author} {\bibfnamefont {Z.}~\bibnamefont
  {Davoudi}},\ and\ \bibinfo {author} {\bibfnamefont {T.~C.}\ \bibnamefont
  {Luu}},\ }\bibfield  {title} {\bibinfo {title} {{Two-Nucleon Systems in a
  Finite Volume: (I) Quantization Conditions}},\ }\href
  {https://doi.org/10.1103/PhysRevD.88.034502} {\bibfield  {journal} {\bibinfo
  {journal} {Phys. Rev. D}\ }\textbf {\bibinfo {volume} {88}},\ \bibinfo
  {pages} {034502} (\bibinfo {year} {2013})},\ \Eprint
  {https://arxiv.org/abs/1305.4903} {arXiv:1305.4903 [hep-lat]} \BibitemShut
  {NoStop}%
\bibitem [{\citenamefont {Edwards}\ and\ \citenamefont
  {Jo\'o}(2005)}]{Edwards:2004sx}%
  \BibitemOpen
  \bibfield  {author} {\bibinfo {author} {\bibfnamefont {R.~G.}\ \bibnamefont
  {Edwards}}\ and\ \bibinfo {author} {\bibfnamefont {B.}~\bibnamefont {Jo\'o}}
  (\bibinfo {collaboration} {SciDAC, LHPC, UKQCD}),\ }\bibfield  {title}
  {\bibinfo {title} {{The Chroma software system for lattice QCD}},\ }\href
  {https://doi.org/10.1016/j.nuclphysbps.2004.11.254} {\bibfield  {journal}
  {\bibinfo  {journal} {Nucl. Phys. B Proc. Suppl.}\ }\textbf {\bibinfo
  {volume} {140}},\ \bibinfo {pages} {832} (\bibinfo {year} {2005})},\ \Eprint
  {https://arxiv.org/abs/hep-lat/0409003} {arXiv:hep-lat/0409003} \BibitemShut
  {NoStop}%
\bibitem [{\citenamefont {Pochinsky}()}]{qlua}%
  \BibitemOpen
  \bibfield  {author} {\bibinfo {author} {\bibfnamefont {A.}~\bibnamefont
  {Pochinsky}},\ }\href@noop {} {\bibinfo {title} {Qlua}},\ \bibinfo
  {howpublished} {\url{https://usqcd.lns.mit.edu/qlua}}\BibitemShut {NoStop}%
\bibitem [{\citenamefont {Clark}\ \emph {et~al.}(2010)\citenamefont {Clark},
  \citenamefont {Babich}, \citenamefont {Barros}, \citenamefont {Brower},\ and\
  \citenamefont {Rebbi}}]{Clark:2009wm}%
  \BibitemOpen
  \bibfield  {author} {\bibinfo {author} {\bibfnamefont {M.~A.}\ \bibnamefont
  {Clark}}, \bibinfo {author} {\bibfnamefont {R.}~\bibnamefont {Babich}},
  \bibinfo {author} {\bibfnamefont {K.}~\bibnamefont {Barros}}, \bibinfo
  {author} {\bibfnamefont {R.~C.}\ \bibnamefont {Brower}},\ and\ \bibinfo
  {author} {\bibfnamefont {C.}~\bibnamefont {Rebbi}} (\bibinfo {collaboration}
  {QUDA}),\ }\bibfield  {title} {\bibinfo {title} {{Solving Lattice QCD systems
  of equations using mixed precision solvers on GPUs}},\ }\href
  {https://doi.org/10.1016/j.cpc.2010.05.002} {\bibfield  {journal} {\bibinfo
  {journal} {Comput. Phys. Commun.}\ }\textbf {\bibinfo {volume} {181}},\
  \bibinfo {pages} {1517} (\bibinfo {year} {2010})},\ \Eprint
  {https://arxiv.org/abs/0911.3191} {arXiv:0911.3191 [hep-lat]} \BibitemShut
  {NoStop}%
\bibitem [{\citenamefont {Babich}\ \emph {et~al.}(2011)\citenamefont {Babich},
  \citenamefont {Clark}, \citenamefont {Joo}, \citenamefont {Shi},
  \citenamefont {Brower},\ and\ \citenamefont {Gottlieb}}]{Babich:2011np}%
  \BibitemOpen
  \bibfield  {author} {\bibinfo {author} {\bibfnamefont {R.}~\bibnamefont
  {Babich}}, \bibinfo {author} {\bibfnamefont {M.~A.}\ \bibnamefont {Clark}},
  \bibinfo {author} {\bibfnamefont {B.}~\bibnamefont {Joo}}, \bibinfo {author}
  {\bibfnamefont {G.}~\bibnamefont {Shi}}, \bibinfo {author} {\bibfnamefont
  {R.~C.}\ \bibnamefont {Brower}},\ and\ \bibinfo {author} {\bibfnamefont
  {S.}~\bibnamefont {Gottlieb}} (\bibinfo {collaboration} {QUDA}),\ }\bibfield
  {title} {\bibinfo {title} {{Scaling lattice QCD beyond 100 GPUs}},\ }in\
  \href {https://doi.org/10.1145/2063384.2063478} {\emph {\bibinfo {booktitle}
  {{International Conference for High Performance Computing, Networking,
  Storage and Analysis}}}}\ (\bibinfo {year} {2011})\ \Eprint
  {https://arxiv.org/abs/1109.2935} {arXiv:1109.2935 [hep-lat]} \BibitemShut
  {NoStop}%
\bibitem [{\citenamefont {Clark}\ \emph {et~al.}(2016)\citenamefont {Clark},
  \citenamefont {Jo\'o}, \citenamefont {Strelchenko}, \citenamefont {Cheng},
  \citenamefont {Gambhir},\ and\ \citenamefont {Brower}}]{Clark:2016rdz}%
  \BibitemOpen
  \bibfield  {author} {\bibinfo {author} {\bibfnamefont {M.~A.}\ \bibnamefont
  {Clark}}, \bibinfo {author} {\bibfnamefont {B.}~\bibnamefont {Jo\'o}},
  \bibinfo {author} {\bibfnamefont {A.}~\bibnamefont {Strelchenko}}, \bibinfo
  {author} {\bibfnamefont {M.}~\bibnamefont {Cheng}}, \bibinfo {author}
  {\bibfnamefont {A.}~\bibnamefont {Gambhir}},\ and\ \bibinfo {author}
  {\bibfnamefont {R.~C.}\ \bibnamefont {Brower}} (\bibinfo {collaboration}
  {QUDA}),\ }\bibfield  {title} {\bibinfo {title} {{Accelerating lattice QCD
  multigrid on GPUs using fine-grained parallelization}},\ }in\ \href
  {https://doi.org/10.5555/3014904.3014995} {\emph {\bibinfo {booktitle}
  {{International Conference for High Performance Computing, Networking,
  Storage and Analysis}}}}\ (\bibinfo {year} {2016})\ \Eprint
  {https://arxiv.org/abs/1612.07873} {arXiv:1612.07873 [hep-lat]} \BibitemShut
  {NoStop}%
\bibitem [{\citenamefont {Baghdadi}\ \emph {et~al.}(2020)\citenamefont
  {Baghdadi}, \citenamefont {Debbagh}, \citenamefont {Abdous}, \citenamefont
  {Benhamida}, \citenamefont {Renda}, \citenamefont {Frankle}, \citenamefont
  {Carbin},\ and\ \citenamefont {S.}}]{tiramisu}%
  \BibitemOpen
  \bibfield  {author} {\bibinfo {author} {\bibfnamefont {R.}~\bibnamefont
  {Baghdadi}}, \bibinfo {author} {\bibfnamefont {A.~N.}\ \bibnamefont
  {Debbagh}}, \bibinfo {author} {\bibfnamefont {K.}~\bibnamefont {Abdous}},
  \bibinfo {author} {\bibfnamefont {F.~Z.}\ \bibnamefont {Benhamida}}, \bibinfo
  {author} {\bibfnamefont {A.}~\bibnamefont {Renda}}, \bibinfo {author}
  {\bibfnamefont {J.~E.}\ \bibnamefont {Frankle}}, \bibinfo {author}
  {\bibfnamefont {M.}~\bibnamefont {Carbin}},\ and\ \bibinfo {author}
  {\bibfnamefont {A.}~\bibnamefont {S.}},\ }\href@noop {} {\bibinfo {title}
  {{Tiramisu: A polyhedral compiler for dense and sparse deep learning}}}
  (\bibinfo {year} {2020}),\ \Eprint {https://arxiv.org/abs/2005.04091}
  {arXiv:2005.04091 [cs.DC]} \BibitemShut {NoStop}%
\bibitem [{\citenamefont {Harris}\ \emph {et~al.}(2020)\citenamefont {Harris},
  \citenamefont {Millman}, \citenamefont {van~der Walt}, \citenamefont
  {Gommers}, \citenamefont {Virtanen}, \citenamefont {Cournapeau},
  \citenamefont {Wieser}, \citenamefont {Taylor}, \citenamefont {Berg},
  \citenamefont {Smith}, \citenamefont {Kern}, \citenamefont {Picus},
  \citenamefont {Hoyer}, \citenamefont {van Kerkwijk}, \citenamefont {Brett},
  \citenamefont {Haldane}, \citenamefont {del R{\'{i}}o}, \citenamefont
  {Wiebe}, \citenamefont {Peterson}, \citenamefont {G{\'{e}}rard-Marchant},
  \citenamefont {Sheppard}, \citenamefont {Reddy}, \citenamefont {Weckesser},
  \citenamefont {Abbasi}, \citenamefont {Gohlke},\ and\ \citenamefont
  {Oliphant}}]{harris2020array}%
  \BibitemOpen
  \bibfield  {author} {\bibinfo {author} {\bibfnamefont {C.~R.}\ \bibnamefont
  {Harris}}, \bibinfo {author} {\bibfnamefont {K.~J.}\ \bibnamefont {Millman}},
  \bibinfo {author} {\bibfnamefont {S.~J.}\ \bibnamefont {van~der Walt}},
  \bibinfo {author} {\bibfnamefont {R.}~\bibnamefont {Gommers}}, \bibinfo
  {author} {\bibfnamefont {P.}~\bibnamefont {Virtanen}}, \bibinfo {author}
  {\bibfnamefont {D.}~\bibnamefont {Cournapeau}}, \bibinfo {author}
  {\bibfnamefont {E.}~\bibnamefont {Wieser}}, \bibinfo {author} {\bibfnamefont
  {J.}~\bibnamefont {Taylor}}, \bibinfo {author} {\bibfnamefont
  {S.}~\bibnamefont {Berg}}, \bibinfo {author} {\bibfnamefont {N.~J.}\
  \bibnamefont {Smith}}, \bibinfo {author} {\bibfnamefont {R.}~\bibnamefont
  {Kern}}, \bibinfo {author} {\bibfnamefont {M.}~\bibnamefont {Picus}},
  \bibinfo {author} {\bibfnamefont {S.}~\bibnamefont {Hoyer}}, \bibinfo
  {author} {\bibfnamefont {M.~H.}\ \bibnamefont {van Kerkwijk}}, \bibinfo
  {author} {\bibfnamefont {M.}~\bibnamefont {Brett}}, \bibinfo {author}
  {\bibfnamefont {A.}~\bibnamefont {Haldane}}, \bibinfo {author} {\bibfnamefont
  {J.~F.}\ \bibnamefont {del R{\'{i}}o}}, \bibinfo {author} {\bibfnamefont
  {M.}~\bibnamefont {Wiebe}}, \bibinfo {author} {\bibfnamefont
  {P.}~\bibnamefont {Peterson}}, \bibinfo {author} {\bibfnamefont
  {P.}~\bibnamefont {G{\'{e}}rard-Marchant}}, \bibinfo {author} {\bibfnamefont
  {K.}~\bibnamefont {Sheppard}}, \bibinfo {author} {\bibfnamefont
  {T.}~\bibnamefont {Reddy}}, \bibinfo {author} {\bibfnamefont
  {W.}~\bibnamefont {Weckesser}}, \bibinfo {author} {\bibfnamefont
  {H.}~\bibnamefont {Abbasi}}, \bibinfo {author} {\bibfnamefont
  {C.}~\bibnamefont {Gohlke}},\ and\ \bibinfo {author} {\bibfnamefont {T.~E.}\
  \bibnamefont {Oliphant}},\ }\bibfield  {title} {\bibinfo {title} {Array
  programming with {NumPy}},\ }\href
  {https://doi.org/10.1038/s41586-020-2649-2} {\bibfield  {journal} {\bibinfo
  {journal} {Nature}\ }\textbf {\bibinfo {volume} {585}},\ \bibinfo {pages}
  {357} (\bibinfo {year} {2020})}\BibitemShut {NoStop}%
\bibitem [{\citenamefont {Bezanson}\ \emph {et~al.}(2017)\citenamefont
  {Bezanson}, \citenamefont {Edelman}, \citenamefont {Karpinski},\ and\
  \citenamefont {Shah}}]{Julia-2017}%
  \BibitemOpen
  \bibfield  {author} {\bibinfo {author} {\bibfnamefont {J.}~\bibnamefont
  {Bezanson}}, \bibinfo {author} {\bibfnamefont {A.}~\bibnamefont {Edelman}},
  \bibinfo {author} {\bibfnamefont {S.}~\bibnamefont {Karpinski}},\ and\
  \bibinfo {author} {\bibfnamefont {V.~B.}\ \bibnamefont {Shah}},\ }\bibfield
  {title} {\bibinfo {title} {Julia: A fresh approach to numerical computing},\
  }\href {https://doi.org/10.1137/141000671} {\bibfield  {journal} {\bibinfo
  {journal} {SIAM {R}eview}\ }\textbf {\bibinfo {volume} {59}},\ \bibinfo
  {pages} {65} (\bibinfo {year} {2017})}\BibitemShut {NoStop}%
\bibitem [{\citenamefont {Mogensen}\ and\ \citenamefont
  {Riseth}(2018)}]{mogensen2018optim}%
  \BibitemOpen
  \bibfield  {author} {\bibinfo {author} {\bibfnamefont {P.~K.}\ \bibnamefont
  {Mogensen}}\ and\ \bibinfo {author} {\bibfnamefont {A.~N.}\ \bibnamefont
  {Riseth}},\ }\bibfield  {title} {\bibinfo {title} {Optim: A mathematical
  optimization package for {Julia}},\ }\href
  {https://doi.org/10.21105/joss.00615} {\bibfield  {journal} {\bibinfo
  {journal} {Journal of Open Source Software}\ }\textbf {\bibinfo {volume}
  {3}},\ \bibinfo {pages} {615} (\bibinfo {year} {2018})}\BibitemShut {NoStop}%
\bibitem [{\citenamefont {{Wolfram Research Inc.}}()}]{Mathematica}%
  \BibitemOpen
  \bibfield  {author} {\bibinfo {author} {\bibnamefont {{Wolfram Research
  Inc.}}},\ }\href@noop {} {\bibinfo {title} {Mathematica, {V}ersion 12.2}},\
  \bibinfo {note} {{\tt https://www.wolfram.com/mathematica}}\BibitemShut
  {NoStop}%
\bibitem [{\citenamefont {Hunter}(2007)}]{Hunter:2007}%
  \BibitemOpen
  \bibfield  {author} {\bibinfo {author} {\bibfnamefont {J.~D.}\ \bibnamefont
  {Hunter}},\ }\bibfield  {title} {\bibinfo {title} {Matplotlib: A 2d graphics
  environment},\ }\href {https://doi.org/10.1109/MCSE.2007.55} {\bibfield
  {journal} {\bibinfo  {journal} {Computing in Science \& Engineering}\
  }\textbf {\bibinfo {volume} {9}},\ \bibinfo {pages} {90} (\bibinfo {year}
  {2007})}\BibitemShut {NoStop}%
\end{thebibliography}%
